\documentclass[twoside,11pt]{article}
\usepackage{booktabs}       
\usepackage{amsfonts}       
\usepackage{nicefrac}       
\usepackage{microtype}      
\usepackage{verbatim}
\usepackage{subfig}
\usepackage{xcolor}
\usepackage{tabstackengine}
\usepackage{setspace}
\usepackage{amsthm}
\usepackage{amsmath}
\usepackage{graphicx}
\usepackage{cancel}
\usepackage{float}
\usepackage{setspace}
\usepackage{epstopdf}
\usepackage{geometry}
\usepackage{graphics}
\usepackage{booktabs}
\restylefloat{table}
\usepackage{tikz}
\usetikzlibrary{fit,positioning,arrows,automata}
\usetikzlibrary{matrix}
\usepackage{bm}
\usepackage{tikz-cd}
\usepackage{tcolorbox}
\usepackage{mathtools}
\usepackage{algorithm}
\usepackage{algpseudocode}
\usepackage{quiver}
\usepackage{enumitem}

\usepackage[abbrvbib, preprint]{jmlr2e}

\jmlrheading{1}{2022}{1-48}{4/00}{10/00}{meila00a}{Sudipto Banerjee}

\ShortHeadings{Dynamic Bayesian Learning}{Banerjee}
\firstpageno{1}

\begin{document}

\title{Dynamic Bayesian Learning for Spatiotemporal Mechanistic Models}

\author{\name Sudipto Banerjee \email sudipto@ucla.edu \\
       \addr Department of Biostatistics\\
       University of California, Los Angeles\\
       Los Angeles, CA 90025, USA
       \AND
       \name Xiang Chen \email xiangchen@ucla.edu \\
       \addr Department of Biostatistics\\
       University of California, Los Angeles\\
       Los Angeles, CA 90025, USA
       \AND
       \name Ian Frankenburg \email ian.frankenburg@ucla.edu \\
       \addr Department of Biostatistics\\
       University of California, Los Angeles\\
       Los Angeles, CA 90025, USA
       \AND
       \name Daniel Zhou \email hunanzhou@g.ucla.edu\\
       \addr Department of Biostatistics\\
       University of California, Los Angeles\\
       Los Angeles, CA 90025, USA
       }

\editor{Debdeep Pati}

\maketitle

\begin{abstract}
We develop an approach for Bayesian learning of spatiotemporal dynamical mechanistic models. Such learning consists of statistical emulation of the mechanistic system that can efficiently interpolate the output of the system from arbitrary inputs. The emulated learner can then be used to train the system from noisy data achieved by melding information from observed data with the emulated mechanistic system. This joint melding of mechanistic systems employ hierarchical state-space models with Gaussian process regression. Assuming the dynamical system is controlled by a finite collection of inputs, Gaussian process regression learns the effect of these parameters through a number of training runs, driving the stochastic innovations of the spatiotemporal state-space component. This enables efficient modeling of the dynamics over space and time. This article details exact inference with analytically accessible posterior distributions in hierarchical matrix-variate Normal and Wishart models in designing the emulator. This step obviates expensive iterative algorithms such as Markov chain Monte Carlo or variational approximations. We also show how emulation is applicable to large-scale emulation by designing a dynamic Bayesian transfer learning framework. Inference on $\bm \eta$ proceeds using Markov chain Monte Carlo as a post-emulation step using the emulator as a regression component. We demonstrate this framework through solving inverse problems arising in the analysis of ordinary and partial nonlinear differential equations and, in addition, to a black-box computer model generating spatiotemporal dynamics across a graphical model.
\end{abstract}


\section{Introduction}
Probabilistic learning of spatial-temporal processes governed by physical or mechanistic systems is of interest in a variety of scientific disciplines. Such models are sometimes represented through a system of equations derived from physical or scientific principles, or they may be arbitrarily complex computer programs that simulate a phenomenon. These models are broadly referred to as \emph{computer models}\footnote{In the rest of the article, we occasionally refer to the mechanistic system which we seek to model, learn, and estimate parameters from field data as a {computer model} to be consistent with terminology in statistics for referring to deterministic models.} and, at least in the scope of this paper, are deterministic in the sense that rerunning the model with a specific set of inputs always produce identical value of the outputs. In what is considered a seminal manuscript in statistical science, \cite{Sacks1989} describes a framework for modeling the output as a realization of a stochastic process and laying the statistical foundations for designing computer experiments for efficient prediction. This field has blossomed into a substantial area within machine learning and statistical science with the incorporation of ideas from signal processing, dynamical systems, inverse problems, functional data analysis, non-parametric models, Gaussian processes, spatial statistics and several other fields \citep[while a comprehensive review is not the aim of this article, we refer to][and references therein for diverse perspectives]{Santner2019, Fang2005, kangDeng2020wires}. 

Analyzing data from computer experiments refer to interpolating or predicting the output at new inputs after training the model using runs of the computer model without assuming mathematical tractability of the computer model. This is referred to as \emph{emulation} \citep[see][with the latter offering an excellent perspective of relative recent literature]{Sacks1989, conti2010bayesian} and comprises the construction of a statistical model that mimics the behavior of the computer model. Rather than assuming an explicit functional relationship between the inputs and outputs, a Gaussian process is used as a prior on the unknown function \citep[see, e.g.,][]{OHagan1992, Haylock1996, GP_ML} and the outputs are assumed to be a realization of this process.  

A second problem of interest is to learn about the parameters in the mechanistic system from field observations while also accounting for the information available from runs of the computer experiment. This is related to the problem of \emph{calibration} using field data \citep[see, e.g.,][for calibration using field observations in diverse contexts]{Kennedy2001, Oakley2004, Higdon2004, Bayarri2007, Higdon2008, Bayarri2009} and can be regarded as an inverse problem seeking optimal values of unknown parameters to be learned from observational field data. We depart somewhat from the paradigm of traditional calibration, where an ``optimal'' (perhaps unknown) value of the input is assumed. Instead pursue \emph{melding} \citep{Poole2000, GelEtAl2004} and synthesize information from mechanistic systems with field observations. Emulation and calibration becomes challenging when the computer model or posited functional relationship is complex in nature or expensive to compute and any practicable learning framework must account for the scale of the problem.            

\begin{figure}[t]
    \centering
    \includegraphics[scale=.6]{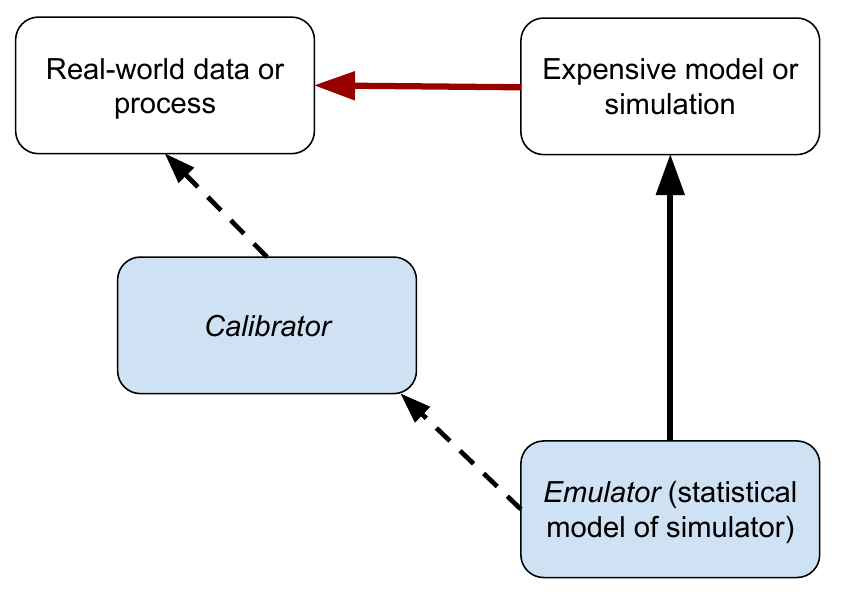}
    \caption{A graphical illustration of the emulation/calibration paradigm. A statistical model emulates the mechanistic system from experimental runs to enable efficient calibration based on observed data. We build the emulator and calibrator modules through a combination of Gaussian and non-Gaussian state-space methods and Gaussian process regression.}
    \label{fig:paradigm}
    \vspace{-0.25in}
\end{figure}

We focus upon emulating and inferring on spatial-temporal mechanistic systems using a Bayesian hierarchical modeling framework. Figure~\ref{fig:paradigm} depicts emulation and calibration in our melding context. A hierarchical description of a physical process consists of a layered approach, whereby simpler conditional dependence structures specify complex relationships. Probabilistic learning proceeds from the posterior distribution given by  
\begin{equation}\label{eq: generic_paradigm}
    \begin{split}
    \left[\text{process, parameters}\mid \text{data}\right]&\propto [\text{data}\mid\text{process}, \text{parameters}]\\
    &\qquad\times [\text{process} \mid \text{parameters}] \times [\text{parameters}]\;.
    \end{split}
\end{equation}
At the top level, the probability model specifies the distribution of the data conditional on the physical process and any other parameters needed to describe the data generating mechanism. The next level represents the underlying mechanistic system as a realization of a stochastic process capturing physical or mechanistic knowledge \citep[as promulgated by][]{Sacks1989}. The last level models uncertainty about parameters. 

Probabilistic machine learning using \eqref{eq: generic_paradigm} adopts Bayesian inference of all unknown quantities given the data available to the modeler. Bayesian inference for deterministic systems has been developed and explored in settings that resemble computer experiment settings. Examples, by no means exhaustive, include Bayesian melding developed by \cite{Poole2000} that pursues inference from deterministic simulation methods and has witnessed substantial use in climate modeling \citep[see][for one application in forecasting problems]{GelEtAl2004}, Bayesian data assimilation \citep{WikleBerliner2007} and Bayesian state space models formed by finite difference approximations of differential equations \citep[see, e.g.,][for applications encompassing ecology, climate and industrial hygiene]{wikle2003hierarchical, Stroud2010, WikleHooten2010, AbdallaEtAl2020}. 

In this manuscript we specifically focus upon spatial-temporal mechanistic systems and builds upon a rapidly evolving literature in Bayesian learning for computer models \citep[see][and references therein]{LiuWest2009, Farah2014, conti2010bayesian, gu2016parallel, gu2019robustgasp, gu2020generalized, gu2022gaussian, gu2022robustcalibration}. Notably, \cite{gu2019robustgasp} discusses software development and fast implementation of the methodology in \cite{gu2016parallel} using experiments and field observations for vector-valued outputs while also focusing upon accelerating computations. Other work on combining Gaussian processes and state-space models may be found in the machine learning literature \citep[see, e.g.,][]{Turner2010, Eleftheriadis2017}, where the process is used to model the transition function rather than stochastic innovations.  

 Following \cite{conti2010bayesian} and \cite{gu2020generalized}, we use matrix-variate distributions with rows and columns corresponding to inputs and locations, but we incorporate temporal emulation over discrete epochs leading to matrix-variate Bayesian dynamic models. In this regard, we differ from \cite{gu2020generalized} who use continuous time Kalman filters. While sacrificing some richness in statistical inference, we aim to harness exact analytically tractable distribution theory to generate exact posterior samples without resorting to iterative algorithms such as Markov chain Monte Carlo \citep[MCMC,][]{Robert2005, Gamerman2006book}, Variational Bayes \citep{renBanerjeeEtAl2011csda, bleiEtAl2017jasaReview} or Laplace approximations \citep{Rue2009}. We devise a matrix-variate Forward Filter Backward Sampling (FFBS) algorithm \citep{Carter1994, fruhwirth2006} to emulate dynamically evolving spatial random fields using exact sampling from matrix-normal and Wishart families. Closed form expressions for the necessary statistical distributions (also using hyper-T distributions) are derived as are expressions for log point-wise predictive or posterior predictive distributions that are used for model selection. 

For scaling up emulation, we adapt the FFBS algorithm to Bayesian transfer learning for emulating dynamically evolving mechanistic fields with very large amounts of spatial locations or time points. Here, too, we depart from much of the statistical literature on high-dimensional inference with computer models \citep{Higdon2008, gu2016parallel, gu2022gaussian} where dimension-reduction is achieved using low-dimensional projections embedded within the statistical models (such as orthogonal latent factor models). Our approach here resembles data partitioning strategies such as treed Gaussian processes \citep{Gramacy2007}, meta-kriging \citep{guhaniyogi2018meta, guhaniyogi2017divide} and predictive stacking over data subsets \citep{presicce2024bayesian}, but is simpler than in geostatistical settings because the dynamic emulator can be designed executed over regular spatial coordinates and time points to enable sequential updating of the FFBS algorithm over the subsets of the data. For probabilistic learning of the model inputs, we regress the observed field on the emulated field and the mechanistic system parameters are estimated with uncertainty quantification using MCMC but without requiring any additional emulation \citep[using ``modularized'' inference as described in][]{Bayarri2007,Bayarri2007tech,Bayarri2009}.

The structure of this article is as follows. Section~\ref{DLMs} develops a conjugate family of Bayesian state space models for space-time mechanistic systems emphasizing exact conjugate matrix-variate models to emulate the mechanistic system at arbitrary inputs. Section~\ref{sec:mod_compare} provides Bayesian model comparison metrics to compare different models for emulating the mechanistic system. Section~\ref{sec:bayestransferlearning} devises the Bayesian transfer learning approach for scaling up emulation over large spatial fields. Section~\ref{sec:calibration} develops probabilistic learning of mechanistic parameters using field-data and Section~\ref{sec:app_emulation}. Section~\ref{sec:app_calibration} provide a set of illustrations showing the applications of our dynamic framework using diverse mechanistic systems.   

\section{Bayesian State-Space Models for Mechanistic Systems}\label{DLMs}
We offer a brief review of Bayesian state-space models (SSMs) -- also known as Bayesian dynamic models (DLM) -- closely following developments in \cite{HarrisonWest1997} and \cite{Petris2009}, but adapting or extending familiar distribution theory in these texts to a general multivariate DLM context for our subsequent developments. 
\subsection{Exact (multivariate) Bayesian inference}\label{exact_bayesian_dlm_multivariate}
We consider mechanistic systems depending on space and time. Examples include broad classes of space-time partial differential equations and, even more generally, all deterministic computer models or agent-based models whose inputs include space and time. We build dynamic emulator models that will allow us to introduce spatial dependence in the output of the mechanistic system. Let $\mathcal{X}=\{\bm{x}_1,\ldots,\bm{x}_N\}$ be a set of $N$ mechanistic system inputs, where each $\bm{x}_i$ is a $d$-dimensional vector. Let $\mathcal{S}=\{\bm s_1,\ldots,\bm s_S\}$ denote the set of observed spatial locations. For any combination $(\bm x, \bm s) \in \mathcal{X}\times \mathcal{S}$, we denote ${y}_t(\bm x, \bm s)$ as the output of the mechanistic system at time point $t$.

A conjugate Matrix-Normal-Inverse-Wishart Bayesian model for space-time settings is 
\begin{equation}
    \label{basic_dlm_matrix-variate}
    \begin{split}
        \bm Y_t &= \bm F_t \bm{\Theta}_t + \bm{\mathcal{E}}_t, \quad \bm{\mathcal{E}}_t\stackrel{\text{ind.}}{\sim}\mathcal{MN}_{N\times S}(\bm O, \bm V_t, \bm\Sigma), \quad t=1,2,\ldots,T \\
        \bm{\Theta}_{t} &= \bm{G}_t\bm{\Theta}_{t-1} + \bm\Gamma_t, \quad \bm\Gamma_t\stackrel{\text{ind.}}{\sim} \mathcal{MN}_{p\times S}(\bm O, \bm W_t, \bm\Sigma), \quad t=1,2,\ldots,T\\
        \bm\Theta_0\mid\bm\Sigma &\sim \mathcal{MN}_{p\times S}(\bm m_0, \bm M_0, \bm\Sigma),\quad \bm\Sigma \sim \mathcal{IW}(n_0, \bm D_0),
    \end{split}
\end{equation}
where $\bm{Y}_t$ is $N\times S$ whose $(i,j)$-th element is ${y}_t(\bm x_i, \bm s_j)$, $\bm F_t$ and $\bm G_t$ are $N\times p$ and $p\times p$, respectively, with entries completely known, $\bm{\Theta}_t$ is $p\times S$ comprising latent effects over space and $\bm V_t$ and $\bm W_t$ are correlation matrices of order $N\times N$ and $p\times p$, respectively, for each $t=1,2,\ldots,T$. The important distinction from the more conspicuous univariate DLM setting is that $\bm{\mathcal{E}}_t$ and $\bm{\Gamma}_t$ are now matrix-variate normal distributions with zero matrices as their means, $\bm{V}_t$ and $\bm W_t$ are known correlation matrices modeling dependence across the rows for the respective distributions, and $\bm\Sigma$ is an $S\times S$ covariance matrix modeling the spatial dependence among the columns in $\bm{\mathcal{E}}_t$ and in $\bm{\Gamma}_t$. The joint prior density $p(\bm\Theta_0, \bm\Sigma)$ in \eqref{basic_dlm_matrix-variate} is a Matrix-Normal-Inverse-Wishart, denoted by $\mathcal{MNIW}(\bm\Theta_0, \bm\Sigma \mid \bm m_0, \bm M_0, n_0, \bm D_0)$, with density function 
\begin{equation}
    \label{matrix_normal_wishart_prior}
    \begin{split}
        p(\bm\Theta_0, \bm\Sigma)
        &\propto \underbrace{\frac{\left(\det(\bm D_0)\right)^{n_0/2}\exp\left(-\frac{1}{2}\mbox{tr}\left[\bm D_0\bm\Sigma^{-1}\right]\right)}{2^{n_0S/2}\Gamma_{S}(\frac{n_0}{2})\left(\det(\bm\Sigma)\right)^{\frac{n_0 + S + 1}{2}}}}_{p(\bm\Sigma) = \mathcal{IW}(\bm\Sigma \mid n_0, \bm D_0)} \\
        &\qquad \qquad \times \underbrace{\frac{\exp\left(-\frac{1}{2}\mbox{tr}\left[(\bm\Theta_0 - \bm m_0)^{\top}\bm M_0^{-1}(\bm \Theta_0 - \bm m_0)\bm\Sigma^{-1}\right]\right)}{\left(2\pi\right)^{pS/2}\left(\det(\bm M_0)\right)^{S/2}\left(\det(\bm\Sigma)\right)^{p/2}}}_{p(\bm\Theta_0\mid \bm\Sigma)=\mathcal{MN}(\bm\Theta_0 \mid \bm m_0, \bm M_0, \bm\Sigma)}\;,
    \end{split}
\end{equation}
where $\Gamma_S$ is the $S$-variate multivariate Gamma function and $\mbox{tr}(\cdot)$ is the trace function of a matrix. The matrix-normal density of a random matrix corresponds to a multivariate normal density for the vectorized columns of the matrix. For example, the above density $p(\bm\Theta_0 \mid \bm\Sigma)$ in \eqref{matrix_normal_wishart_prior} is equivalent to $\mbox{vec}\left(\bm\Theta_0\right) \mid \bm\Sigma \sim \mathcal{N}_{pS}\left(\mbox{vec}(\bm m_0), \bm\Sigma\otimes \bm M_0\right)$, where $\mbox{vec}(\bm \Theta_0)$ is the $pS\times 1$ vector of the stacked columns (first to last) of $\bm \Theta_0$ and $\otimes$ is the Kronecker product \citep[see, e.g.,][for properties of the vec operator and its connections to the Kronecker product]{banerjee2014linear}. Standard calculations for multivariate normal distributions yield
\begin{equation}\label{matrix_normal_wishart_posterior}
    \begin{split}
        p(\bm\Theta_{t}, \bm\Sigma \mid \bm Y_{1:t}) &= \underbrace{\mathcal{IW}(\bm\Sigma \mid n_t, \bm D_t)}_{p(\bm\Sigma\mid \bm Y_{1:t})} \times \underbrace{\mathcal{MN}(\bm \Theta_t \mid \bm m_t, \bm M_t, \bm\Sigma)}_{p(\bm\Theta_t \mid \bm Y_{1:t}, \bm\Sigma)} \\
        &= \mathcal{MNIW}(\bm\Theta_{t}, \bm\Sigma \mid \bm m_t, \bm M_t, n_t, \bm D_t)\;,
    \end{split}
\end{equation}
where $\bm m_t$, $\bm M_t$, $n_t$ and $\bm D_t$ can be calculated recursively over $t$ as we describe shortly. Sampling $[\bm\Theta_{t}, \bm\Sigma]\sim \mathcal{MNIW}(\bm m_t, \bm M_t, n_t, \bm D_t)$ proceeds by first drawing $\bm\Sigma \sim \mathcal{IW}(n_t, \bm{D}_t)$ and then drawing $\bm\Theta_{t} \sim \mathcal{MN}(\bm m_t, \bm M_t, \bm\Sigma)$ using the drawn value of $\bm\Sigma$. 

Algorithm~\ref{alg:sampMN} describes the procedure to sample from the matrix-normal distribution using, as input, the mean and the lower-triangular Cholesky factors of the two covariance matrices. We indicate the complexity in terms of the number of floating point operations (flops) for each step in the right-hand column of Algorithm~\ref{alg:sampMN}. Drawing $\bm{Z}\sim\mathcal{MN}_{p \times S}(\bm{O}, \bm{I}_p, \bm{I}_S)$ in Line~4 costs $\sim O(pS)$, while Line~5 costs $\sim O(pS^2 + p^2S)$, so the total cost is $O(pS + pS^2 + p^2S) \sim O(pS^2)$ since $p \leq S$ in our subsequent applications.

\begin{algorithm}[t]
    \caption{Sampling from matrix normal distribution}\label{alg:sampMN}
    \begin{algorithmic}[1]
    \State \textbf{Input:} $p \times S$ mean matrix $\bm{m}$, $p \times p$ lower-triangular Cholesky factor $\bm{L}_{\bm{M}}$ of $\bm M$ and $S \times S$ lower-triangular Cholesky factor $\bm{L}_{\bm\Sigma}$ of ${\bm\Sigma}$
    \State \textbf{Output:} Sample from $\mathcal{MN}_{p \times S}(\bm{m}, \bm{M}, \bm{\Sigma})$
    \Function{\texttt{SampleMatrixNormal}}{$\bm{m},\bm{L}_{\bm{M}}, \bm{L}_{\bm{\Sigma}}$}
    \State Draw $\bm{Z}\sim\mathcal{MN}_{p \times S}(\bm{O}, \bm{I}_p, \bm{I}_S)$
    \Comment{$\bm{I}$ is identity matrix, $O(pS)$}
    \State $\bm{\Theta}=\bm{m}+\bm{L}_{\bm{M}}\bm{Z}\bm{L}_{\bm{\Sigma}}^{\top}$
    \Comment{$O(pS^2 + p^2S)$}
    \State \Return $\bm{\Theta}$
    \EndFunction
    \Comment{$O(pS^2)$}
\end{algorithmic}
\end{algorithm}

Closely related to equations \eqref{matrix_normal_wishart_prior} and \eqref{matrix_normal_wishart_posterior} is the Hyper-T distribution. We obtain it by integrating $\bf\Sigma$ from the Matrix-Normal-Inverse-Wishart \citep{Gupta_Nagar_1999,NIPS2007_061412e4}. Doing so for, say, equation \eqref{matrix_normal_wishart_posterior}, which we denote by $\mathcal{HT}(\bm{\Theta}_t\vert \bm{m}_t, \bm{M}_t, n_t, \bm{D}_t)$, gives us:

\begin{equation}\label{eq:hyper_t}
    \begin{split}
        p(\bm{\Theta}_t \vert \bm{Y}_{1:t}) &= \int_{\bm{\Sigma} > 0}\mathcal{MNIW}(\bm{\Theta}_t, \bm{\Sigma}\mid \bm{m}_t, \bm{M}_t, n_t, \bm{D}_t)d\bm{\Sigma}\\
        &= \frac{\Gamma_{S}\left(\frac{n_{t}+p}{2}\right)(\det(\bm{I}_{S} + \bm{D}_{t}^{-1}(\bm \Theta - \bm{m}_{t})^{\top}\bm{M}^{-1}_{t}(\bm \Theta - \bm{m}_{t})))^{-\frac{n_{t} + p}{2}}}{\pi^{pS/2}\Gamma_{S}\left(\frac{n_{t}}{2}\right)(\det(\bm{M}_t))^{S/2}(\det(\bm{D}_{t}))^{p/2}}\\
        &= \mathcal{HT}(\bm{\Theta}_t\vert \bm{m}_t, \bm{M}_t, n_t, \bm{D}_t)
    \end{split}
\end{equation}

\begin{algorithm}[H]
    \caption{Sampling from hyper-T distribution}\label{alg:sampHT}
    \begin{algorithmic}[1]
    \State \textbf{Input:} $p \times S$ matrix $\bm{m}$, row-covariance matrix $p\times p$ matrix $\bm{M}$, positive integer degrees of freedom $n$, positive symmetric definite $S\times S$ scale matrix $\bm{D}$
    \State \textbf{Output:} Sample from $\mathcal{HT}_{p \times S}(\bm{m}, \bm{M},n,\bm{D})$
    \Function{\texttt{SampleHyperT}}{$\bm{m},\bm{M},n,\bm{D}$}
    \State Draw $\bm{\Sigma} \sim \mathcal{IW}_{S\times S}(n, \bm{D})$
    \Comment{$O(S^3)$}
    \State $\bm{L}_{\bm{M}} \gets \texttt{Cholesky}(\bm{M})$
    \Comment{$O(p^3)$}
    \State $\bm{L}_{\bm{\Sigma}} \gets \texttt{Cholesky}(\bm{\Sigma})$
    \Comment{$O(S^3)$}
    \State $\bm{\Theta} \gets$ \texttt{SampleMatrixNormal}($\bm{m}, \bm{L}_{\bm{M}}, \bm{L}_{\bm{\Sigma}}$)
    \Comment{Algorithm \ref{alg:sampMN}}
    \State \Return $\bm{\Theta}$
    \EndFunction
    \Comment{$O(S^3)$}
\end{algorithmic}
\end{algorithm}

The canonical form given in \citep{Gupta_Nagar_1999} or \citep{NIPS2007_061412e4} can be obtained by setting the degrees of freedom for the inverse-Wishart $n_{t} \leftarrow n_{t} + S - 1$, implying that the hyper-T is the matrix-T distribution with degrees of freedom $n_t - S + 1$. (It is thus important to point out that the degrees of freedom of the inverse-Wishart $n_t$ is not automatically equal to the degrees of freedom for the canonical form of the matrix-T.) Each entry of the matrix-T random variable is a univariate t-distribution with degrees of freedom $n_t - S + 1$, mean $m_{t,ij}$, and scale term $D_{t,ii}M_{t,jj}/n_t$.

Algorithm~\ref{alg:sampHT} describes the procedure to sample from the $\mathcal{HT}_{p \times S}(\bm{m}, \bm{M},n,\bm{D})$. As in Algorithm~\ref{alg:sampMN}, the right-hand side column provide the complexity in flops. Lines-4,~5,~and~6 together cost $\sim O(p^3 + 2S^3)$ and Line-7 costs $\sim O(pS^2)$, which brings the total cost to $\sim O(p^3 + 2S^3 + pS^2)\sim O(S^3)$ since $p \leq S$ in our subsequent applications.

Also important to consider is a matrix-normal $\bm{\Theta}_t$, but where $\Sigma = \sigma^{2}\bm{R}$, where $\sigma^{2}$ is distributed as inverse-gamma ($\mathcal{IG}(n_t, d_t)$) and $\bm{R}$ being a given matrix. We denote this density by $\mathcal{MNIG}(\bm{\Theta}_t, \sigma^{2} \mid \bm{m}_t, \bm{M}_t, n_t, d_t, \bm{R})$. Next, we marginalize out the $\sigma^{2}$ to obtain a multivariate t-distribution, but one in which the mean arguments are matrices rather than vectors. We call this the hyper-T-scalar and denote it by $\mathcal{HT}_{s}(\bm{\Theta}_t\mid \bm{m}_t, \bm{M}_t, n_t, d_t, \bm{R})$:
\begin{equation}\label{eq:hyper-T-scalar}
    \begin{split}
        p(\bm{\Theta}_t\mid \bm{Y}_{1:t}, \bm{R}) =& \int_{\sigma^{2} > 0} \mathcal{MNIG}(\bm{\Theta}_t, \sigma^{2} \mid \bm{m}_t, \bm{M}_t, n_t, d_t, \bm{R})d\sigma^{2}\\
        =& \frac{\Gamma(\frac{pS}{2}+n_t)}{\Gamma(n_t)}(2\pi d_t)^{-\frac{pS}{2}}\lvert\bm{M}_t\rvert^{-\frac{S}{2}}\lvert\bm{R}\rvert^{-\frac{p}{2}}\times\\
        &\left| 1 + \frac{1}{2d_t}\mathrm{tr}\left[\bm{R}^{-1}(\bm{\Theta}_t - \bm{m}_t)^{\top}\bm{M}_{t}^{-1}(\bm{\Theta}_t - \bm{m}_t)\right]\right|^{-\left(\frac{pS}{2} + n_t\right)}\\
        =& \mathcal{HT}_{s}(\bm{\Theta}_t\mid \bm{m}_t, \bm{M}_t, n_t, d_t, \bm{R})
    \end{split}
\end{equation}

\begin{algorithm}[t]
    \caption{Kalman (forward) filter}\label{alg:KF}
    \begin{algorithmic}[1]
    \State \textbf{Input:} Data $\bm Y_{1:T}$, hyperparameters $n_0$, $\bm D_0$, $\bm{m}_0$, $\bm{M}_0$, correlation matrices $\bm{V}_{1:T}$, $\bm{W}_{1:T}$,
    \State {\color{white}\textbf{Input:}} observation and state transition matrices $\bm{F}_{1:T}$ and $\bm{G}_{1:T}$.
    \State \textbf{Output:} Filtering distribution parameters at time $t=0,\ldots, T$
    \Function{\texttt{Filter}}{$\bm Y_{1:T}, n_0, \bm D_0,\bm m_0, \bm M_0,\bm G_{1:T},\bm F_{1:T}, \bm V_{1:T}, \bm W_{1:T}$}
    \For{$t=1$ to $T$}
        \State \textit{\# Compute prior distribution} $[\bm\Theta_t, \bm\Sigma_t \mid \bm{Y}_{1:t-1}]\sim \mathcal{MNIW}(\bm{a}_t, \bm{A}_t, n_t^*, \bm D_t^*):$
        \State $\bm a_t\gets \bm G_t\bm m_{t-1}, \bm A_t\gets \bm G_t\bm M_{t-1}\bm G_t^\top+\bm W_t$, $n_t^*\gets n_{t-1}, \bm D_t^*\gets \bm D_{t-1}$
        \Comment{$O(p^3 + p^2S)$}
        \State \textit{\# Compute one-step-ahead forecast}   $p(\bm{Y}_t|\bm{Y}_{1:t-1})\sim\mathcal{HT}(\bm{F}_{t}\bm{a}_{t}, \bm{F}_{t}\bm{A}_{t}\bm{F}_{t} + \bm{V}_{t}, n_{t}^{*}, \bm{D}_{t}^{*})$
        \State $\bm q_t\gets \bm{F}_t\bm{a}_t,  \boldsymbol Q_t\gets\boldsymbol{F}_t\boldsymbol{A}_t\boldsymbol{F}_t^\top+\boldsymbol V_t$
        \Comment{$O(pSN + p^2N + N^2)$}
        \State \textit{\# Compute filtering distribution} $p(\bm{\Theta}_{t}, \bm\Sigma_t \mid \bm{Y}_{1:t})\sim\mathcal{MNIW}(\bm{m}_t,\bm M_t, n_t, \bm D_t):$
        \State  $\bm m_t\gets\bm{a}_t+\bm{A}_t\bm{F}_t^\top\bm{Q}_t^{-1}(\bm{Y}_{t}-\bm{q}_t), \ \bm M_t\gets\bm{A}_t - \bm{A}_t\bm{F}_t^\top\bm{Q}_t^{-1}\bm{F}_t\bm{A}_t^\top$ \\
        \Comment{$O(pN^2 + p^2N + p^3 + pSN + N^3)$}
        \State $n_t\gets n_t^* + N, \ \bm D_t \gets \bm D_t^* + (\bm{Y}_t-\bm{q}_t)^\top\bm{Q}_t^{-1}(\bm{Y}_t-\bm{q}_t)$
        \Comment{$O(SN^2 + S^2N)$}
    \EndFor
    \State \Return $\left\{n_t, \bm D_t,\bm{a}_t,\bm{A}_t,\bm{m}_t,\bm{M}_t\right\}_{t=0}^T$
    \EndFunction
    \Comment{$O(T(p^2S + S^2N))$}
    \end{algorithmic}
\end{algorithm}

The matrix-variate SSM in \eqref{basic_dlm_matrix-variate} delivers posterior and predictive distributions in closed form. Exact Bayesian inference is possible by sampling from the posterior distribution $p(\bm\Theta_{0:T}, \bm\Sigma\mid \bm Y_{1:T})$ using the Forward Filter Backward Sampling algorithm \citep{Carter1994, fruhwirth2006}. The basic idea is fairly simple. We first move forward in time up to $t=T$ and draw one instance of $(\bm\Sigma,\bm\Theta_T)$ from $p(\bm\Theta_T, \bm\Sigma \mid \bm Y_{1:T})$. This is the ``forward filtering'' step (``FF''). Next, we execute ``backward sampling'' (BS): for each $t=T-1,T-2,\ldots, 0$, we draw one instance of $\bm\Theta_t$ from $p(\bm\Theta_t \mid \bm\Sigma, \bm\Theta_{t+1}, \bm Y_{1:t})$. This entire process yields one instance of $(\bm\Sigma, \bm\Theta_{0:T})$ from $p(\bm\Theta_{0:T}, \bm\Sigma \mid \bm Y_{1:T})$. Backward sampling makes use of the Markovian structure in \eqref{basic_dlm_matrix-variate}, which implies that
\begin{equation}\label{backwardsampling}
    p(\bm\Theta_{0:T}, \bm\Sigma \mid \bm Y_{1:T}) = p(\bm\Theta_T, \bm\Sigma \mid \bm Y_{1:T})\prod_{t=0}^{T-1} p(\bm\Theta_{t} \mid \bm\Sigma, \bm\Theta_{t+1}, \bm Y_{1:t})\;.
\end{equation}
The ``FF'' followed by the ``BS'' steps complete one iteration of the FFBS algorithm yielding one draw from $p(\bm\Theta_{0:T}, \bm\Sigma\mid \bm Y_{1:T})$. In essence, the FFBS algorithm first applies the Kalman filter of Algorithm \ref{alg:KF} before initializing the backward sampling through the Kalman or Rauch–Tung–Striebel smoother \citep{rauch1965maximum,Sarkka2013}. Repeating the FFBS cycle $L$ times yields $L$ posterior samples of $\bm{\Theta}_{0:T}$ and $\bm\Sigma$. In object-oriented computing paradigms, we can execute $L$ draws of $(\bm\Sigma,\bm\Theta_T)$ at the end of FF. Then, for each drawn $(\bm\Sigma,\bm\Theta_T)$ we execute $T-1$ draws of the BS to obtain $L$ posterior samples.   

\begin{algorithm}[t]
    \caption{Backward sampler}\label{alg:BS}
    \begin{algorithmic}[1]
    \State \textbf{Input:} Filtering parameters and inputs from Algorithm \ref{alg:KF}
    \State \textbf{Output:} $L$ posterior samples from $p(\bm\Theta_{0:T},\bm\Sigma\mid \bm Y_{1:T})$
    \Function{\texttt{BackwardSample}}{$\left\{n_t, \bm D_t,\bm{a}_t,\bm{A}_t,\bm{m}_t,\bm{M}_t, \bm G_t\right\}_{t=0}^T$, $L$}
    \State Draw $L$ samples from $\bm\Sigma\sim\mathcal{IW}(n_T,\bm D_T)$
    \Comment{$O(LS^3)$}
    \State Draw $L$ samples from $\bm\Theta_T\sim\mathcal{MN}(\bm{m}_T, \bm M_T, \bm\Sigma)$
    \Comment{$O(LS^3)$}
    \State $\bm h_T \gets \bm m_T, \bm H_T \gets \bm M_T$
    \For{$t=T-1$ to $1$}
    \
    \State $\bm h_t \gets \bm{m}_t+\bm{M}_t\bm G_{t+1}^\top\bm{A}_{t+1}^{-1}\left(\bm{h}_{t+1}-\bm{a}_{t+1}\right)$
    \Comment{$O(p^3 + p^2S)$}
    \State $\bm{H}_t \gets \bm{M}_t-\bm{M}_t\bm G_{t+1}^\top\bm{A}_{t+1}^{-1}\left(\bm{A}_{t+1}-\bm{H}_{t+1}\right)\bm{A}_{t+1}^{-1}\bm G_{t+1}\bm{M}_t$
    \Comment{$O(p^3)$}
    \State Draw $L$ samples from $\bm\Theta_t\sim\mathcal{MN}(\bm{h}_t, \bm H_t, \bm\Sigma)$
    \Comment{$O(LS^3)$}
    \EndFor
    \State \Return $\{\bm{h}_{1:T}, \bm{H}_{1:T}\}$ and $L$ samples of $\{\bm\Theta_{0:T}, \bm\Sigma\}$
    \EndFunction
    \Comment{$O(TLS^3)$}
\end{algorithmic}
\end{algorithm}

The specific FFBS algorithm to estimate \eqref{basic_dlm_matrix-variate} accomplishes the forward filtering step by recursively computing the following quantities for each $t=1,\ldots,T$,
\begin{equation}\label{FF:conj1}
    \begin{split}
        & \bm{a}_t = \bm{G}_t\bm{m}_{t-1};\quad \bm{A}_t = \bm{G}_t\bm{M}_{t-1}\bm{G}_t^{\top} + \bm{W}_t;\quad \bm{q}_t = \bm{F}_t\bm{a}_t;\quad \bm{Q}_t = \bm{F}_t\bm{A}_t\bm{F}_t^{\top} + \bm{V}_t;\\
        & \bm{m}_t = \bm{a}_t + \bm{A}_t\bm{F}_t^{\top}\bm{Q}_t^{-1}\left(\bm{Y}_t - \bm{q}_t\right);\quad \bm{M}_t = \bm{A}_t - \bm{A}_t\bm{F}_t^{\top}\bm{Q}_t^{-1}\bm{F}_t\bm{A}_t;\\
        n_t &= n_{t-1} + N;\quad \mbox{and}\quad \bm D_t = \bm D_{t-1} + \left(\bm{Y}_t - \bm{q}_t\right)^{\top}\bm{Q}_t^{-1}\left(\bm{Y}_t - \bm{q}_t\right)
    \end{split}
\end{equation}
Algorithm~\ref{alg:KF} outlines the steps for forward filtering. The right-hand column provides the computational complexity in terms of flops. Lines-7,~9,~11,~and~13 together complete one iteration of the for-loop, with a combined cost of $\sim O(p^3 + p^2(S+N) + p(SN + N^2) + S^2N + SN^2 + N^3)$. In particular, if $p, N \leq S$, then the combined cost is $\sim O(p^2S + S^2N)$ for each $t$. This yields a total cost of $\sim O(T(p^2S + S^2N))$. We provide the costs in our subsequent algorithms using $p, N \leq S$. This is true for all our applications except the predator-prey system in Section~\ref{sec:app_emulation_LV}.

We then draw one instance of $(\bm{\Theta}_T, \bm\Sigma)\mid \bm{Y}_{1:T} \sim \mathcal{MNIW}(\bm{m}_T, \bm{M}_T, n_T, \bm D_T\cdot)$, where $\bm a_t$ and $\bm m_t$ are both $p\times S$ and $\bm q_t$ is $N\times S$. For the inverse-Wishart parameters, $n_t = n_{t-1} + N$ is a scalar and $\bm D_t = \bm D_{t-1} + (\bm Y_t - \bm q_t)^{\top}\bm Q_t^{-1}(\bm Y_t - \bm q_t)$ is $S\times S$. An exact FFBS algorithm proceeds by forward filtering to $t=T$ to draw $L$ samples from $p(\bm\Theta_{T}, \bm\Sigma \mid \bm Y_{1:T})$ from \eqref{matrix_normal_wishart_posterior}. Next, for backward sampling we exploit standard Bayesian probability calculations to note that $(\bm{\Theta}_t, \bm\Sigma) \mid \bm{Y}_{1:T}) \sim \mathcal{MNIW}(\bm{h}_t, \bm{H}_t, n_T, \bm{D}_T)$, where $\bm{h}_T = \bm{m}_T$, $\bm{H}_T = \bm{M}_T$, and 
\begin{equation}\label{BS:conjMNIG}
    \begin{split}
        \bm{h}_t &= \bm{m}_t + \bm{M}_t\bm{G}_{t+1}^{\top}\bm{A}_{t+1}^{-1}(\bm{h}_{t+1} - \bm{a}_{t+1})\quad \mbox{and}\\ 
        \bm{H}_t &= \bm{M}_t - \bm{M}_t\bm{G}_{t+1}^{\top}\bm{A}_{t+1}^{-1}\left(\bm{A}_{t+1} - \bm{H}_{t+1}\right)\bm{A}_{t+1}^{-1}\bm{G}_{t+1}\bm{M}_t\;.
    \end{split}
\end{equation}
Therefore, for each $t=T-1,\ldots,0$ we draw one value of $\bm{\Theta}_t \sim \mathcal{MN}(\bm{h}_t, \bm{H}_t, \bm \Sigma)$ for each of the $L$ values of $\bm\Sigma$ at the end of the ``FF'' step, where $\bm{h}_t$ and $\bm{H}_t$ are computed using \eqref{BS:conjMNIG}. The end of the FFBS yields $L$ posterior samples from $p(\bm\Theta_{0:T},\bm\Sigma \mid \bm Y_{1:T})$. 

\begin{algorithm}[ht]
    \caption{Forward-filter-backward-sampler algorithm}\label{alg:FFBS}
    \begin{algorithmic}[1]
    \State \textbf{Input:} Data $\bm Y_{1:T}$ and Kalman filter starting values $n_0$, $\bm{D}_0$, $\bm{m}_0$, $\bm{M}_0$
    \State {\color{white}\textbf{Input:}} Observation and state transition matrices $\bm{F}_{1:T}$ and $\bm{G}_{1:T}$
    \State {\color{white}\textbf{Input:}} Correlation matrices $\bm{V}_{1:T}$ and $\bm{W}_{1:T}$, number of samples $L$
    \State \textbf{Output:} Sample from posterior $p(\bm\Theta_{0:T},\bm\Sigma|\bm{Y}_{1:T})$
    \Function{\texttt{FFBS}}{$\bm Y_{1:T},  n_0, \bm{D}_0, \bm{m}_0, \bm{M}_0, \bm F_{1:T}, \bm G_{1:T}, \bm V_{1:T}, \bm W_{1:T}, L$}
    \State $
    \texttt{FF}_\texttt{Out}\gets \texttt{Filter}(\bm Y_{1:T}, n_0, \bm D_0,\bm m_0, \bm M_0, \bm G_{1:T}, \bm F_{1:T}, \bm V_{1:T},\bm W_{1:T}$)
    \Comment{Algorithm~\ref{alg:KF}}
    \State $
    \texttt{BS}_\texttt{Out}\gets \texttt{BackwardSample}(\{n_t, \bm D_t,\bm{a}_t,\bm{A}_t,\bm{m}_t,\bm{M}_t, \bm G_t\}_{t=0}^T, L$)
    \Comment{Algorithm~\ref{alg:BS}}
    \State \Return $ \texttt{BS}_\texttt{Out} =\{\bm{h}_{1:T}, \bm{H}_{1:T}, \{\bm\Theta_{0:T}^{(l)},\bm\Sigma^{(l)}\}_{l=1}^L\}$ 
    \EndFunction
    \Comment{$O(TLS^3)$}
\end{algorithmic}
\end{algorithm}

Algorithm~\ref{alg:BS} outlines the steps for backward sampling, with the computational complexity detailed in the right-hand column in terms of flops. Lines-4~and~5 each draw $L$ samples of $\bm\Sigma$ and $\bm\Theta_T$, respectively, with a computational cost of $\sim O(LS^3)$ for both. Meanwhile, Lines-8,~9,~and~10 together constitute one iteration of the for-loop, with a cost of $\sim O(p^3 + p^2S + LS^3) \sim O(LS^3)$. This results in a total cost of $\sim O(TLS^3)$ assuming $p, N \leq S$. Algorithm~\ref{alg:FFBS} presents the complete forward-filter-backward-sampler algorithm, incorporating both Algorithm~\ref{alg:KF} and Algorithm~\ref{alg:BS}. The overall computational cost is $\sim O(T(p^2S + S^2N + LS^3)) \sim O(TLS^3)$.

Bayesian predictive inference is also convenient if we seek to predict or impute the values in a new $\tilde{N}\times S$ matrix $\tilde{\bm Y_t}$ corresponding to a new set of mechanistic inputs $\tilde{\mathcal{X}}$, given the $\tilde{N}\times p$ matrix $\tilde{\bm F_t}$ for any given $t$, where $\tilde{N}$ represents the number of new mechanistic system inputs. The predictive distribution for $\tilde{\bm Y_t}$ follows from the augmented model,
\begin{equation}\label{eq:basic_dlm_matrix_variate_predictive}
    \begin{bmatrix} \bm Y_t \\  \tilde{\bm Y_t}  \end{bmatrix} = \begin{bmatrix}
        \bm F_t \\ \tilde{\bm F_t}
    \end{bmatrix} \bm{\Theta_t} + \begin{bmatrix}
       \bm{\mathcal{E}_t} \\ \tilde{\bm{\mathcal{E}}_t}
    \end{bmatrix},  \quad \begin{bmatrix}
       \bm{\mathcal{E}}_t \\ \tilde{\bm{\mathcal{E}}_t}
    \end{bmatrix} \stackrel{\text{ind.}}{\sim}\mathcal{MN}_{(N+\tilde{N})\times S}\left(\begin{bmatrix}
        \bm O_{N\times S} \\ \bm O_{\tilde{N}\times S}
    \end{bmatrix}, \begin{bmatrix}
        \bm V_t & \bm J_t \\ {\bm J_t}^{\top} & \tilde{\bm V_t}
    \end{bmatrix}, \bm\Sigma\right)\;,\\ 
\end{equation}
for $t=1,2,\ldots,T$. Therefore, for each drawn value of $\{\bm \Theta_t,\bm\Sigma\} \sim p(\bm\Theta_{0:T}, \bm\Sigma \mid \bm Y_{1:T})$ from the FFBS algorithm, we draw one instance of $\tilde{\bm Y_t}$ from the conditional predictive density
\begin{equation}\label{eq: conditional_posterior_predictive_density}
p(\tilde{\bm Y_t}\mid \bm Y_{1:T}, \bm{\Theta_t}, \bm\Sigma) = \mathcal{MN}(\tilde{\bm Y_t} \mid \tilde{\bm F_t}\bm\Theta_t + \bm J_t^{\top}\bm V_t^{-1}(\bm Y_t - \bm F_t\bm\Theta_t), \tilde{\bm V_t} - \bm J_t^{\top}\bm V_t^{-1}\bm J_t, \bm\Sigma)\;.
\end{equation}

Repeating this for all the posterior samples of $\{\bm\Theta_{0:T},\bm\Sigma\}$ will yield samples of $\tilde{\bm Y}_{1:T}$ from the posterior predictive distribution $p(\tilde{\bm Y}_{1:T} \mid \bm Y_{1:T})$. Therefore, predictive inference can be carried out using \eqref{eq: conditional_posterior_predictive_density} for arbitrary inputs in $\tilde{\mathcal{X}}$ using stored posterior samples from the training data. The full expression for the posterior predictive density is itself a Hyper-T, with the following expression: 

\begin{equation}\label{eq:postpredictive}
    \begin{split}
    p(\tilde{\bm{Y}}_{t}\mid \bm{Y}_{1:T}) =& \mathcal{HT}(\tilde{\bm{Y}}_t\mid \tilde{\bm{F}}_t \bm{h}_t + \bm{J}_{t}^{\top}\bm{V}_{t}^{-1}(\bm{Y}_t -\bm{F}_t \bm{h}_t), \tilde{\bm{F}}_t \bm{H}_t \tilde{\bm{F}}_{t}^{\top} + \tilde{\bm{V}}_t - \\
    &(\bm{F}_t \bm{H}_t \tilde{\bm{F}}_{t}^{\top} + \bm{J}_{t})^{\top} (\bm{F}_{t} \bm{H}_{t} \bm{F}_{t}^{\top} + \bm{V}_{t})^{-1} (\bm{F}_{t} \bm{H}_{t}\tilde{\bm{F}}_{t} + \bm{J}_t) , n_T, \bm{D}_T)
    \end{split}
\end{equation}

The same conclusions we draw as for equation \eqref{eq:hyper_t} lead us to conclude that each entry of $\tilde{\bm{Y}}_t$, $\tilde{\bm{Y}}_{t,ij}$, is a univariate-t, with mean $\tilde{\bm{F}}_{t,i:}\bm{h}_{t,:j} + \bm{J}_{t,:i}^{\top}\bm{V}_{t}^{-1}(\bm{Y}_{t,:j} - \bm{F}_{t}\bm{h}_{t,:j})$ and scale parameter $(\tilde{\bm{F}}_{t,j:} \bm{H}_t \tilde{\bm{F}}_{t,j:}^{\top} + \tilde{\bm{V}}_{t,jj} - (\bm{F}_{t} \bm{H}_t \tilde{\bm{F}}_{t,:j}^{\top} + \bm{J}_{t,:j})^{\top} (\bm{F}_{t} \bm{H}_{t} \bm{F}_{t}^{\top} + \bm{V}_{t}) (\bm{F}_{t} \bm{H}_{t}\tilde{\bm{F}}_{t,j:}^{\top} + \bm{J}_{t,:j}))(\bm{D}_{T,ii})/n_{T}$. 

We do not require fresh training of the model for predictive inference provided the distribution in \eqref{eq:basic_dlm_matrix_variate_predictive} is valid---a matter we turn to next.


\subsection{Mechanistic emulation using Gaussian Processes}\label{GPR}
Emulating the mechanistic system requires that we train the model in \eqref{basic_dlm_matrix-variate} from the inputs in $\mathcal{X}$ using the FFBS algorithm and, subsequently, to predict or interpolate the outcome matrix for possibly new inputs in $\tilde{\mathcal{X}}$. For predictive inference, we must ensure that the distribution in \eqref{eq:basic_dlm_matrix_variate_predictive} is well-defined, which, in turn, requires that $\displaystyle \begin{bmatrix} \bm V_t & \bm J_t \\ {\bm J_t}^{\top} & \tilde{\bm V}_t \end{bmatrix}$ is positive definite for arbitrary mechanistic inputs $\mathcal{X} \cup\tilde{\mathcal{X}}$. This will be ensured if we ensure that each column of the augmented outcome matrix $\displaystyle \begin{bmatrix} \bm Y_t \\  \tilde{\bm Y_t}  \end{bmatrix}$ is a realization of a valid stochastic process over $\mathcal{X} \cup\tilde{\mathcal{X}}$. A customary choice for mechanistic emulation is the Gaussian process.

The Gaussian process acts as a prior on of $\displaystyle \begin{bmatrix} \bm Y_t \\  \tilde{\bm Y_t}  \end{bmatrix}$. Thus, for each of the $S$ locations (or column indices), we assume $y_t(\bm{x},\bm{s}) \overset{ind}{\sim} \mathcal{GP}\left(\mu_t(\bm{x},\bm{s}), {C}(\cdot, \cdot;\bm\beta)\right)$, where the mean function $\mu_t(\bm{x},\bm{s}) = \bm{f}_t(\bm x, \bm s)^{\top}\bm\Theta_t(\bm{s})$ and a positive definite correlation function $C(\bm x, \bm x'; \bm\beta)$. This yields, $\bm V_t = (C(\bm x_i, \bm x_j; \bm\beta))$ for pairs of inputs in $\mathcal{X}$, $\tilde{\bm V}_t = (C(\tilde{\bm x}_i, \tilde{\bm x}_j; \bm\beta))$ for pairs of inputs in $\tilde{\mathcal{X}}$ and $\bm J_t = (C(\bm x_i, \tilde{\bm x}_j; \bm\beta))$ for $\bm x_i \in \mathcal{X}$ and $\tilde{\bm x}_j$ in $\tilde{\mathcal{X}}$. This specification ensures that \eqref{eq:basic_dlm_matrix_variate_predictive} is a valid probability distribution by virtue of $C(\bm x, \bm x'; \bm\beta)$ being a valid correlation function. Several options for valid correlation functions are available, including the rich Mat\'ern class used widely in spatial statistics. In mechanistic emulation, we are mostly concerned with interpolation so any valid correlation function of a simpler form will suffice. For our current illustrations, we use the squared exponential function 
\begin{equation}
    \label{corr}
    {C}(\bm{x},\bm{x}';\bm\beta)=\exp\left(-\sum_{i=1}^d\beta_i(x_i- x'_i)^2\right),
\end{equation}
where the range parameter $\beta_i$ controls the decay of correlation along the $i$-th dimension of $\bm\beta$. In particular, let us denote $\bm Y_t$ and $\tilde{\bm Y}_t$ as $\bm Y_t(\mathcal{X})$ and $\tilde{\bm Y}_t(\tilde{\mathcal{X}})$, respectively. Then, the predictive mean $\mathbb{E}[\bm Y_t(\tilde{\mathcal{X}}) \mid \bm Y_t(\mathcal{X}), \bm{\Theta}_{0:T}, \bm\Sigma]$ in \eqref{eq: conditional_posterior_predictive_density} acts as an interpolator over the mechanistic inputs. More precisely, if $\tilde{\mathcal{X}} \subseteq \mathcal{X}$, then the rows of $\bm J_t^{\top}$ are a subset of the rows of $\bm V_t$, which implies that $\bm J_t^{\top}\bm V_t^{-1} = \bm I(\tilde{\mathcal{X}})$ (since $\bm V_t\bm V_t^{-1} = \bm I(\mathcal{X})$), where $\bm I(\tilde{\mathcal{X}})$ is the subset of rows of the identity matrix indexed by $\tilde{\mathcal{X}}$. Therefore, $\bm J_t^{\top}\bm V_t^{-1}(\bm Y_t - \bm F_t\bm\Theta_t) = \bm Y_t(\tilde{\mathcal{X}}) - \tilde{\bm F}_t\bm\Theta_t$ and     
\begin{multline*}
  \mathbb{E}[\bm Y_t(\tilde{\mathcal{X}}) \mid \bm Y_t(\mathcal{X}), \bm{\Theta}_{0:T}, \bm\Sigma] =  \tilde{\bm F_t}\bm\Theta_t + \bm J_t^{\top}\bm V_t^{-1}(\bm Y_t - \bm F_t\bm\Theta_t) = \tilde{\bm F}_t\bm\Theta_t + \bm Y_t(\tilde{\mathcal{X}}) - \tilde{\bm F}_t\bm\Theta_t = \bm Y_t(\tilde{\mathcal{X}})\;. 
\end{multline*}
Furthermore, this interpolation is deterministic because   
\begin{multline*}
    \mbox{Var}[\bm Y_t(\tilde{\mathcal{X}}) \mid \bm Y_t(\mathcal{X}), \bm{\Theta}_{0:T}, \bm\Sigma] = \tilde{\bm V_t} - \bm J_t^{\top}\bm V_t^{-1}\bm J_t = \tilde{\bm V_t} - \bm I(\tilde{\mathcal{X}})\bm J_t = \tilde{\bm V_t} - \tilde{\bm V_t} = \bm O\;,  
\end{multline*}
where the second to last equation follows from the fact that $\bm I(\tilde{\mathcal{X}})\bm J_t = \tilde{\bm V_t}$ when $\tilde{\mathcal{X}} \subseteq \mathcal{X}$. 

\subsection{Limitations}\label{limitations}
If we are able to evaluate and store the output from a space-time mechanistic system over a fixed set of $S$ locations and $T$ time points, then we are able to learn about the temporal evolution of the process and the association over the spatial locations (through $\bm \Sigma$) using a conjugate Matrix-variate dynamic linear model framework. A clear advantage of this approach is that inference from the FFBS samples from the exact posterior and posterior predictive distributions and there is no need to resort to more computationally expensive iterative algorithms (e.g., MCMC, INLA, VB) that will require diagnosis of convergence. 

The relative simplicity of this learning framework, however, comes with some severe limitations with respect to spatial modeling. First, and perhaps most obviously, the above approach uses an unstructured $\bm \Sigma$ to model the associations across space, whereas it could be argued that one should model $\bm \Sigma$ further using conventional geostatistical kernels that more explicitly model association as a function of the locations (or distances between them). Second, since the dimension of $\bm \Sigma$ is fixed, learning of the mechanistic system occurs only over the fixed set of $S$ spatial locations that have been determined at the design stage to run the system but precludes predictions at arbitrary new locations. Third, it does not accommodate the possibility that for each spatial location, we may not have the same inputs (breaking the $N\times S$ matrix structure) which is the likely scenario when our learning framework includes information from field data that we use to calibrate the mechanistic system.  

A more flexible approach builds a learning system that models both the mechanistic system and the spatial associations using stochastic processes with Gaussian processes being the customary choice here. Now $\bm \Sigma$ is constructed using a spatial covariance kernel which leads to a sparser parametrization, easier interpretation, and accommodates predictions over space. While the exact distribution theory for the $S\times S$ covariance matrix $\bm \Sigma$ is lost, we arrive at a richer and more flexible framework that also eliminates the requirement for the dimension of $\bm Y_t$ to be the same for each $t$. Therefore, we construct
\begin{equation}
    \label{dlm_gp_double}
    \begin{split}
        \bm Y_{t} &= \bm F_{t} \bm{\Theta}_{t} + \bm{\mathcal{E}}_{t}, \quad \bm{\mathcal{E}}_{t}\stackrel{\text{ind.}}{\sim}{\sim}\mathcal{MN}_{N_t\times S}(\bm O,\bm V_{t}, \sigma^2\bm R),\\
        \bm{\Theta}_{t} &= \bm{G}_{t}\bm{\Theta}_{t-1} + \bm\Gamma_{t}, \quad \bm\Gamma_{t}\stackrel{\text{ind.}}{\sim} \mathcal{MN}_{p_t\times S}(\bm O, \bm W_{t}, \sigma^2 \bm R),\\
        \bm \Theta_0 \mid \sigma^2 &\sim \mathcal{MN}_{p_t\times S}(\bm m_0, \bm M_0, \sigma^2 \bm R),\quad \sigma^2 \sim \mathcal{IG}(n_0, d_0), 
    \end{split}
\end{equation}
where $\bm Y_t$ and $\bm \Theta_t$ are $N_t \times S$ and $p_t\times S$, respectively, $\bm F_t$ and $\bm G_t$ are $N_t\times p_t$ and $p_t\times p_{t-1}$, respectively, $\bm R$ is $S\times S$ spatial correlation matrices whose $(i,j)$-th elements are values of some spatial correlation function $\rho_t(\bm s_i, \bm s_j)$ with fixed process parameters. Fixing $\bm V_t$, $\bm W_t$, and $\bm R$ in \eqref{dlm_gp_double} will, again, yield conjugate Bayesian posterior distributions obtained using an FFBS algorithm.  

The FFBS algorithm to estimate \eqref{dlm_gp_double} adapts Algorithms~\ref{alg:KF},~\ref{alg:BS}~and~\ref{alg:FFBS} by recursively computing the following quantities for each $t=1,\ldots,T$,
\begin{equation}\label{FF:conj1MNIG}
    \begin{split}
        & \bm{a}_t = \bm{G}_t\bm{m}_{t-1};\quad \bm{A}_t = \bm{G}_t\bm{M}_{t-1}\bm{G}_t^{\top} + \bm{W}_t;\quad \bm{q}_t = \bm{F}_t\bm{a}_t;\quad \bm{Q}_t = \bm{F}_t\bm{A}_t\bm{F}_t^{\top} + \bm{V}_t;\\
        & \bm{m}_t = \bm{a}_t + \bm{A}_t\bm{F}_t^{\top}\bm{Q}_t^{-1}\left(\bm{Y}_t - \bm{q}_t\right);\quad \bm{M}_t = \bm{A}_t - \bm{A}_t\bm{F}_t^{\top}\bm{Q}_t^{-1}\bm{F}_t\bm{A}_t;\\
        n_t &= n_{t-1} + \frac{NS}{2};\quad \mbox{and}\quad d_t = d_{t-1} + \frac{1}{2}\mbox{tr}\left[\left(\bm{Y}_t - \bm{q}_t\right)^{\top}\bm{Q}_t^{-1}\left(\bm{Y}_t - \bm{q}_t\right)\bm {R}^{-1}\right]\;.
    \end{split}
\end{equation}
For the inverse-Gamma parameters, $n_t$ is the shape and $d_t$ is the rate. As earlier, we use forward filtering up to $t=T$ to draw $L$ samples from $p(\bm\Theta_{T}, \sigma^2 \mid \bm Y_{1:T})$ from $\mathcal{MNIG}(\bm{m}_T, \bm{M}_T, n_T, d_T)$, where $\bm a_t$ and $\bm m_t$ are both $p_t\times S$ and $\bm q_t$ is $N_t\times S$. For backward sampling, we note that $(\bm{\Theta}_t, \sigma^2) \mid \bm{Y}_{1:T}) \sim \mathcal{MNIG}(\bm{h}_t, \bm{H}_t, n_T, d_T)$, where $\bm{h}_T = \bm{m}_T$, $\bm{H}_T = \bm{M}_T$, and 
\begin{equation}\label{BS:conj}
    \begin{split}
        \bm{h}_t &= \bm{m}_t + \bm{M}_t\bm{G}_{t+1}^{\top}\bm{A}_{t+1}^{-1}(\bm{h}_{t+1} - \bm{a}_{t+1})\quad \mbox{and}\\ 
        \bm{H}_t &= \bm{M}_t - \bm{M}_t\bm{G}_{t+1}^{\top}\bm{A}_{t+1}^{-1}\left(\bm{A}_{t+1} - \bm{H}_{t+1}\right)\bm{A}_{t+1}^{-1}\bm{G}_{t+1}\bm{M}_t\;.
    \end{split}
\end{equation}
Therefore, for each $t=T-1,\ldots,0$ we draw one value of $\bm{\Theta}_t \sim \mathcal{MN}(\bm{h}_t, \bm{H}_t, \sigma^2 \bm R)$ for each of the $L$ values of $\sigma^2$ at the end of the ``FF'' step, where $\bm{h}_t$ and $\bm{H}_t$ are computed using \eqref{BS:conj}. The end of the FFBS yields $L$ posterior samples from $p(\bm\Theta_{0:T},\sigma^2 \mid \bm Y_{1:T})$.

\section{Bayesian Model Comparisons} \label{sec:mod_compare}
\subsection{Widely applicable information criteria} \label{sec:WAIC}

We evaluate predictive accuracy of our models using the Watanabe-Akaike (Widely Applicable) Information Criteria (WAIC) defined as

\begin{equation}\label{waic}
    \mbox{WAIC} = -2(\mbox{lppd} - p_{\mathrm{WAIC}})
\end{equation}
where $\mbox{lppd} = \sum_{t=1}^{T}\log\mathbb{E}_{\bm{\Theta}_t\mid \bm{Y}_{1:T}}\left[p({\bm{Y}}^{*}_{t}\mid \bm{\Theta}_t)\right]$ and $p_{\mathrm{WAIC}} = \sum_{t=1}^{T}\mathrm{Var}_{\bm{\Theta}_t\mid\bm{Y}_{1:T}}\left[\log p({\bm{Y}}^{*}_{t}\mid \bm{\Theta}_t)\right]$, where we use ${\bm{Y}}^{*}_{t}$ for the posterior predictive density $p({\bm{Y}}^{*}_{t}\mid \bm{Y}_{1:T})$. The ``lppd'' is the ``log pointwise predictive density'' measuring how well the model fits the data, while $p_{\mathrm{WAIC}}$ is the sum of the posterior variance of the log-predictive density and estimates the effective number of parameters and serves as a penalty for the model. The difference between lppd and $p_{\mathrm{WAIC}}$ is multiplied by $-2$ to be on the deviance scale \citep{Vehtari_Gelman_Gabry_2017, ChanGolston_Banerjee_Handcock_2020}. Hence, lower WAIC values indicate preferred models.

Since our densities are predominantly matrix-normal, with $\bm{\Sigma}\sim \mathcal{IW}(n_T, \bm{D}_T)$ and $\bm{\Sigma} = \sigma^{2}\bm{R}$ being integrable from the $\mathcal{MNIW}$ and $\mathcal{MNIG}$ respectively, the lppd can be computed analytically for each. Without loss of generality, we address the case with $\bm{\Sigma} \sim \mathcal{IW}(n_T, \bm{D}_T)$, with moments computed with Algorithm~\ref{alg:FFBS}: $\mathbb{E}_{\bm{\Theta}_t\mid \bm{Y}_{1:T}}\left[p(\bm{Y}^{*}_t\mid \bm{\Theta}_t)\right] = p(\bm{Y}^{*}_t\mid \bm{Y}_{1:T}) = \mathcal{HT}(\bm{Y}^{*}_t\mid \bm{F}_t\bm{h}_t, \bm{F}_t\bm{H}_t\bm{F}^{\top}_t + \bm{V}_t, n_T, \bm{D}_T)$. The log-density follows from the expression for the Hyper-T in equation \eqref{eq:hyper_t}:
\begin{equation}\label{eq:lppd_analytic}
    \begin{split}
    \mbox{lppd} =& -\frac{NST}{2}\log\pi + \sum_{t=1}^{T}\left\{\log\Gamma_{S}\left(\frac{n_T + N}{2}\right) - \log\Gamma_{S}\left(\frac{n_T}{2}\right)\right.\\
     &\left.- \frac{S}{2}\log\det(\bm{F}_t\bm{H}_t\bm{F}_t^{\top} + \bm{V}_t) - \frac{N}{2}\log\det(\bm{D}_T)\right.\\
     &\left.- \frac{n_T + N}{2}\log\det\left(\bm{I}_{S} + \bm{D}^{-1}_{T}(\bm{Y}_t - \bm{F}\bm{h}_t)^{\top}(\bm{F}_t\bm{H}_t\bm{F}_t^{\top} + \bm{V}_t)^{-1} (\bm{Y}_t - \bm{F}_{t}\bm{h}_t)\right)\right\} 
    \end{split}
\end{equation}
Sampling is required to compute the $p_{\mathrm{WAIC}}$, with $L$ samples of $\bm{\Theta}_t \mid \bm{Y}_{1:T}$ taken to get the empirical $p_{\mathrm{WAIC}}$: $\hat{p}_{\mathrm{WAIC}} = \sum_{t=1}^{T}\mathrm{Var}\left[\left\{\log p\left(\bm{Y}^{*}_t\mid \bm{\Theta}^{(l)}_t\right)\right\}_{l = 1,\ldots,L}\right]$.

Patterning after \citep{Vehtari_Gelman_Gabry_2017}, we compute the standard error of the WAIC by taking the sample variance across time. Denote the term-wise WAIC by the following:
\begin{equation}\label{eq:waic_t}
    \mbox{WAIC}_t = -2\left(\log\mathbb{E}_{\bm{\Theta}_t\mid \bm{Y}_{1:T}}[p(\bm{Y}_{t}^{*}\mid \bm{\Theta}_t)] - \mathrm{Var}_{\bm{\Theta}_t\mid \bm{Y}_{1:T}}[\log p(\bm{Y}_{t}^{*}\mid \bm{\Theta}_t)]\right)
\end{equation}

Then the standard error of the WAIC can be computed as follows:

\begin{equation}
    \label{eq:waic_se}
    \mbox{se}(\mbox{WAIC}) = \sqrt{T\mathrm{Var}\left(\{\mbox{WAIC}_t\}_{t=1,\ldots,T}\right)}
\end{equation}

\subsection{Posterior predictive loss criteria} \label{sec:post_loss_criteria}

This criteria based on \citep{Gelfand_Ghosh_1998} involves sampling independent replicates of the model outcome based on the distribution's computed moments. Denote one such set of replicates by $\bm{Y}_{t}^{\text{(rep)}}$, where $\bm{Y}_{t}^{\text{(rep)}}$ is a ``future'' observation which is replicated from the distribution of $\bm{Y}_{t}$. However, since we account for the actual observations $\bm{Y}_{1,\ldots,T}$ in our criteria, the replicate we sample is actually $\bm{Y}_{t}^{\text{(rep)}}\mid \bm{Y}_{1,\ldots,T}$. For $L$ such replicates, we sample $\{\bm{Y}_{t}^{\text{(rep)},l}\}_{l=1,\ldots,L}$. Then the sample mean of the $L$ samples is $\hat{\bm{\mu}}^{\text{(rep)},L}_t = \frac{1}{L}\sum_{l=1}^{L}\bm{Y}_{t}^{\text{(rep)},l}$ and its per-coordinate variance is $\hat{\sigma}^{2}_{t}(\bm{x}_i, \bm{s}_j) = \mathrm{Var}((y_{t}(\bm{x}_i,\bm{s}_j )^{\text{(rep)},l})_{l=1,\ldots,L})$. These two moments are used to define a $D$-score \cite{Gelfand_Ghosh_1998} as the sum of a goodness of fit measure ($G$) and a penalization term ($P$). Here, $D=G+P$, where $G$ and $P$ are defined as
\begin{equation}
    \label{eq:GPD_coords}
    G = \sum_{t}\sum_{i,j}(y_{t}(\bm{x}_i,\bm{s}_j) - \hat{\mu}_{t}^{\text{(rep),L}}(\bm{x}_i, \bm{s}_j))^{2}; \qquad P = \sum_{t}\sum_{i,j}\hat{\sigma}^{2}_{t}(\bm{x}_i, \bm{s}_j)
\end{equation}
Due to the normality inherent in the DLM, both of our moments, and hence the $G$ and $P$ scores, can be computed analytically without requiring samples. Hence, $\bm{\mu}_{t}^{\text{(rep)}} = \mathbb{E}_{\bm{Y}^{\text{(rep)}}_{t}\mid\bm{Y}_{1,\ldots,T}}\hat{\bm{\mu}}_t^{\text{(rep)},L} = \bm{F}_{t}\bm{h}_{t}$ (the values of $G$ do not depend on the covariance structure $\bm{\Sigma}$), while $\sigma^{2}_{t}(\bm{x}_i,\bm{s}_j) = \mathbb{E}_{\bm{Y}^{\text{(rep)}}_{t}\mid\bm{Y}_{1,\ldots,T}}\hat{\sigma}^{2}_{t}(\bm{x}_i, \bm{s}_j)$ takes the following forms for each model:
\begin{equation}
    \label{eq:P_terms}
    \sigma^{2}_{t,\mathrm{IW}} (\bm{x}_i, \bm{s}_j) = \frac{H_{t,ii}D_{T,jj}}{n_T - 2}; \qquad 
    \sigma^{2}_{t,\mathrm{IG}} (\bm{x}_i, \bm{s}_j) = \frac{H_{t,ii}d_T R_{jj}}{n_T - 1};\qquad \sigma^{2}_{t,\mathrm{id}} (\bm{x}_i, \bm{s}_j) = H_{t,ii};
\end{equation}
where $\sigma^{2}_{t,\mathrm{IW}}$, $\sigma^{2}_{t,\mathrm{IG}}$ and $\sigma^{2}_{t,\mathrm{id}}$ denote individual terms in $P$ corresponding to $\bm{\Sigma}$ being inverse-Wishart, $\bm{\Sigma} = \sigma^{2}\bm{R}$ and $\bm{\Sigma} = \bm{I}$, respectively. Summed over the $N$ inputs $\bm{x}_i$ and $S$ spatial coordinates $\bm{s}_j$, we write $G = \sum_{t}\lvert\lvert\bm{Y}_{t} - \bm{\mu}_{t}^{\left(rep\right)}\rvert\rvert^{2}_{F}$ ($\|\cdot\|_F$ is the Frobenius norm) and $P$ as
\begin{equation}
    \label{eq:P_sums}
   P_{\mathrm{IW}} = \frac{\mathrm{tr} (\bm{D}_{T})}{n_T - 2}\sum_{t}\mathrm{tr}(\bm{H}_{t}); \qquad 
    P_{\mathrm{IG}} = \frac{d_T \mathrm{tr}(\bm{R})}{n_T - 1}\sum_{t}\mathrm{tr}(\bm{H}_t); \qquad P_{\mathrm{id}} = S\sum_{t}\mathrm{tr}(\bm{H}_t)
\end{equation}
As in \eqref{eq:P_terms}, each of the terms $P_{\cdot}$ in \eqref{eq:P_sums} denotes the $P$ for the model corresponding to the variance structure $\bm{\Sigma}$.
The lower the D-score of the distribution, the better the fit of the data to the model. Table \ref{table:GPD} shows the results of this statistic for the FFBS applied to the data.

\section{A Bayesian Transfer Learning Approach for BIG DATA settings}\label{sec:bayestransferlearning}
The problem of emulating high-dimensional computer model outputs has been comprehensively addressed by \cite{Higdon2008, gram14, gu2016parallel, gramacy2022deepGP}. Most of these developments have revolved around models that achieve dimension reduction using scalable derivations of Gaussian processes. Even a cursory review reveals a significant literature on statistical methods for massive spatial datasets, which is too vast to be summarized here \citep[see, e.g.,][and references therein]{Banerjee2017, Heaton2019}. Within the Bayesian setting, inference proceeds from spatial processes that scale massive data sets. Examples range from reduced-rank processes or subsets of regression approaches \citep[][and references therein]{Quinoner2005, snelson2005sparse, cressie2008frk, Banerjee2008, Wikle2011, wilson2015kernel}, multi-resolution approaches \citep{nychka2015, katzfussmultires}, and graph-based models inducing sparsity \citep{Datta2016, katzfuss2021general, Krock2021modeling, dey2022graphical, gramacy2022deepGP, cao2023vChol}.

We depart from the above model-based approaches and adopt a Bayesian transfer learning approach that will divide and conquer a dataset of possibly massive dimensions, construct a sequence of smaller datasets, and then apply the FFBS algorithm described in Algorithm~\ref{alg:FFBS} to this sequence. In spirit, this is similar to spatial meta-kriging or predictive stacking \citep[see][and references therein]{guhaniyogi2017divide, guhaniyogi2018meta, presicce2024bayesian}. To elucidate further, let us use the analogy of a streaming show where the data at each $t=1,\ldots,T$ is referred to as a ``season'' for the show and each season consists of $K$ subsets of the data, which we refer to as ``episodes''. This scheme is depicted in Figure~\ref{fig:bigdata}, where the data (season) at time $t$ is partitioned into $K$ subsets (episodes) denoted by $\mathcal{D}_{1t}, \ldots, \mathcal{D}_{Kt}$. Each $\mathcal{D}_{kt} = \{\bm Y_{k,t}, \bm F_{k,t}, \bm G_{k,t}, \bm V_{k,t}, \bm W_{k,t}\}$, where each $\bm Y_{k,t}$ is $r\times c$, $\bm F_{k,t}$ is $r\times p$, $\bm G_{k,t}$ is $p\times p$, $\bm V_{k,t}$ and $\bm W_{k,t}$ are positive definite covariance matrices of dimension $r\times r$ and $p\times p$, respectively. One convenient specification emerges naturally by treating each of the $K$ matrices $\bm Y_{k,t}$ as submatrices of $\bm Y_{t}$. We let $K = k_1k_2$, where $rk_1 = N$ and $ck_2 = S$ where we choose $r$ and $c$ to ensure that we can execute Algorithm~\ref{alg:FFBS} on each $\mathcal{D}_{k,t}$ as a sequence of datasets over $KT$ time points.    

The preceding divide and conquer method conveniently retains the conjugate distribution theory underlying Algorithm~\ref{alg:FFBS}. The transfer learning mechanism scales the emulation of mechanistic systems when either $N$ or $S$ is large. Specifying $\bm V_{k,t}$ using Gaussian processes, as described in Section~\ref{GPR}, allows interpolation of $\bm Y_{k,t}(\tilde{\bm\chi})$ for arbitrary units $\tilde{\bm\chi}$. However, inference on the complete spatial field involving $S$ spatial locations is precluded as the $\bm\Sigma$ matrix only captures the dependence among the $c$ columns of $\bm Y_{k,t}$ in each episode. If full inference on the spatial field is desired, then we model the $S$ columns of the full dataset as a spatial covariance matrix from which we extract the $c\times c$ matrix corresponding to each episode in the transfer learning algorithm. More specifically, we consider the model 
\begin{equation}
    \label{eq:dlm_sp}
    \begin{split}
        \bm Y_{k,t} &= \bm F_{k,t} \bm{\Theta}_{k,t} + \bm{\mathcal{E}}_{k,t}, \quad \bm{\mathcal{E}}_{k,t}\stackrel{\text{ind.}}{\sim}\mathcal{MN}_{r\times c}(\bm O,\bm V_{k,t},\bm \Sigma)\\
        \bm{\Theta}_{k,t} &= \bm{G}_{k,t}\bm{\Theta}_{\mbox{pa}[k,t]} + \bm\Gamma_{k,t}, \quad \bm\Gamma_{k,t}\stackrel{\text{ind.}}{\sim} \mathcal{MN}_{p\times c}(\bm O, \bm W_{k,t}, \bm \Sigma),
    \end{split}
\end{equation}
where the indices $\{k,t\}$ refer to the $k$-th episode in season $t$, $\bm V_{k,t}$ and $\bm W_{k,t}$ are matrices with rows and columns corresponding to the episode $k$ within season $t$ and $\mbox{pa}[k,t]$ denotes the index that immediately precedes (is the ``parent'' of) the index $\{k,t\}$ and defined as  
\[
    \mbox{pa}[k,t] = \left\{\begin{array}{l}
	\{k-1,t\}\; \mbox{ if }\; k=2,\ldots,K\;;\quad t=1,2,\ldots,T \\
	\{K,t-1\}\; \mbox{ if }\; k=1\;;\quad t=2,\ldots,T \\
    \{0,0\}  \; \mbox{ if }\; k=1,t=1\;.
	\end{array} \right. \;.
\]
Algorithm~\ref{alg:FFBS} requires that we maintain a shared covariance matrix among the columns of $\bm Y_{k,t}$ for each $k$ and $t$, which is possible only if each episode has the same number of columns. Since each $Y_{k,t}$ is $r\times c$, it follows that the $c\times c$ covariance matrix $\bm\Sigma$ models the covariance among the columns for every episode in the entire stream.

To analyze the computational complexity of Algorithm ~\ref{alg:FFBS} under the transfer learning framework, we replace the number of locations with $c$ and the number of time points with $K\times T$. This substitution yields a computational cost of $\sim O(KTLc^3)$. The primary benefit of transfer learning in this context is the reduction of the cubic term in complexity from $S^3$ to $Kc^3$, which can substantially alleviate the computational burden.

\begin{figure}[t]
    \centering
    \begin{tikzpicture}[
baseline={(current bounding box.center)},
node distance = 1 mm and 1 mm,
nobox/.style = {text width=.3cm},
noboxbig/.style = {minimum size=1.3 cm},
box/.style = {rectangle, draw=black!100, thick, minimum size=1.3 cm},]

\node[box] (Y1) {$\bf{Y}_1$};
\node[nobox] (Xlabel1) [left=of Y1] {$\chi$};
\node[nobox] (Slabel1) [above=of Y1] {$\mathcal{S}$};
\node[noboxbig] (D1s) [below=0mm of Y1,xshift=-0mm] {$\mathcal{D}_{11}\cup \cdots \cup \mathcal{D}_{K1}$};

\node[nobox] (Xlabel2) [right=1 cm of Y1] {$\chi$};
\node[box] (Y2) [right=of Xlabel2] {$\bf{Y}_2$};
\node[nobox] (Slabel2) [above=of Y2] {$\mathcal{S}$};
\node[noboxbig] (D2s) [below=0mm of Y2,xshift=0mm] {$\mathcal{D}_{12}\cup \cdots \cup \mathcal{D}_{K2}$};

\node[noboxbig] (cdots) [right=1 cm of Y2] {$\cdots$};
\node[noboxbig] (subset) [below=0mm of cdots] {};

\node[nobox] (XlabelT) [right=1 cm of cdots] {$\chi$};
\node[box] (YT) [right=of XlabelT] {$\bf{Y}_T$};
\node[nobox] (SlabelT) [above=of YT] {$\mathcal{S}$};
\node[noboxbig] (DTs) [below=0mm of YT,xshift=0mm] {$\mathcal{D}_{1T}\cup \cdots \cup \mathcal{D}_{KT}$};

\draw[->] (Y1) -- (Xlabel2) node[midway,below=1.05cm] () {};
\draw[->] (Y2) -- (cdots);
\draw[->] (cdots) -- (XlabelT);

\end{tikzpicture}
    \caption{The transfer learning scenario. At each time point, a $\chi \times S$ matrix is recorded. Within each $Y_t$, information propagates from subdomains within the grid $\mathcal{D}_{1t}\cup \cdots \cup \mathcal{D}_{Kt}$, before propagating into subdomains for the next matrix $Y_{t+1}$.}
    \label{fig:bigdata}
\end{figure}

\section{Applications for Emulation}\label{sec:app_emulation}

\subsection{Predator-prey analysis}\label{sec:app_emulation_LV}

We apply our methodology to a multivariate dynamical system specified through a set of coupled ordinary differential equations. This may be considered a multivariate statistical modeling problem in lieu of spatial. The numerical solution to this dynamical system - which we call our computer model - is cheap to evaluate. To motivate the ecological computer model, we overview the Lotka-Volterra equations, which models the interaction between the populations of a single predator species and a single prey species:
\begin{equation}
\begin{split}
    \frac{du_t}{dt} &= \eta_1 u_t - \eta_2 u_t v_t\\
    \frac{dv_t}{dt} &= -\eta_3 v_t + \eta_4 u_t v_t\;,
\end{split}
\label{eq:lotka_volterra}    
\end{equation}
where $\eta_1$ is the growth rate of the prey population and $\eta_3$ is the rate of population loss within the predator population independent of the interactions the two populations would have with one another; $\eta_2$ and $\eta_4$ represent the respective rates at which the prey population shrinks and the predator population grows
, which is represented through the product of their respective populations. The parameters $\eta_1$, $\eta_2$, $\eta_3$, and $\eta_4$ must be positive; furthermore, the values of $\eta_1, \ldots, \eta_4$ must correspond to the units of the quantities they multiply. Thus, $\eta_2$ and $\eta_4$, in particular, should be one or two orders of magnitude smaller than $\eta_1$ and $\eta_3$, since they correspond to the product of the populations of the predators and prey.

\begin{figure}[t]
    \centering
    \includegraphics[width=0.32\linewidth]{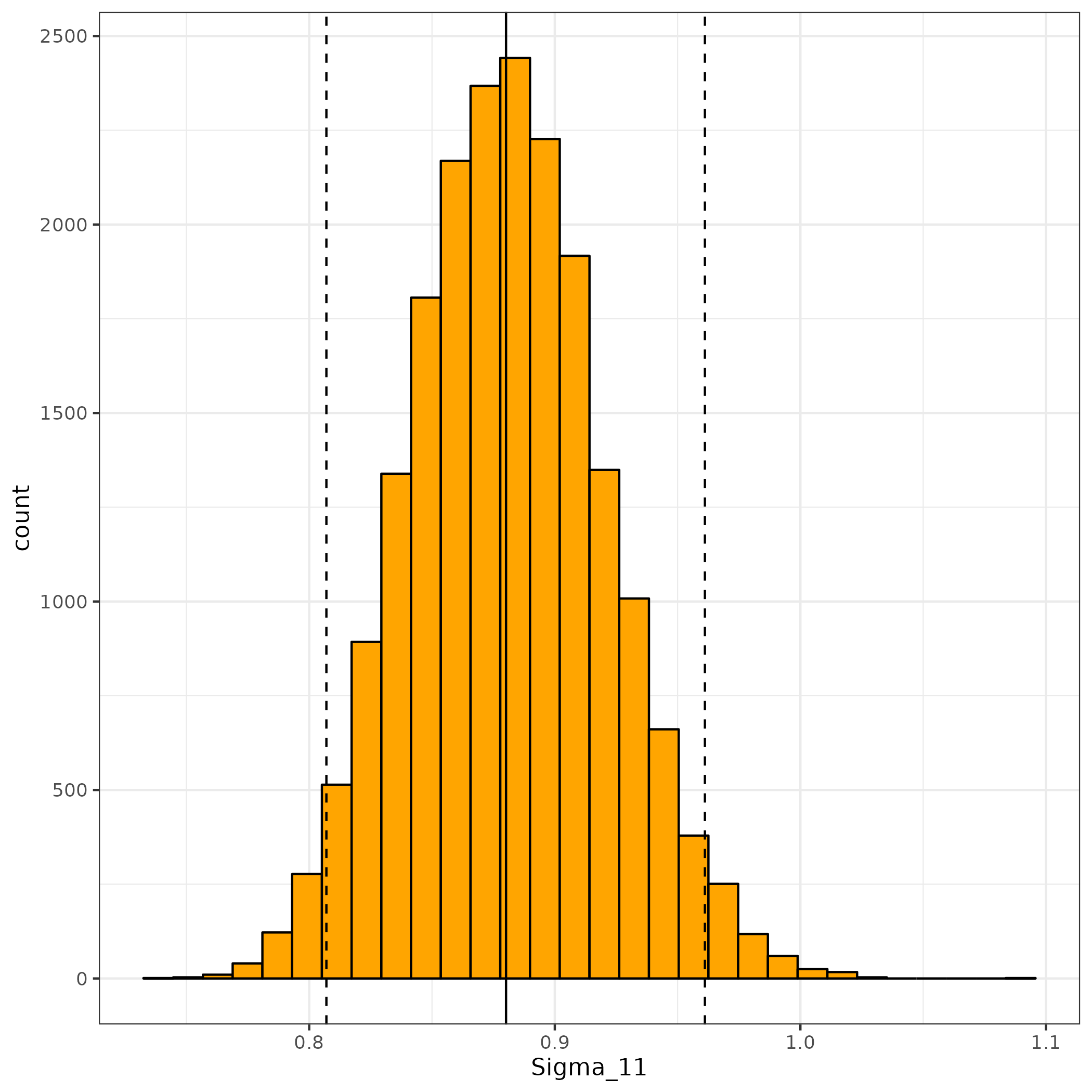}
    \includegraphics[width=0.32\linewidth]{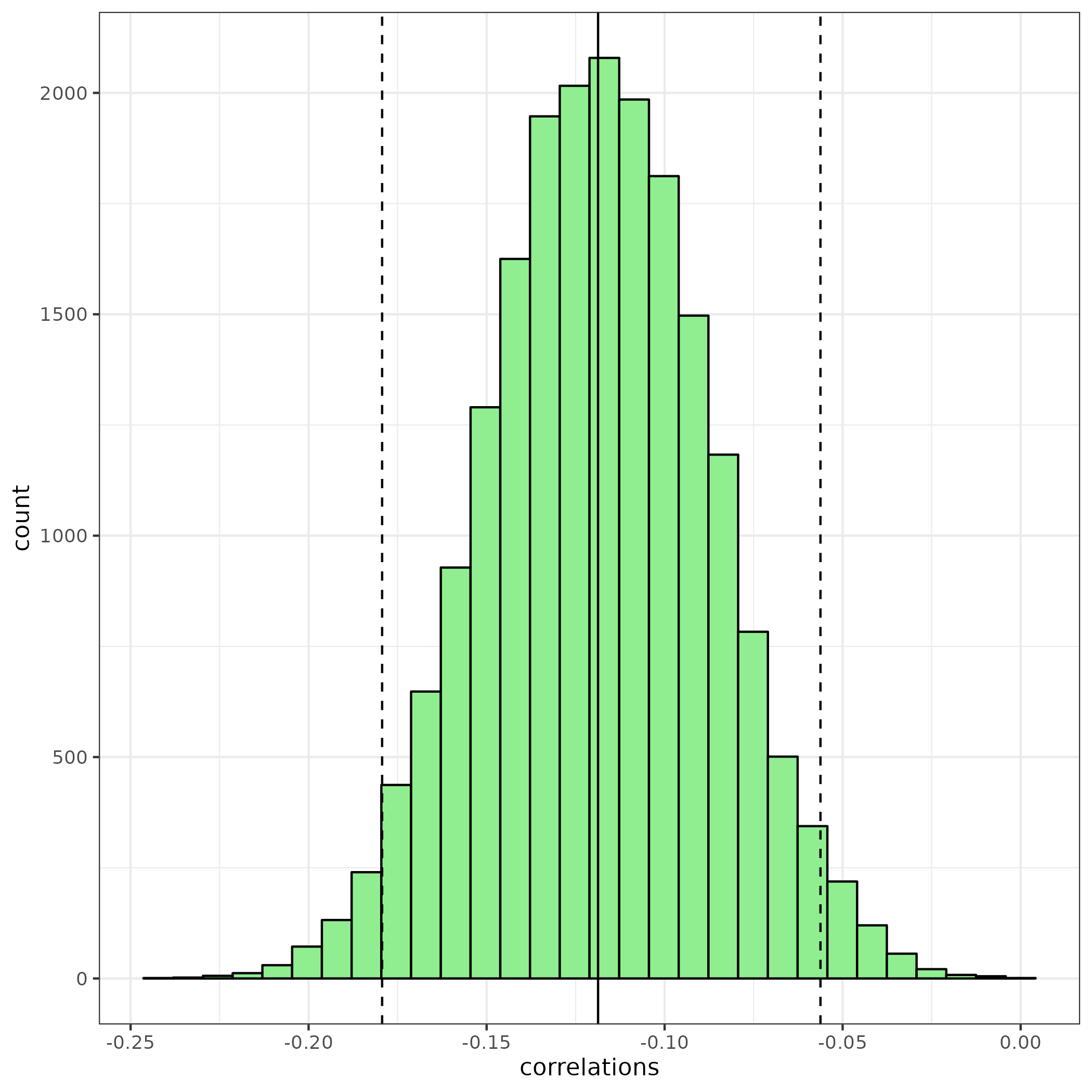}
    \includegraphics[width=0.32\linewidth]{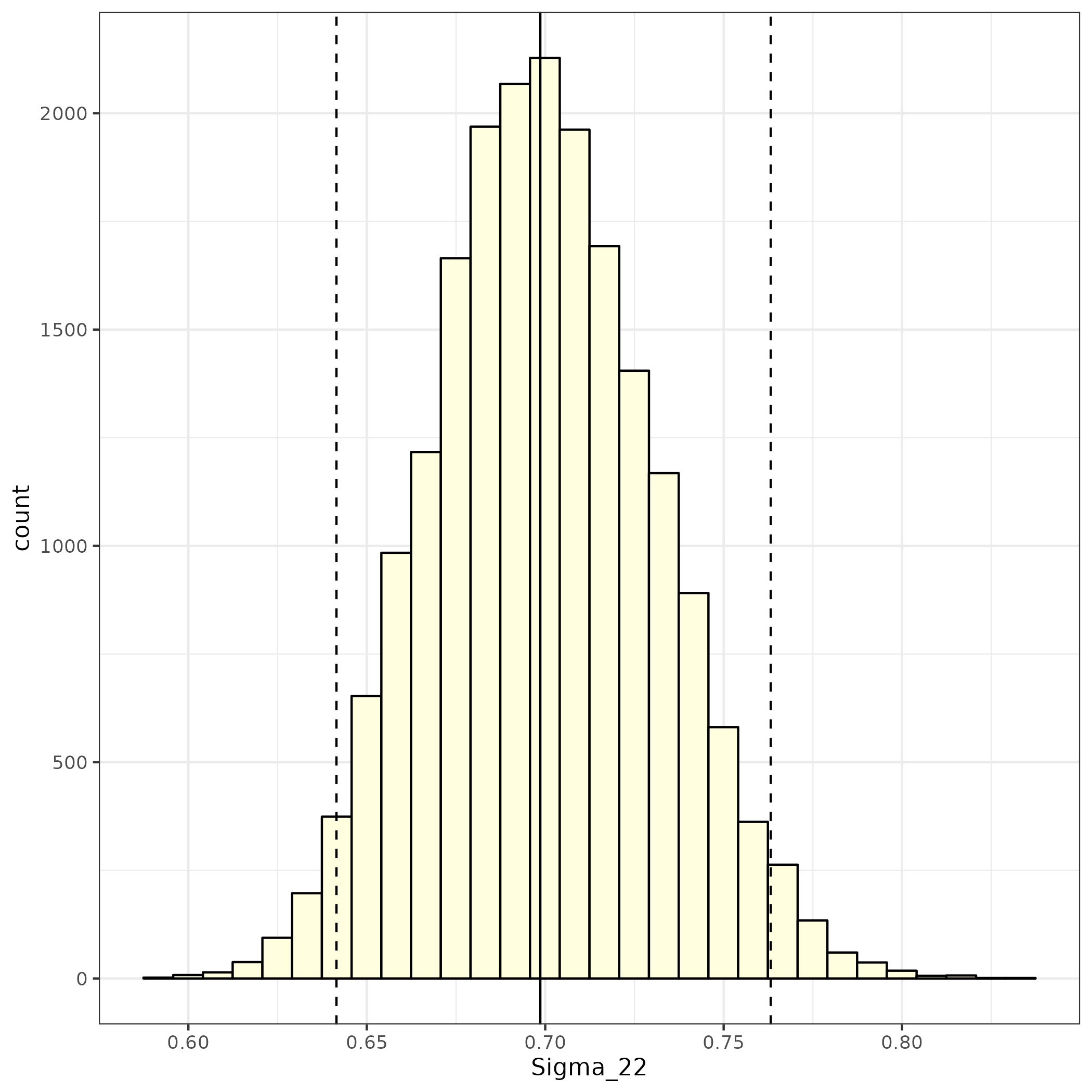}
    \caption{Posterior distributions of the variances of the prey (left), the predator (right), and the correlation between the two populations (center). The 95\% credible interval of the correlation is $(-0.179, -0.056)$, capturing the negative relation between the (log-transformed) predator and prey populations.}
    \label{fig:LV_corr_hists}
\end{figure}

To train the FFBS model using Algorithm \ref{alg:FFBS}, we generate $N = 50$ sets of lognormal sampled parameters, so that $\eta_i \sim \exp\left(\mathcal{N}(\mu_i, \sigma_i)\right)$ for $i=1,\ldots,4$
, where $\mu_i$ and $\sigma_i$ are the respective means and standard deviations of the normal random variables prior to exponentiation. The training data is generated over a Latin hypercube design from the multivariate lognormal with $\mu_1 = \mu_3 = 0$ and $\mu_2 = \mu_4 = -3$, and $\sigma_1 = \sigma_2 = \sigma_3 = \sigma_4 = 0.5$, so that $\eta_1$ and $\eta_3$ will be close to 1 and $\eta_2$ and $\eta_4$, which control the prey's population decline and the predator's population growth respectively, will be close to 0.05.

We let $\bm Y_t$ be the $N \times 2$ matrix, where row $i$ corresponds to the log-transformed solutions $[u_t, v_t]$ of \eqref{eq:lotka_volterra} for the $i$th training parameter $\bm\eta_i$, $i=1,\ldots,N$. The design matrix for the model approximation is the data matrix from the prior time point in an AR(1) setup, so that $\bm F_t = \bm Y_{t-1}$ for $t = 1,\ldots,T$ \citep[see, e.g.,][for details of such formulations]{HarrisonWest1997}. 
We set $\bm Y_0$ to the log-population of the Canadian lynx from the data in the year 1900 and emulate $\bm Y_t$ from \eqref{eq:lotka_volterra} over $T=20$ time points. We also generate $\bm V_t = (C(\bm\eta_i; \bm\eta_j; \beta))$ using $\beta_i = \beta$ in \eqref{corr} with $\beta = \frac{3}{0.5\times d_{\max}}$, where $d_{\max}$ is the maximum distance between $\bm\eta_i$'s, and we let $\bm G_t$ and $\bm W_t$ be identity matrices. 




Also of interest in fitting Algorithm \ref{alg:FFBS} to the data is 
a measure for the covariance between the log-predator and log-prey populations, which enables us to compute a correlation measure to interpret the results we have been given. Figure~\ref{fig:LV_corr_hists} presents the posterior distributions using 20,000 posterior samples of $\bm \Sigma$. The left and right figures are the posterior samples of the diagonal elements, $\Sigma_{11}$ and $\Sigma_{22}$, of $\bm\Sigma$ while the middle panel is the correlation $\Sigma_{12}/\sqrt{\Sigma_{11}\Sigma_{22}}$. The 95\% credible interval for the correlation is $(-0.179, -0.056)$, which complies with the dynamics of the predator and prey populations in the Lotka-Volterra cycle. As the prey population increases, the predator population increases at the expense of the prey population, enough to cause the prey to decrease dramatically. Once the prey become scarce, the predators also begin to die off when there are less prey to eat, allowing for the prey population to grow, after which the cycle repeats.

\begin{figure}[t]
    \centering
    \includegraphics[width=0.25\textwidth]{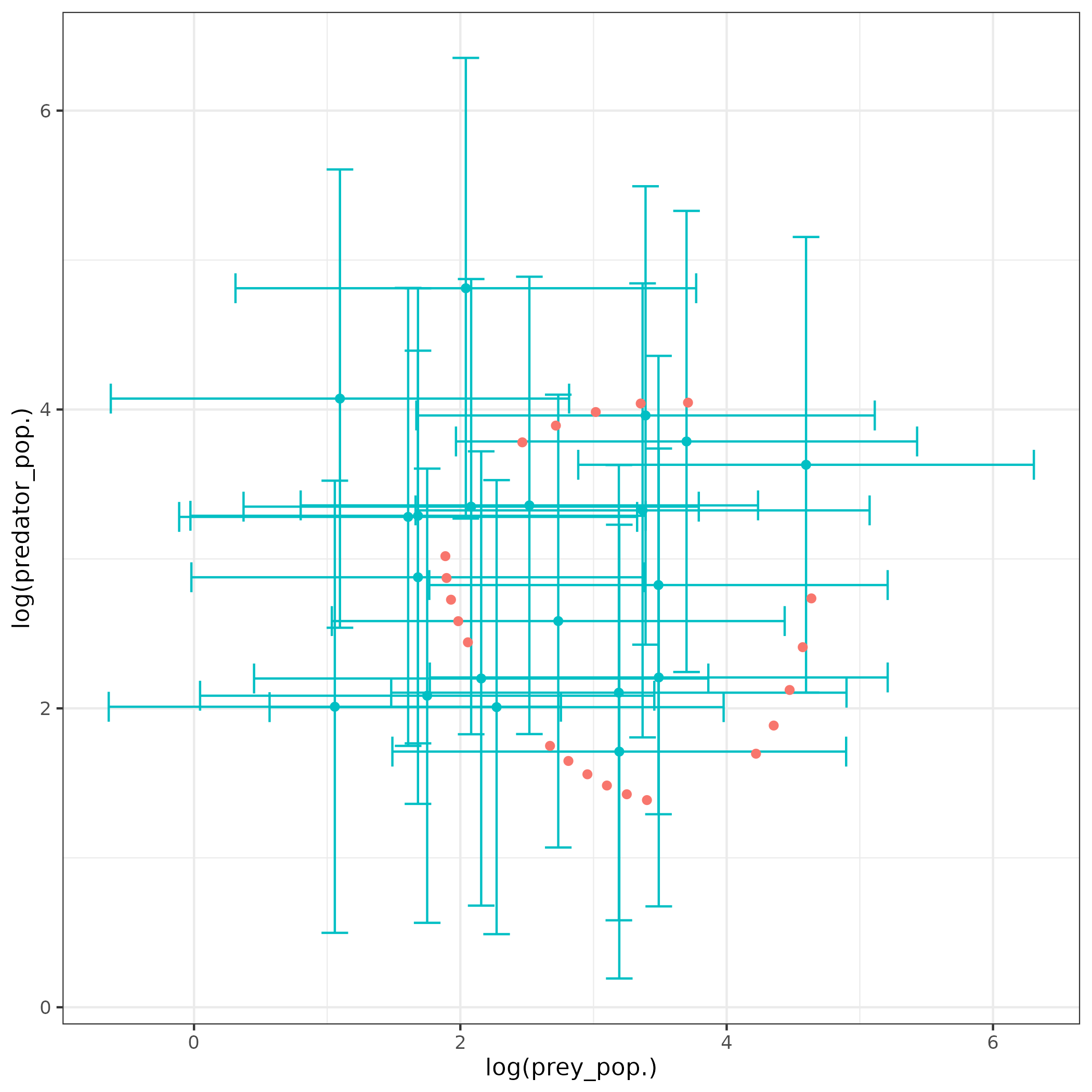}
    \includegraphics[width=0.25\textwidth]{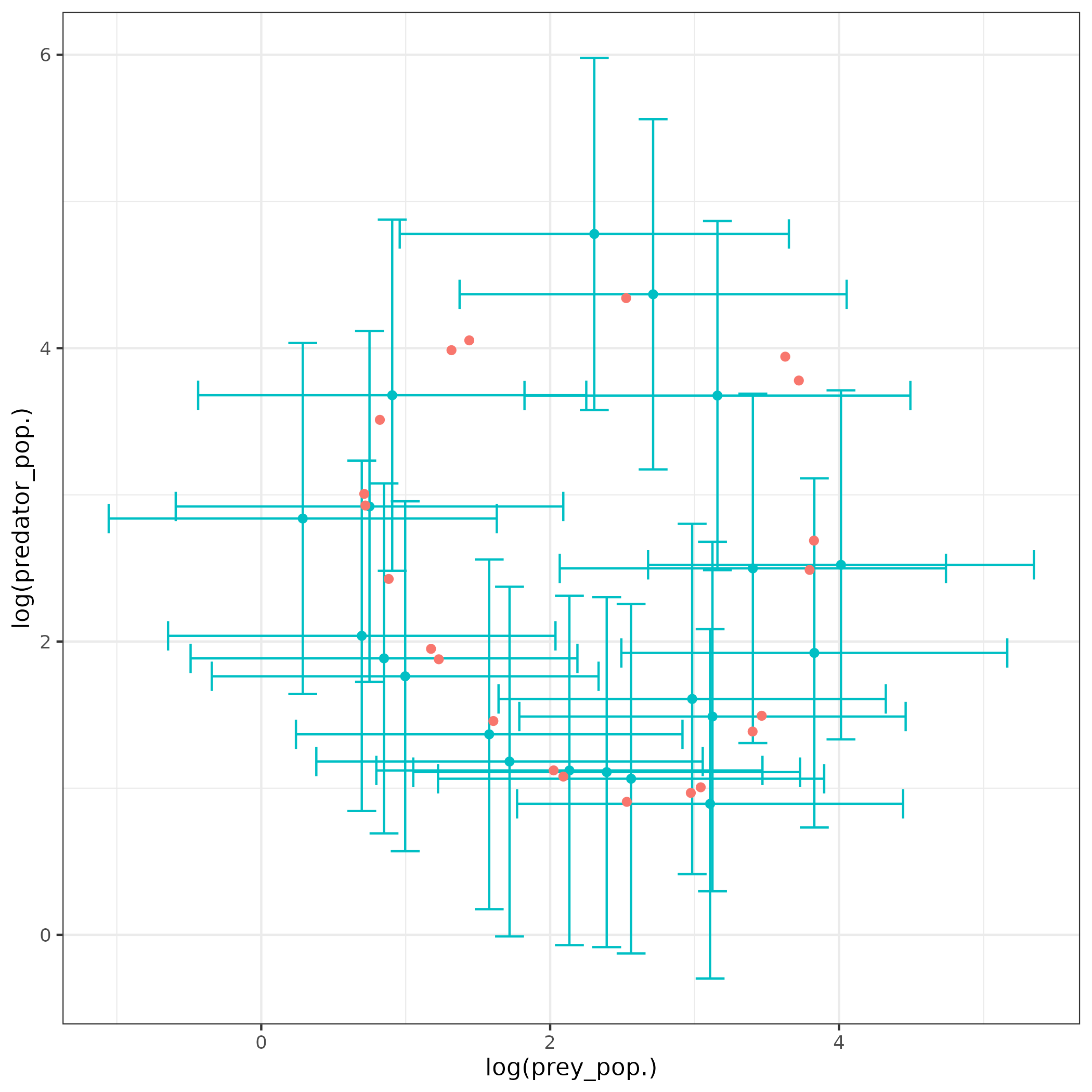}
    \includegraphics[width=0.25\textwidth]{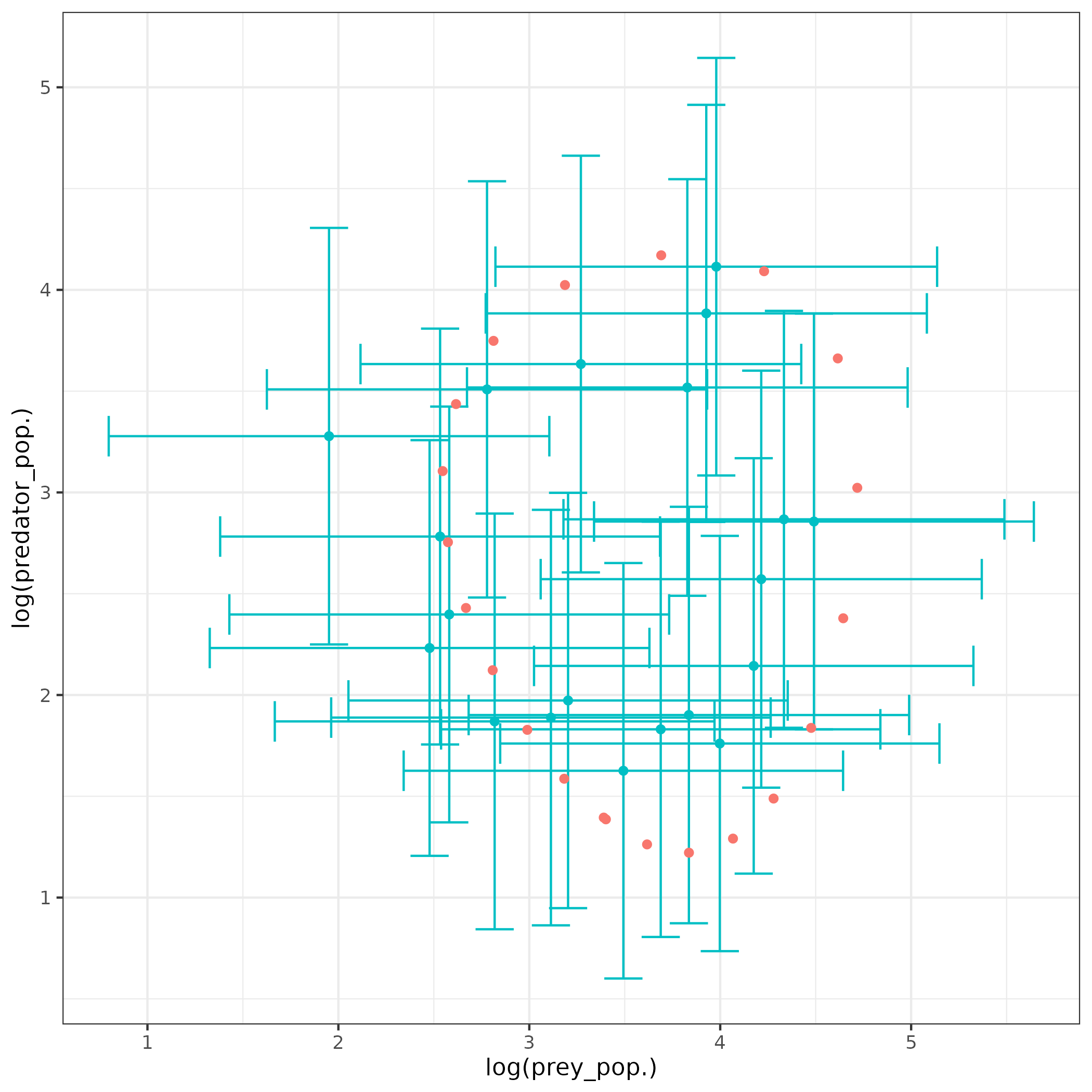}
    \includegraphics[width=0.25\linewidth]{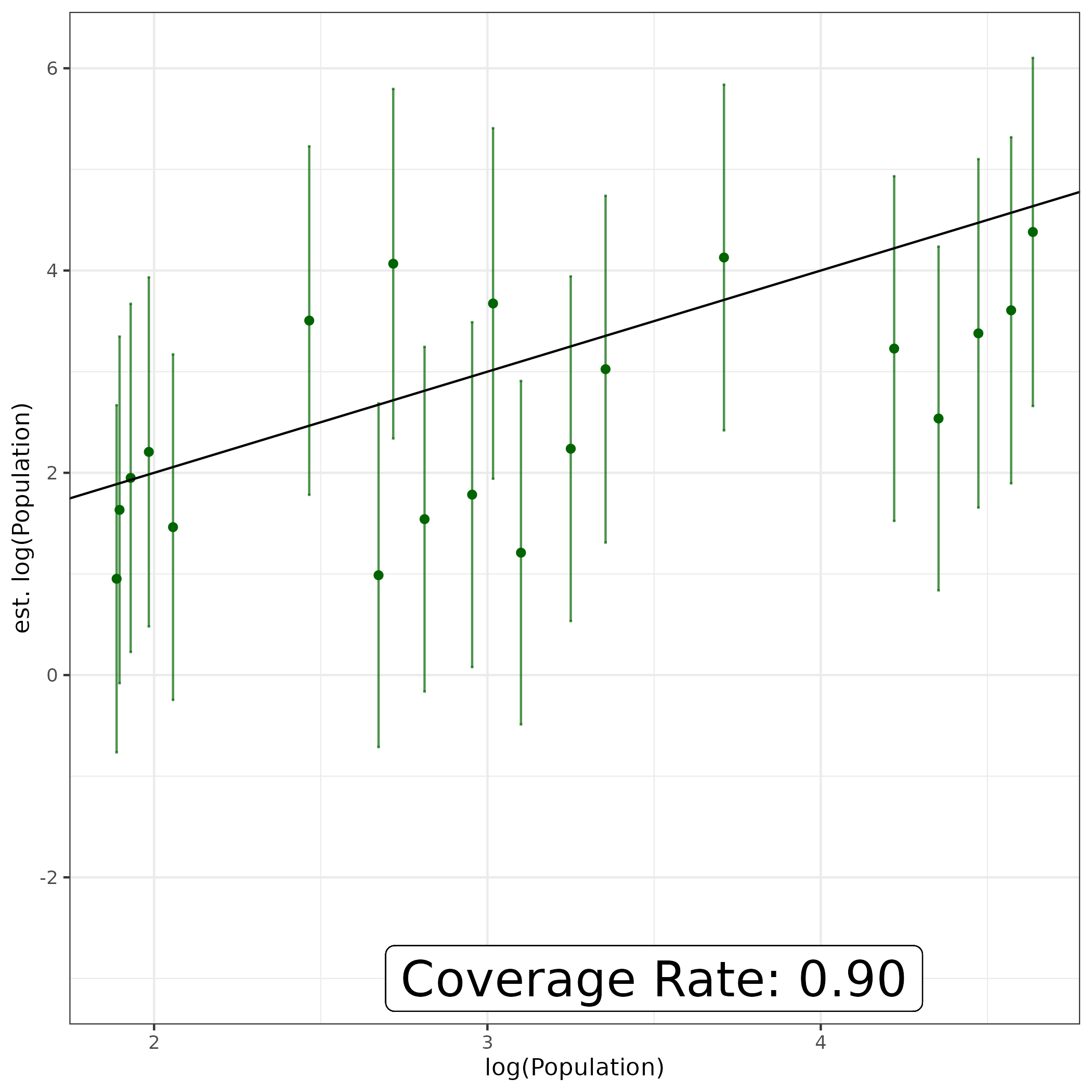}
    \includegraphics[width=0.25\linewidth]{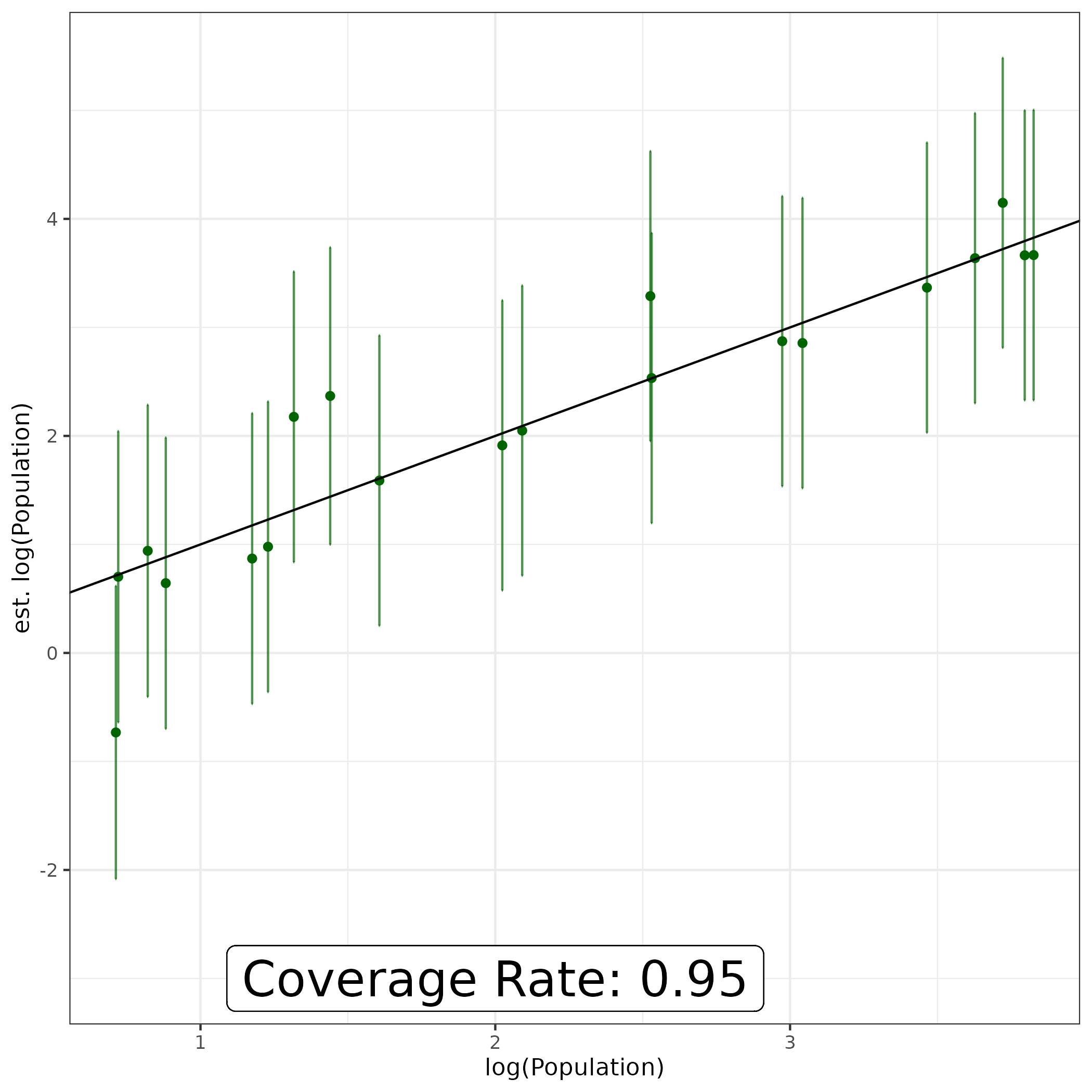}
    \includegraphics[width=0.25\linewidth]{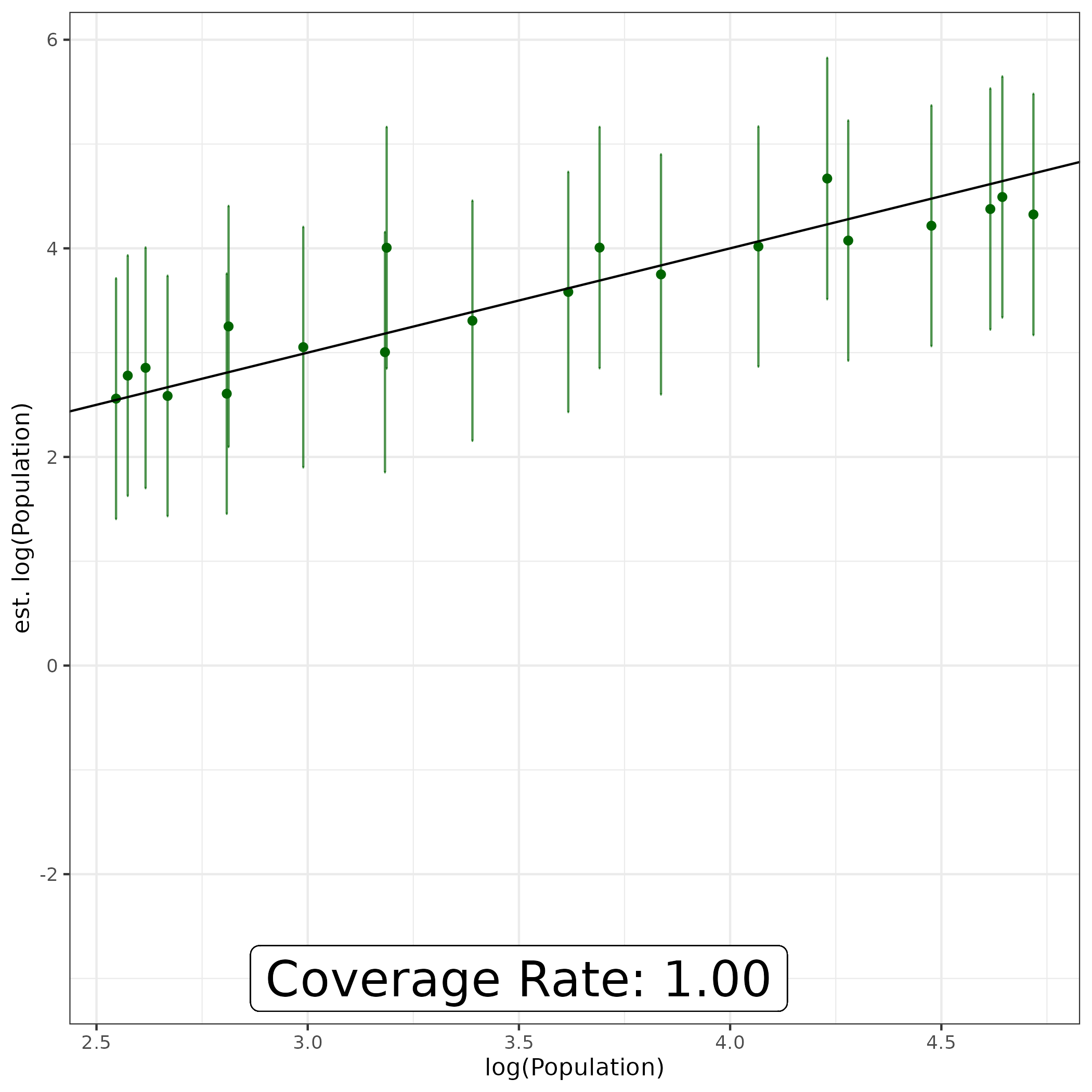}
    \includegraphics[width=0.25\linewidth]{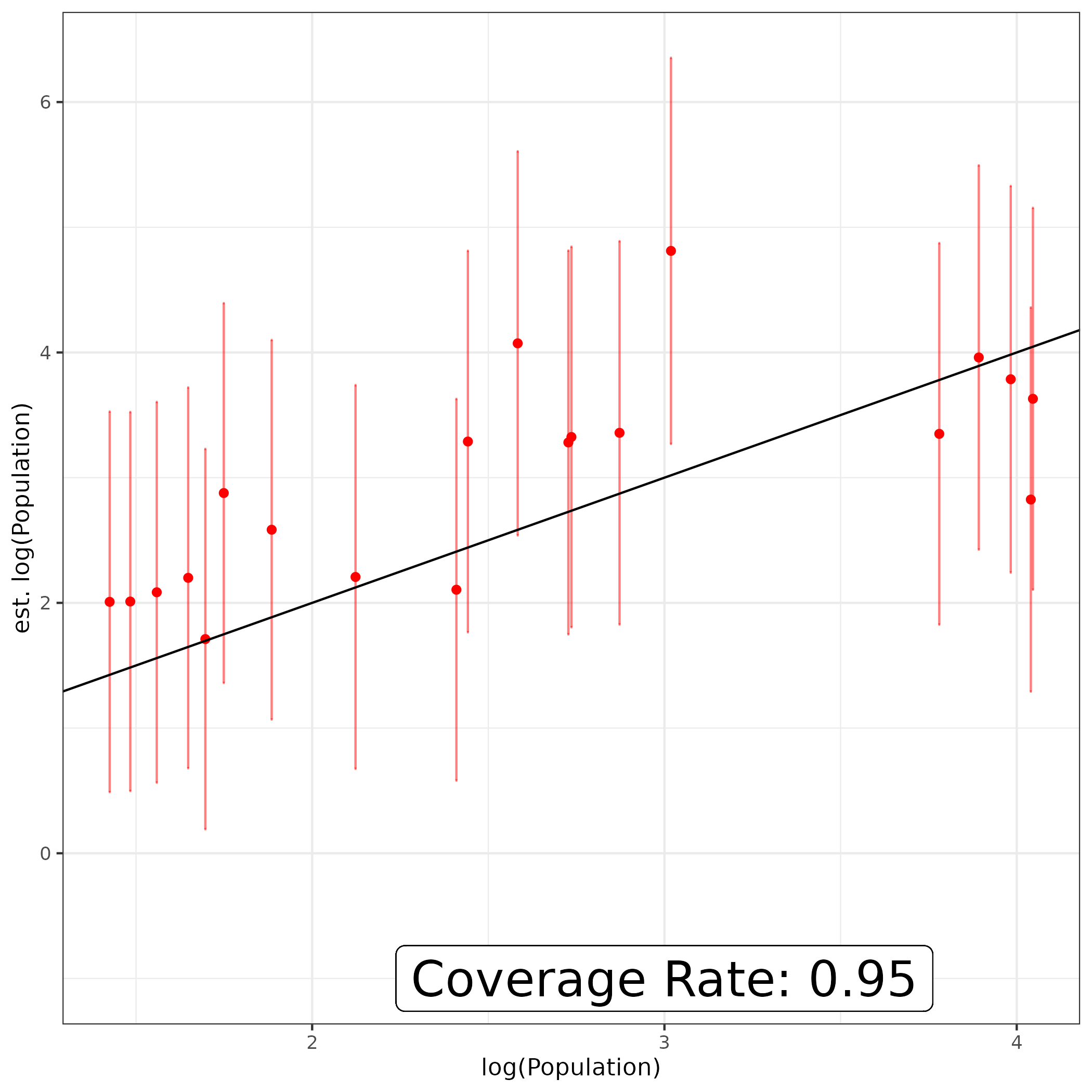}
    \includegraphics[width=0.25\linewidth]{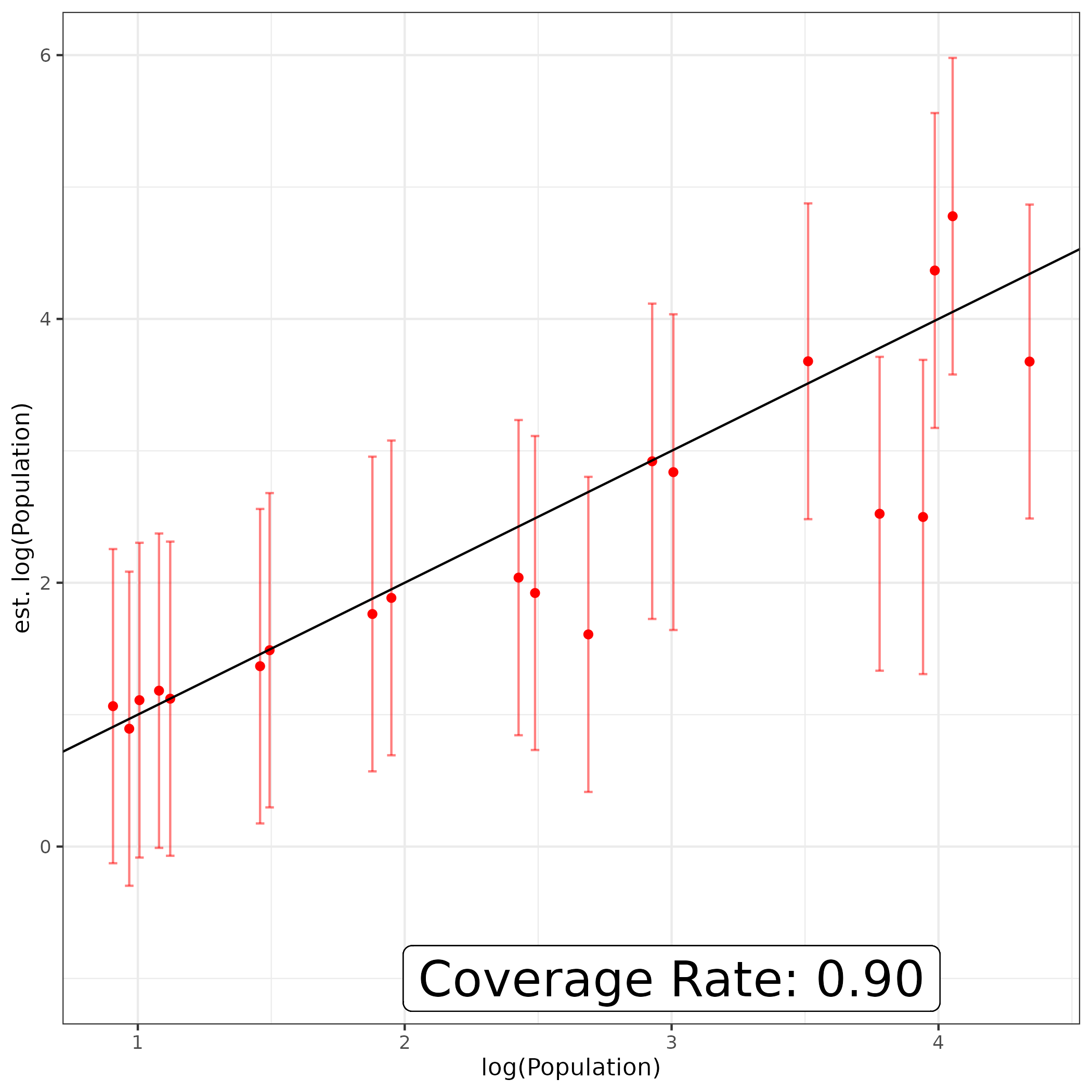}
    \includegraphics[width=0.25\linewidth]{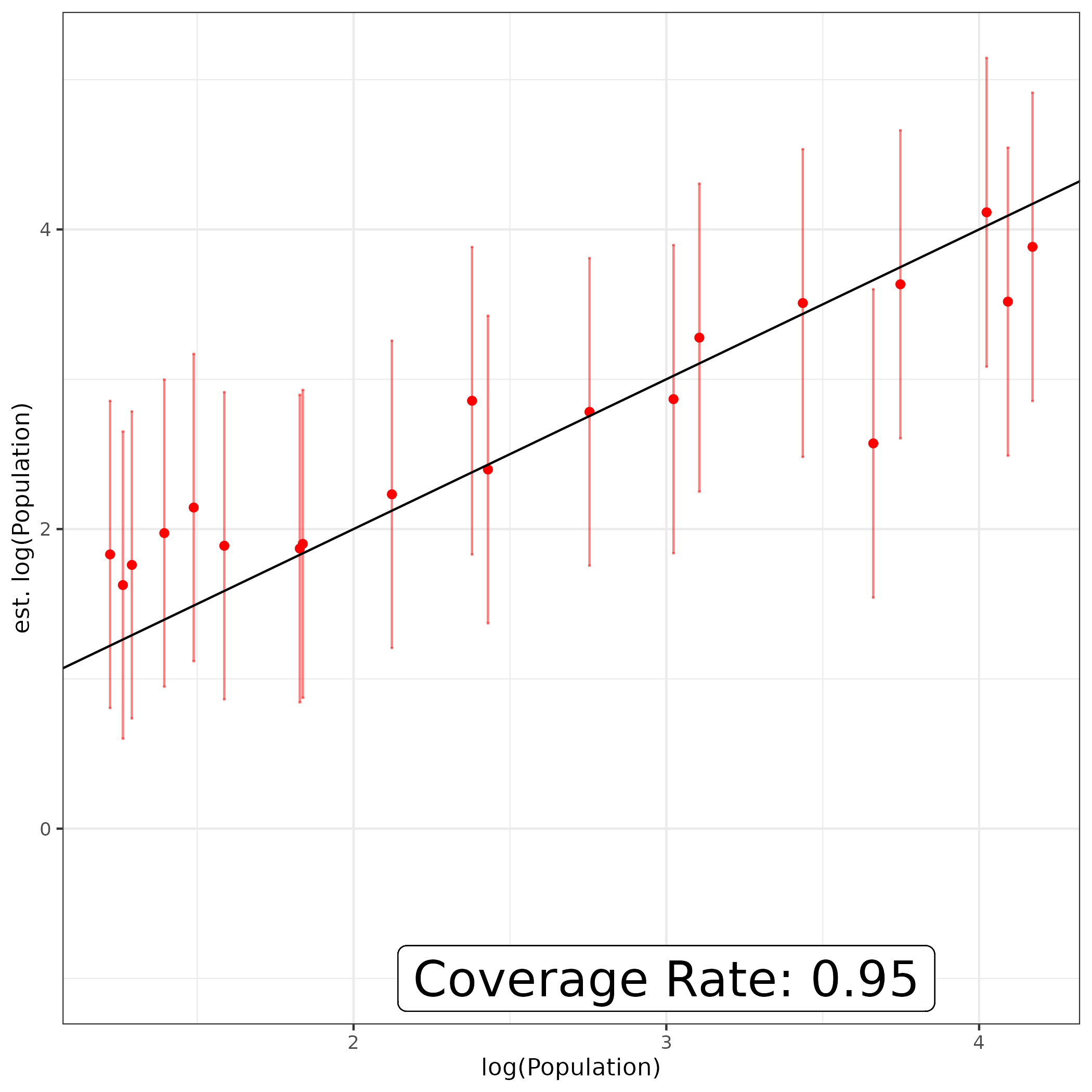}
    \caption{Prey and predator log-populations are plotted against their predicted values with 95\% credible intervals for data corresponding to inputs held out of the training phase. The top three plots contain the point estimates with their credible intervals (turquoise) with the ground truth curves in red. The middle three plots correspond to the log-prey populations and the bottom three correspond to the log-predator populations. The black diagonal line denotes the region where the predicted and true values are equal. Each column corresponds to a distinct value of $\bm \eta$.}
    \label{fig:LV_qq_plots}
\end{figure}

Analytic solutions with their credible intervals are sampled and plotted in Figure \ref{fig:LV_qq_plots}. The bottom two plots feature the comparison of log-population values between the actual values of the log-prey and log-predator populations and their estimates produced by the FFBS emulator respectively. The 95\% credible intervals accurately capture the trajectories of the predator and prey populations; particularly for the middle and right plots.

\subsection{Nonlinear Partial Differential Equation Emulation}\label{sec:app_emulation_pde}
\begin{figure}[t]
    \centering
    \includegraphics[width=0.80\textwidth]{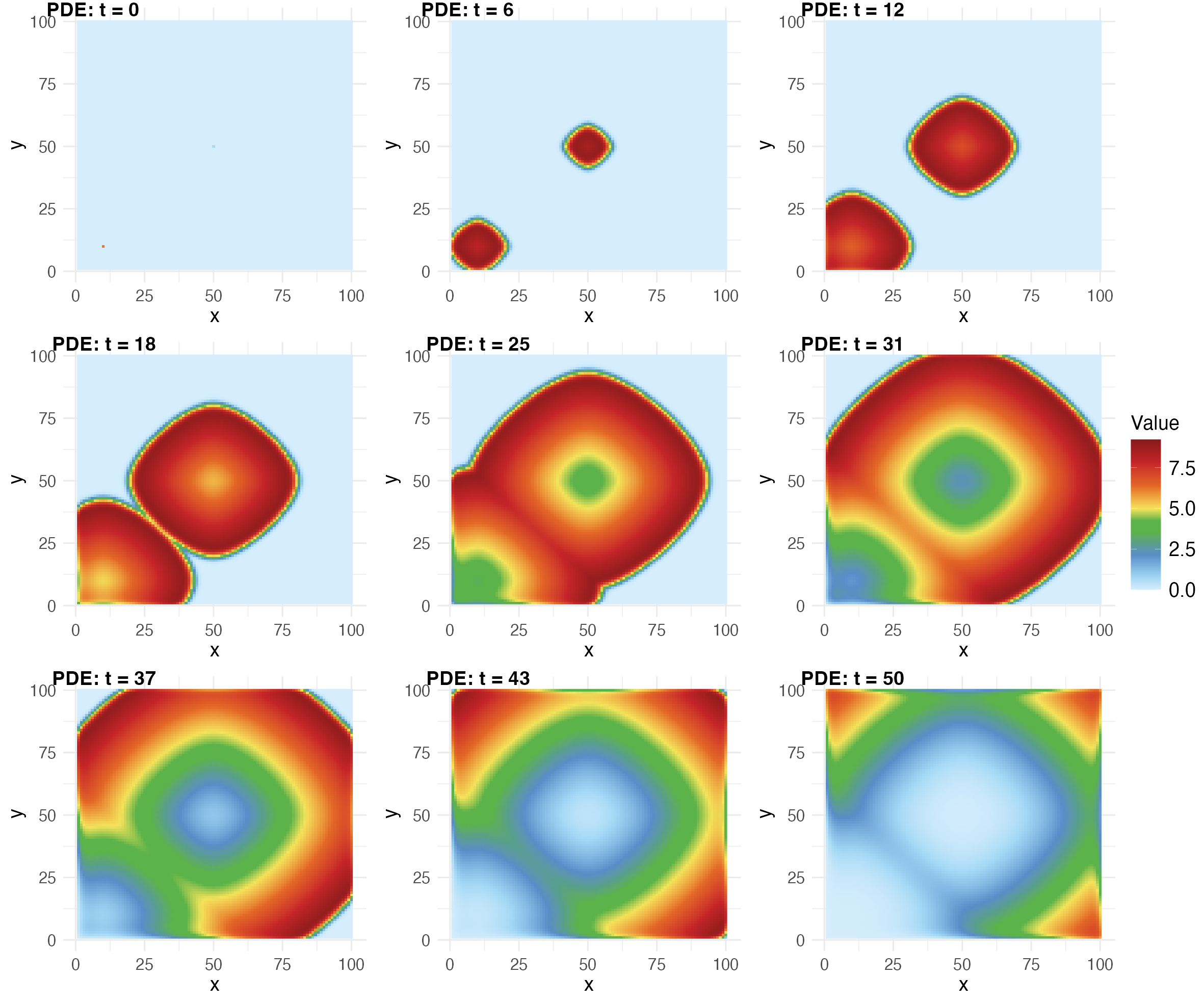}
    \caption{Spatiotemporal dynamics of coupled nonlinear partial differential equations. Darker colors indicate larger function values. Wave-like dynamics radiating outwards from the initial seeding locations are apparent.}
    \label{fig:pde_field}
\end{figure}

We apply our methodology to a spatiotemporal model arising from a set of coupled nonlinear PDEs, a natural way to describe the dynamics of a continuous outcome across space and time. Our model uses the set of coupled PDEs shown in (\ref{eq:sir_pde}).
This mechanistic system is popular within the applied mathematics community to capture qualitative behavior of spreading infection dynamics \citep{Nobel1974, keeling2008}. Literature studying such systems of PDEs often focuses on proving various mathematical properties with no discussion about parameter inference or connections to real-world data or field measurements. Specifically, \cite{Zhang2014} posit the existence of traveling waves in an influenza outbreak for a system such as (\ref{eq:sir_pde}), but omit discussion of parameter inference. See Figure \ref{fig:pde_field} for a visualization of this traveling wave phenomenon when two seeding locations are used as initial sources of the disease. To motivate the system of equations under study, we modify a standard SIR compartment model by including a Laplacian term to induce diffusion of the populations over space. In this model, disease transmission is predominantly a localized process where transmission is most likely between nearby locations. The movement of individuals then facilitates the geographical spread of infectious diseases. Such a process may be captured through the following system
\begin{equation}
\label{eq:sir_pde}
    \begin{split}
        \frac{\partial S_t(\bm s)}{\partial t}&=-\eta_1 \frac{S_t(\bm s)I_t(\bm s)}{N} + \alpha_1\nabla^2S_t(\bm s)\\[4pt]
        \frac{\partial I_t(\bm s)}{\partial t}&=\eta_1 \frac{S_t(\bm s)I_t(\bm s)}{N} -\eta_2 I_t(\bm s)+ \alpha_2\nabla^2I_t(\bm s)\\[4pt]
        \frac{\partial R_t(\bm s)}{\partial t}&=\eta_2 I_t(\bm s) + \alpha_3\nabla^2R_t(\bm s),
    \end{split}
\end{equation}
where $\nabla^2f_t(\bm s)=\sum_{i=1}^d{\partial^2f_t(\bm s)}/{\partial \bm s_i^2}$. In applying our methodology, the training data is generated through the numerical solution of (\ref{eq:sir_pde}) for fixed parameter setting $\bm x=(\eta_1^\prime,\eta_2^\prime,\alpha_1^\prime,\alpha_2^\prime,\alpha_3^\prime)$. The outcome $y_t(\bm s, \bm x)$ denotes the infection counts at a location and time, namely $I_t(\bm s)$. 

\begin{figure}[bt]
    \centering
    \includegraphics[width=0.65\textwidth]{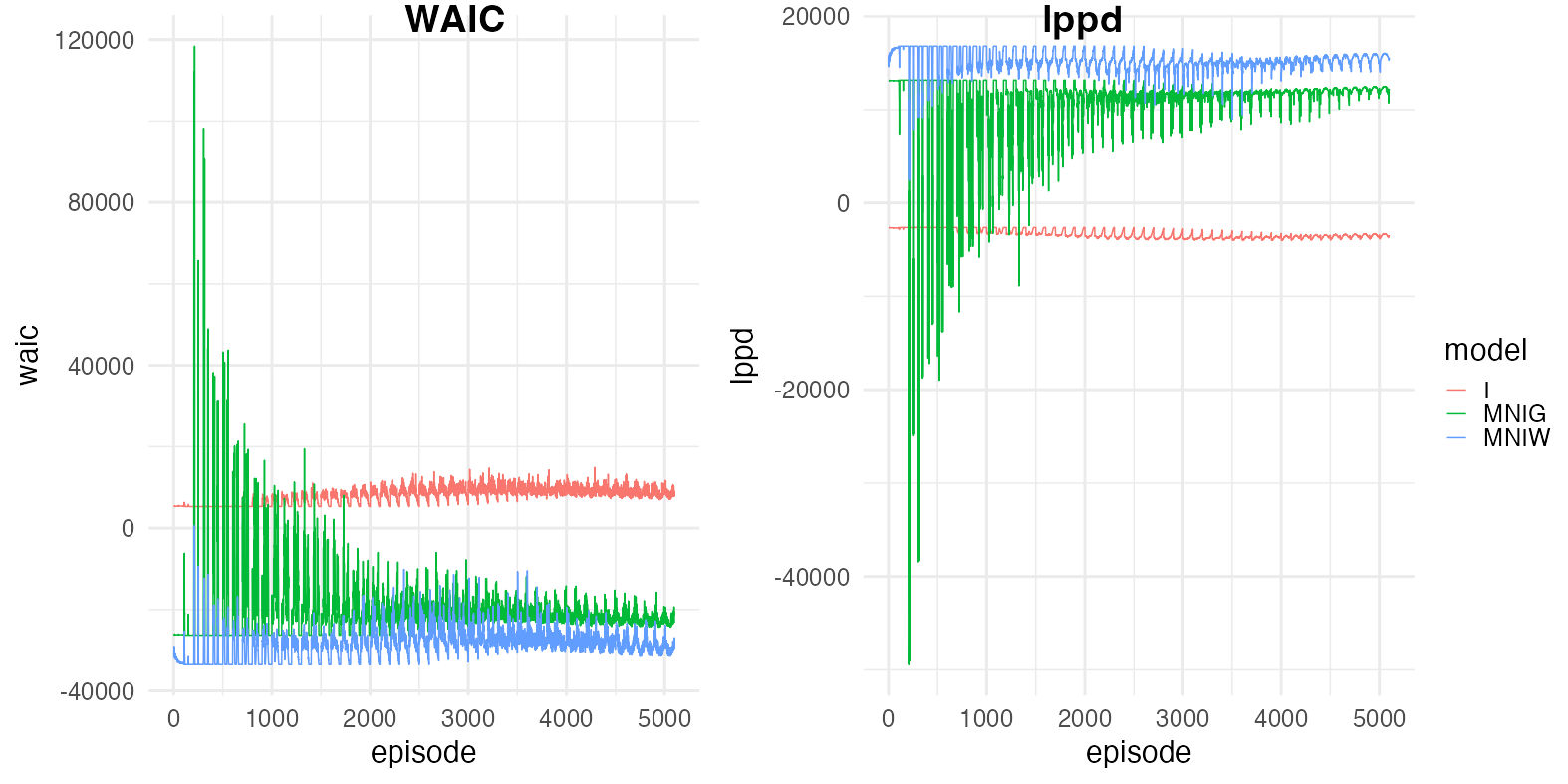}
    \caption{WAIC over episode/time for FFBS models with three different variance structures: inverse-Wishart, inverse-Gamma with spatial correlation matrix $\bm R$, and $\bm{I}$.} 
    \label{fig:waic}
\end{figure}

We employ \texttt{deSolve} (\texttt{R} package version 1.40) \citep{Karline2010} to solve the system over the range of input parameters. Our spatial domain consists of a $100\times100$ grid, resulting in $S = 10,000$ spatial locations. The system's initial state begins with two seeding locations: one in the middle and one in the bottom left corner. We generated $N=50$ sets of $\bm x$ used in equation \eqref{eq:sir_pde}, and for each input generated solutions over $T = 50$ time points from the PDE system, resulting in 51 time points in total including time 0. 

Letting the number of infected people at time $t$ over $N$ inputs and $S$ spatial coordinates be the $N\times S$ matrix $\mathbb{I}_{t}$, we set the data matrix $\bm Y_t = \log(\mathbb{I}_t + 1)$. We further adopt a second-order autoregressive (AR2) structure for the covariate matrix $\bm F_t$, but structure it so that each episode has access to data from the previous two seasons, so that $\bm F_{k,t} = [\bm Y_{k,t-1}, \bm Y_{k,t-2}]$. As with the predator-prey example in the preceding section, we also specify $\bm V_{k,t} = (C(\bm\eta_i, \bm\eta_j; \bm\beta)$, where $\bm\beta$ is defined the same way, as well as $\bm G_t$ and $\bm W_t$ as the identity matrices. We use 40 generated parameters to train the FFBS algorithm (Algorithm \ref{alg:FFBS}), taking $L = 10$ samples in the backward sampling step, and employ a Gaussian Process to emulate the PDE solution. We use the transfer learning parameters specified in Section~\ref{sec:bayestransferlearning} set as $r=40$ and $c=100$, so that we regress on one column of the grid at each step to efficiently emulate our PDE. 

We emulate the space-time field of the PDE using three different variance structures: (i) $\bm{\Sigma}$ follows inverse-Wishart; (ii) $\bm{\Sigma}=\sigma^{2}\bm{R}$; and (iii) $\bm{\Sigma}=\bm{I}$. 
Table \ref{table:waic} shows the WAIC and its component statistics. The inverse-Wishart model outperforms the inverse-Gamma and identity covariance. Figure \ref{fig:waic} shows the details of WAIC over episode/time for FFBS models with three different variance structures. Table \ref{table:GPD} presents the GPD scores ($D=G+P$ from \eqref{eq:GPD_coords}) obtained from independent replicates. The inverse-Wishart model performs slightly better than the inverse-Gamma model, and both outperform the identity model.

Figure \ref{fig:pde_field} shows an example of the dynamics of Equation \eqref{eq:sir_pde} generated by deSolve. The disease spreads over time to cover nearly the entire grid at $t=37$, but the people closest to the source points have already begun to recover. By the final time point of the simulation $t=50$, nearly everyone in the grid has recovered from the disease, with its prevalence being restricted to the corners of the grid, which caught the disease later than the other points.

\begin{figure}[t]
    \centering
    \includegraphics[width=0.8\textwidth]{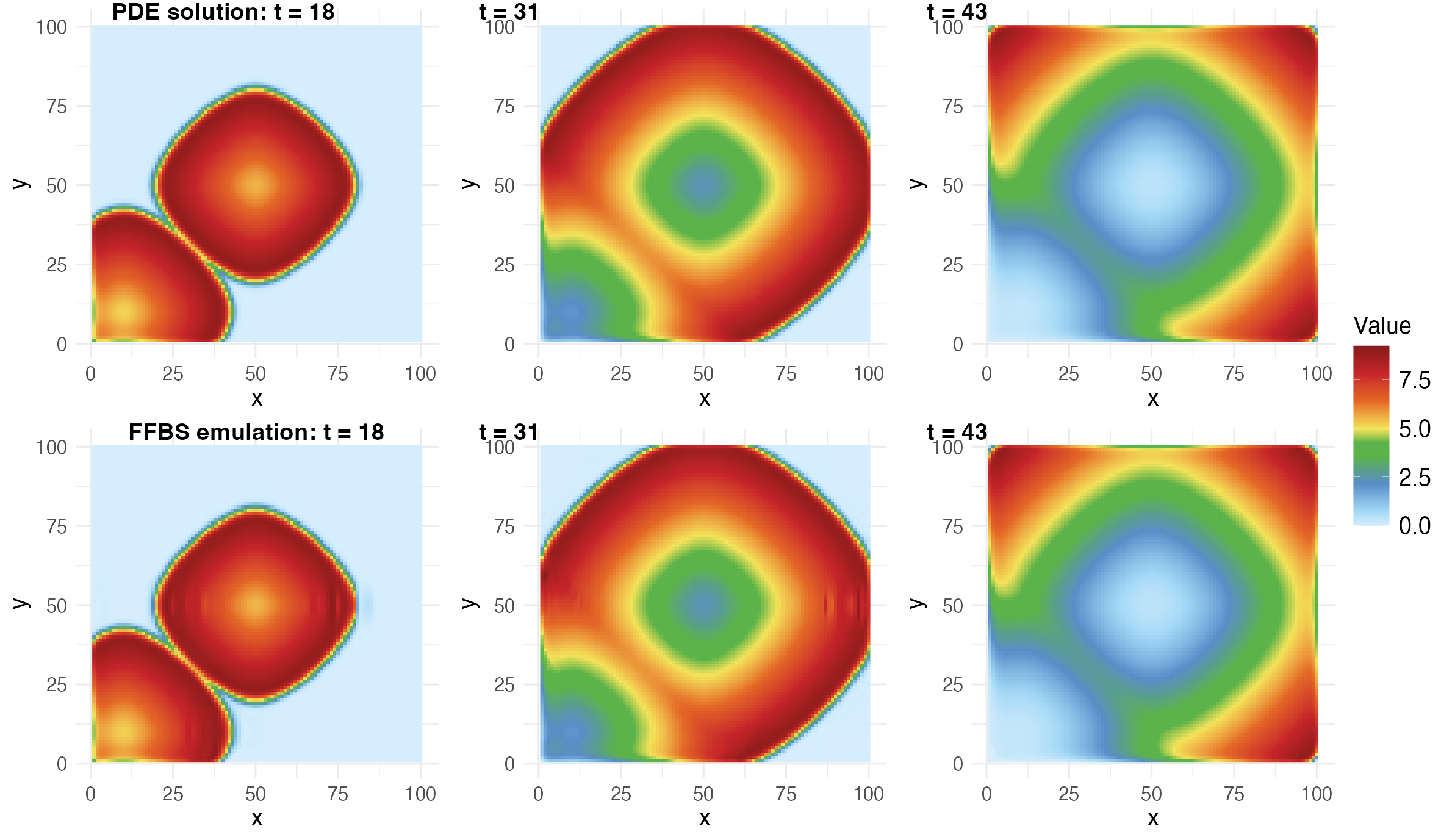}
    \caption{Heatmap comparison of one example of spatiotemporal fields between PDE solutions and FFBS emulations. The parameters are $\eta_1 = 3.549$, $\eta_2 = 0.268$, $\alpha_1 = 0.010$, $\alpha_2 = 0.143$, and $\alpha_3 = 0.170$ at times $t=18, 31, 43$. The first row presents the space-time fields generated by the PDE, while the second row estimates the field for some unobserved inputs held out of the training data.} 	
    \label{fig:heatmap}
\end{figure}

Figure \ref{fig:heatmap} compares the PDE solution and the FFBS emulation results using the $\mathcal{MNIW}$ covariance structure at the same selected time points. The boundaries and overall region of the spread of the disease is accurately emulated, with some inaccuracies taking place within the infected regions. By $t=31$, however, even the number of infected people within the infected region has been accurately emulated, and by $t=43$, the emulation is nearly perfect as revealed by comparing the fields in the second row of Figure~\ref{fig:heatmap} with those in the first row (also corresponding to $t=18, 31, 43$ in Figure~\ref{fig:pde_field}). 

\begin{figure}[t]
    \centering
    \includegraphics[width=0.8\textwidth]{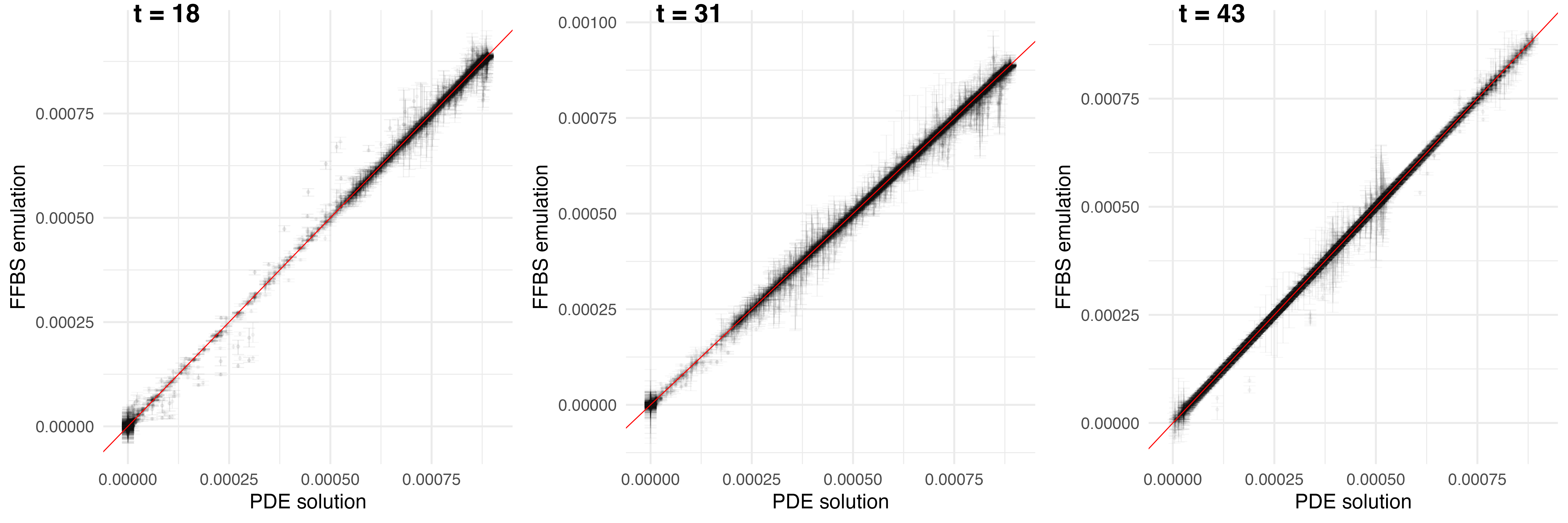}
    \caption{Scatter plot featuring a 95\% credible interval that contrasts PDE solutions and FFBS emulation results corresponding to some unobserved inputs held out of training data. The red line denotes the region where emulation is perfect.}
    \label{fig:errorbar}
\end{figure}

Figure \ref{fig:errorbar} contrasts the PDE solution and FFBS emulated values across all locations, with the same parameters as in Figure \ref{fig:heatmap}. Both figures demonstrate good predictive accuracy of the FFBS algorithm in emulating the nonlinear PDE.

\begin{table}[ht]
\begin{center}
\begin{tabular}{|c|c|c|c|}
\hline
 & $\bm{\Sigma} = \text{inverse-Wishart}$ & $\bm{\Sigma} = \sigma^{2}\bm{R}$  & $\bm{\Sigma} =  \bm{I}$ (reference)\\\hline
    lppd & $77.8 \times 10^{6}$ & $55.7 \times 10^{6}$ & $-17.1 \times 10^{6}$ \\\hline
    lppd (analytic computation) & $79.3\times 10^{6}$ & $57.1\times 10^{6}$ & $-16.2\times 10^{6}$\\\hline
    $p_{\mathrm{WAIC}}$ & $5.2 \times 10^{6}$ & $4.7 \times 10^{6}$ & $3.2 \times 10^{6}$ \\\hline
    $\mbox{WAIC} = -2(\mbox{lppd} - p_{\mathrm{WAIC}})$ & $-145.4\times 10^{6}$ & $-101.9 \times 10^{6}$ & $40.6\times 10^{6}$ \\\hline
\end{tabular}
\caption{The WAIC and its component statistics computed from the PDE emulation data with different variance structure $\bm{\Sigma}$ including inverse-Wishart, $\sigma^{2}\bm{R}$, and $\bm{I}$.}
\label{table:waic}
\end{center}
\end{table}


\begin{table}[ht]
    \centering
    \begin{tabular}{|c|c|c|c|}
    \hline
     & $\bm{\Sigma} = $ inverse-Wishart & $\bm{\Sigma} = \sigma^{2}\bm{R}$ & $\bm{\Sigma} = \bm{I}$ (reference) \\\hline
        $G$ & $3.0\times 10^{4}$ & $3.0\times 10^{4}$ & $3.0\times 10^{4}$ \\\hline
        $P$ & $1.9\times 10^{7}$ & $3.0\times 10^{7}$ & $935.0\times 10^{7}$ \\\hline
        $D=G+P$ & $1.9\times 10^{7}$  & $3.0\times 10^{7}$ & $935.0\times 10^{7}$ \\\hline
    \end{tabular}
    \caption{The model comparison with independent replicates. Mirroring the results from Table \ref{table:waic}, the inverse-Wishart $\bm\Sigma$ has the best fit, followed by $\bm\Sigma = \sigma^{2}\bm R$, and then by $\bm\Sigma = \bm I$.}
    \label{table:GPD}
\end{table}

\section{Estimating mechanistic system parameters using field data}\label{sec:calibration}

The above emulator is now melded with the field observations to infer about the mechanistic parameters. Since the mechanistic system may not adequately describe observed field data over all spatial locations and time points, we include a spatiotemporally-varying discrepancy term that captures the bias between the observed field data and emulator. Our state-space model with dynamic bias correction $\bm u$ is
\begin{equation}
    \begin{split}
    \label{discrep_model}
    z_t(\bm{s})&=y_t(\bm\eta, \bm{s}) + u_t(\bm s) + \varepsilon_t^z(\bm{s}),\quad \varepsilon_t^z(\bm{s})\stackrel{\text{ind.}}{\sim}\mathcal{N}(0, \tau_t^2)\\
    {u_{t}(\bm s)} &= u_{t-1}(\bm s)+ \varepsilon_t^u(\bm s),\quad \varepsilon_t^u(\bm s)\stackrel{\text{ind.}}{\sim}\mathcal{GP}(0,\tau_t^2 C(\cdot, \cdot;\bm\rho))\\
    \tau_t^{2} &\stackrel{\text{ind.}}{\sim} \mathcal{IG}(b^{t}n^z_0, b^{t}d^z_0), \quad t=1,2,\ldots,T
    \end{split}
\end{equation}
where $y_t(\bm\eta, \bm{s})$ denotes the emulator prediction at mechanistic input $\bm \eta$ and location $\bm s$. We assume that the field data are available over a set of $\tilde{S}$ spatial locations in a set $\tilde{\mathcal{S}} \subseteq \mathcal{S}$. It is worth pointing out that in the traditional exercise of calibrating computer models, the discrepancy term is treated as function of the model inputs. One could certainly incorporate $u_t(\bm \eta, \bm s)$ in \eqref{discrep_model}. However, such paradigms assume that the field data are partial realizations of a process with some unknown ``optimal'' $\eta$ that is adjusted by the discrepancy terms and the noise. We do not need such assumptions for inferring about $\bm \eta$. Instead, we treat the problem as one of a mixed nonlinear regression with the field observations regressed by the mechanistic output for some unknown $\bm \eta$. The space-time discrepancy term now serves as a process that attempts to learn from the mechanistic system and the data.

Let $\bm u_t$ be the $\tilde{S}\times 1$ vector with elements $u_t(\bm s_i)$ and $\bm U(\bm\rho)$ be the $\tilde{S}\times\tilde{S}$ correlation matrix built using $C(\cdot, \cdot;\bm\rho)$. The posterior distribution conditional on the collection of field observations, $\bm z_{1:T}$, and the emulated values, $\bm Y_{1:T}$, is proportional to the joint distribution 
\begin{equation}
    \begin{split}
        \label{calibrator}
        &p(\bm\eta, \bm\rho)\times \mathcal {NG}(\bm u_0, \tau_0^{-2}|\bm m_0^z, \bm M_0^z, n_0^z, d_0^z)\times \prod_{t=1}^T \mathcal{NG}(\bm{u}_t, \tau_t^{-2}\mid \bm{u}_{t-1}, \bm U(\bm \rho), b^t n^z_0, b^t d^z_0) \\ 
        &\quad\qquad \times \prod_{t=1}^T \left\{\mathcal{N}(\bm y_t(\bm\eta)\mid \tilde{\bm \mu}_t(\bm\eta), \tilde{\bm \Sigma}(\bm \eta)) \prod_{i=1}^{\tilde{S}} \mathcal N\left(z_{t}(\bm s_i)\mid y_t(\bm\eta, \bm s_i) + u_t(\bm s_i), \tau_t^2\right)\right\},
    \end{split}
\end{equation}
where $\bm y_t(\bm\eta)$ is the $\tilde{S}\times 1$ vector with elements $y_t(\bm\eta,\bm s)$ for each location $\bm s \in \tilde{\mathcal{S}}$, $\tilde{\bm\mu}_t(\bm\eta)^{\top} = \tilde{\bm F}_t(\bm\eta)\bm\Theta_t(\tilde{\mathcal{S}}) + \bm J_t(\bm\eta)^{\top}\bm V_t^{-1}(\bm Y_t(\tilde{\mathcal{S}}) - \bm F_t\bm\Theta_t(\tilde{\mathcal{S}}))$ and $\tilde{\bm\Sigma}_t(\bm\eta) = (1 - \bm J_t(\bm\eta)^{\top}\bm V_t^{-1}\bm J_t(\bm\eta))\bm\Sigma(\tilde{\mathcal{S}})$ are the $1\times \tilde{S}$ mean vector and $\tilde{S}\times\tilde{S}$ covariance matrix, respectively, derived from the conditional predictive distribution in \eqref{eq: conditional_posterior_predictive_density} for $\bm y_t(\bm\eta)$ given the emulated values $\bm Y_{1:T}$, $\tilde{\bm F}(\bm\eta)$ is $1\times p$ with entries equal to the row in $\tilde{\bm F}_t$ corresponding to $\bm\eta$, $\bm\Theta_t(\tilde{\mathcal{S}})$ are $\bm Y_t(\tilde{\mathcal{S}})$ are $p\times \tilde{S}$ and $N\times \tilde{\mathcal{S}}$ consisting of columns corresponding to locations in $\tilde{\mathcal{S}}$ extracted from $\bm\Theta_t$ and $\bm Y_t$, respectively, $\bm J_t(\bm\eta)$ is $N\times 1$ with $j$th element given by the covariance function $C(\bm{x}_i, \bm{\eta};\bm{\beta})$ in \eqref{corr} and $\bm V_t$ is $N\times N$ as defined in Section~\ref{GPR}. The definition of $\bm\Sigma(\tilde{\mathcal{S}})$ depends upon our specific choice of the emulation model. For example, if we use \eqref{basic_dlm_matrix-variate} for emulation and if the field data locations in $\tilde{\mathcal{S}}$ are a subset of the $S$ locations in $\mathcal{S}$ that were used for emulation, then $\bm\Sigma(\tilde{\mathcal{S}})$ is the $\tilde{S}\times \tilde{S}$ sub-matrix of $\bm\Sigma$ corresponding to the locations in $\tilde{\mathcal{S}}$. Alternatively, if one uses \eqref{dlm_gp_double} for emulation, then $\bm\Sigma(\tilde{\mathcal{S}}) = \sigma^2\bm R(\tilde{\mathcal{S}})$, where $\bm R(\tilde{\mathcal{S}})$ is the $\tilde{S}\times \tilde{S}$ spatial correlation matrix formed over the locations in $\tilde{\mathcal{S}}$ using a spatial correlation function; the field data locations need not be a subset of emulator locations.

\begin{algorithm}[!th]
    \caption{Full conditional distributions for $p(\tau^2_{1:T}\mid \cdot)$.}\label{alg:full_cond_tau}
    \begin{algorithmic}[1]    
    \Function{\texttt{TauSq\_post}}{$\bm u_{0:T}, \bm y_{1:T}, \bm z_{1:T}, n_0, d_0, b, \bm U$}
        \For{$t=1$ to $T$}
        \State $Q_1 \gets (\bm u_t - \bm u_{t-1})^{\top}\bm U^{-1}(\bm u_t - \bm u_{t-1})$, $Q_2 \gets (\bm z_t - \bm y_t - \bm u_t)^{\top}(\bm z_t - \bm y_t - \bm u_t)$
        \Comment{$O(\mathcal{\tilde{S}}^2)$}
        \State Sample $\tau^{2}_t \sim \mathcal{IG}(b^{t}n_0 + \tilde{S}, b^{t}d_0 + \frac{1}{2}(Q_1 + Q_2))$
        \EndFor
    \State \Return $\tau^{2}_{1:T}$
    \EndFunction
    \Comment{$O(T\mathcal{\tilde{S}}^2)$}
\end{algorithmic}
\end{algorithm}

We draw samples from $p(\bm{\eta},\bm\rho,\tau_{0:T}^2, \bm y_{1:T}(\bm \eta), \bm u_{0:T}|\bm z_{1:T},\bm{Y}_{1:T})$ given by
\begin{equation}
\label{hardPosterior}
    \begin{split}
    &\int p(\bm{\eta},\bm\rho,\tau_{0:T}^2, \bm u_{0:T}, \bm y_{1:T}(\bm \eta), \bm\Sigma,\bm{\Theta}_{0:T} \mid \bm z_{1:T},\bm{Y}_{1:T})d\bm\Sigma d\bm{\Theta}_{0:T}\\
    &\qquad =\int p(\bm{\eta},\bm \rho, \tau_{0:T}^2, \bm u_{0:T}, \bm y_{1:T}(\bm \eta) \mid \bm\Sigma,\bm{\Theta}_{0:T}, \bm z_{1:T},\bm{Y}_{1:T}) \times{\underbracket{p(\bm\Sigma,\bm{\Theta}_{0:T} \mid \bm{z}_{1:T},\bm{Y}_{1:T}}_{\mathclap{\text{Modularize: }p(\bm\Sigma,\bm{\Theta}_{0:T} \mid \bm{Y}_{1:T})}}})d\bm\Sigma d\bm{\Theta}_{0:T}
    \end{split}
\end{equation} 
where $\bm y_{1:T}(\bm \eta) = (\bm y_1(\bm \eta)^{\top},\ldots,\bm y_T(\bm\eta)^{\top})^{\top}$ is the $\tilde{S}T\times 1$ vector with $\bm y_t(\bm\eta)$ as the $\tilde{S}\times 1$ vector with elements $y_t(\bm \eta, \bm s_i)$ for $i=1,\ldots, \tilde{S}$. A pragmatic approach to sampling from (\ref{hardPosterior}) relies on the posterior in (\ref{calibrator}) and the notion of \textit{Bayesian Modularization} \citep{Bayarri2009}. Modularization replaces the underlined expression in (\ref{hardPosterior}) with the lower dimensional distribution $p(\bm\Sigma,\bm{\Theta}_{0:T}|\bm{Y}_{1:T})$, which is readily sampled through FFBS and Metropolis-Hastings steps of Algorithm \ref{alg:calibration}. For each drawn value of $\{\bm\Sigma,\bm\Theta\}$, we will need to draw from $p(\bm{\eta},\bm \rho, \tau_{0:T}^2, \bm u_{0:T}, \bm y_{1:T}(\bm \eta) \mid \bm\Sigma,\bm{\Theta}_{0:T}, \bm z_{1:T},\bm{Y}_{1:T})$. 


Drawing samples from $p(\bm{\eta},\bm \rho, \tau_{0:T}^2, \bm u_{0:T}, \bm y_{1:T}(\bm \eta) \mid \bm\Sigma,\bm{\Theta}_{0:T}, \bm z_{1:T},\bm{Y}_{1:T})$ employs Gibbs updates using full conditional distributions for $\bm y_{1:T}(\bm \eta)$ and $\tau^2_{0:T}$, FFBS updates for $\bm u_{0:T}$, and Metropolis random-walk updates for $\bm\eta$ and $\bm\rho$. 

The full conditional distributions for $\tau^2_t \mid \cdot$ are $\mathcal{IG}(n^z_t, d^z_t)$ where $n^z_t = \tilde{S} + b^{t}n^z_{0}$ and $d^z_t = b^{t}d^z_{0} + \frac{1}{2}\{(\bm u_t - \bm u_{t-1})^{\top}\bm{U}(\bm\rho)^{-1}(\bm u_t - \bm u_{t-1}) + \sum_{i=1}^{\tilde{S}}(z_t(\bm s_i) - y_t(\bm \eta, \bm s_i) - u(\bm s_i))^{2}\}$ for $t=1,\ldots, T$ and $i=1,\ldots, \tilde{S}$. The full conditional distributions for $\bm{y}_t(\bm\eta) \mid \cdot$ are $\mathcal N(\bm B_t(\bm \eta)\bm b_t(\bm \eta), \bm B_t(\bm \eta))$ where $\bm B_t(\bm \eta)^{-1} = \tilde{\bm \Sigma}_t(\bm\eta)^{-1} + \tau_t^{-2}\bm I_{\tilde{S}}$ and $\bm b_t(\bm \eta) = \tilde{\bm \Sigma}_t(\bm\eta)^{-1}\tilde{\bm\mu}_t(\bm\eta) + \tau_t^{-2}(\bm z_t - \bm u_t)$ with $\tilde{\bm\mu_t}(\bm\eta) = \left(\tilde{\mu}_t(\bm\eta,\bm s_1), \ldots, \tilde{\mu}_t(\bm\eta,\bm s_{\tilde{S}})\right)^{\top}$. Algorithm~\ref{alg:full_cond_tau} provides the steps to compute and draw samples from the full conditional distributions of $\tau^2_{1:T}\mid \cdot$ with a computational cost of $O(T\tilde{S}^{2})$. Algorithm~\ref{alg:full_cond_y} provides the steps to compute and draw samples from the full conditionals for $\bm y_{1:T}\mid \cdot$ with a computational cost of $O(T\tilde{S}^{3})$. 

\begin{algorithm}[th]
    \caption{Full conditional distributions for $p(\bm y_{1:T}\mid \cdot).$} \label{alg:full_cond_y}
    \begin{algorithmic}[1]    
    \Function{\texttt{y\_post}}{$\bm\eta, \tau^{2}_{1:T}, \bm z_{1:T}, \bm u_{1:T}, \tilde{\mathcal{S}}$}
    \For{$t=1$ to $T$}
        \State $\tilde{\bm\mu}_t(\bm\eta)^{\top} \gets \tilde{\bm F}_t(\bm\eta)\bm\Theta_{t}(\tilde{\mathcal{S}}) + \bm{J}_t(\bm\eta)^{\top}\bm V_t^{-1}(\bm{Y}_t(\tilde{\mathcal{S}}) - \bm F_t \bm\Theta_t(\tilde{\mathcal{S}}))$\Comment{$O(pN\tilde{\mathcal{S}})$}
        \State $\tilde{\bm\Sigma}_t(\bm\eta) \gets \left(1- \bm J_t(\bm\eta)^{\top}\bm V_t^{-1}\bm J_t(\bm\eta)\right)\bm\Sigma(\mathcal{\tilde{S}})$
        \Comment{$O(N^2)$}
        \State $\bm B_t(\bm\eta) \gets (\tilde{\bm\Sigma}_t(\bm\eta)^{-1} + \tau_t^{-2}\bm I_{\tilde{S}})^{-1}$, $\bm b_t(\bm\eta) \gets \tilde{\bm\Sigma}_t(\bm\eta)^{-1}\tilde{\bm\mu}_t(\bm\eta) + \tau_t^{-2}(\bm z_t - \bm u_t)$
        \Comment{$O(\mathcal{\tilde{S}}^3)$}
        \State Sample $\bm{y}_t \sim \mathcal N(\bm B_t\bm b_t, \bm B_t)$ 
        \Comment{$O(\mathcal{\tilde{S}}^3)$}
    \EndFor
    \State \Return $\bm y_{1:T}, \bm B_{1:T}, \bm b_{1:T}$
    \EndFunction 
    \Comment{$O(T\mathcal{\tilde{S}}^3)$}
\end{algorithmic}
\end{algorithm}

The model bias process realizations, $\bm u_{0:T}$, are updated from its joint full conditional, $p(\bm u_{0:T}\mid \cdot)$ using the FFBS algorithm. As with emulation, we compute the moments for $\bm u_{t}\mid \cdot, \bm z_{1:t}, \bm y_{1:t}(\bm \eta)$ using the Forward Filter, and then acquire the smoothed moments of $\bm u_{t} \mid \cdot, \bm z_{1:T}, \bm y_{1:T}(\bm \eta)$ using Backwards Sampling. Note that $\bm u_t$ are vectors relevant to Gibbs sampling, we only need one sample of $\bm u_{0:T} \mid \cdot$ per iteration. Algorithm \ref{alg:FFBS_calib} outlines the steps for generating samples of $\bm u_{0:T}\mid \cdot, \bm z_{1:T}, \bm y_{1:T}(\bm\eta)$ using FFBS algorithm. Lines 5-12 describe the Forward Filtering steps, with a computational cost of $\sim O(T\tilde{S}^{2})$. Similarly, Lines 14-21 detail the Backward Sampling steps, also with a cost of $\sim O(T\tilde{S}^{2})$. Thus, the total computational cost is $\sim O(T\tilde{S}^{2})$. 

Finally, we consider the density for $\bm\rho, \bm\eta\mid\cdot$, which we glean from the relevant terms in \eqref{calibrator}. Unfortunately, even when we choose comparatively simple priors for $p(\bm\rho,\bm\eta)$ (e.g. multivariate uniform distributions), there is no recognized density that would enable us to sample from known distributions. Consequently, we update $\bm\rho$ and $\bm\eta$ at each iteration via the Metropolis algorithm. Algorithm~\ref{alg:metrop} gives steps for the Metropolis random walk. The target density is given in Algorithm~\ref{alg:target_density}, which corresponds to terms in \eqref{calibrator}. 

Since our use cases depend on the entries of $\bm\rho$ and $\bm\eta$ being positive, or at the very least nonnegative, we first transform the parameters to the real line and use a normal density for proposing updates in the Metropolis sampler. Let $\bm\phi = (\bm\rho^{\top}, \bm\eta^{\top})^{\top}$ be the concatenation of $\bm\rho$ and $\bm\eta$, $l_{\bm\phi}$ be the total number of elements of $\bm\rho$ and $\bm\eta$, $\bm\delta \gets \mathcal{N}(0,\epsilon_3^2\bm{I})$ denote the proposal update, and $\bm g(\cdot)$ be the function rescaling each element of $\bm\phi$ to the real line. The resulting transformation necessitates the jacobian $\mathcal{J}_{\bm\phi}(\bm g(\bm\phi))$ for both $\bm\phi$ and $\bm\phi^*\gets \bm g^{-1}(\bm g(\bm\phi) + \bm\delta)$, with $\bm g(\bm\phi)$ and $\bm g(\bm\phi^*)$ considered distributed on $\mathcal{N}(0,1)$. 

We can employ Bayesian modularization to prevent the model discrepancy term $\bm u_t$ from interfering with inference on mechanistic parameters. This additional modularization treats the emulator as an unbiased representation of the field data. After fully sampling the model parameters, each update of the dynamic bias term conditions on a draw from the posterior of the parameters to model the discrepancy between the field data and emulator predictive distribution. Algorithm~\ref{alg:calibration} outlines the steps for learning about the mechanistic system parameters using field data. Each iteration updates the parameters $\tau^{2}_{1:T}$, $\bm u_{1:T}$, $\bm \rho$ and $\bm\eta$ (via $\bm\phi$), and $\bm y_{1:T}$ using the updated values in order; Line 7 in particular proposes new values using a Metropolis update. Repeating the process for $L$ iterations results in a total computational cost of $O(LT\mathcal{\tilde{S}}^3)$.

\begin{algorithm}[!t]
    \caption{FFBS for calibration with computer model bias}\label{alg:FFBS_calib}
    \begin{algorithmic}[1]
        \State \textbf{Input:} Field data $\bm z_{1:T}$, emulation results $\bm y_{1:T}(\bm \eta)$, starting values $\bm m^z_0$, $\tau_{0}^{2}$, $\bm M^z_0$, 
        \item[] \qquad\quad\; data scale variance $\tau_{1:T}^{2}$, and correlation matrix $\bm U$.
        \State \textbf{Output:} Calibration samples $\bm u_{0:T}$ 
        \Function{\texttt{FFBS\_calibration}}{$\bm z_{1:T}, \bm y_{1:T}(\bm \eta), \bm m^z_0, \tau_0^{2}\bm M^z_0, \tau_{1:T}^{2}, \bm U$}
        \State\# Forward Filter
        \For{$t=1$ to $T$}
            \State\# Compute prior distribution covariance matrix
            \State $\tilde{\bm A}_t \gets \tau_{t-1}^{2}\bm M^{z}_{t-1} + \tau_{t}^{2}\bm U$
            \Comment{$O(\tilde{S}^{2})$}
            \State\# Compute one-step ahead forecast covariance matrix
            \State $\tilde{\bm Q}_t \gets \tilde{\bm{A}}_{t} + \tau^{2}_t \bm I_{\tilde{S}}$
            \Comment{$O(\tilde{S})$}
            \State\# Compute filtering distribution moments
            \State $\bm m^z_t \gets \bm m^z_{t-1} + \tilde{\bm A}_t \tilde{\bm Q}_t^{-1} (\bm z_{t} - \bm y_{t}(\bm\eta)- \bm m^z_{t-1}); \bm M^z_t \gets \tau_{t}^{-2}(\tilde{\bm A}_t - \tilde{\bm A}_t\tilde{\bm Q}_t^{-1}\tilde{\bm A}_t)$
            \Comment{$O(\tilde{S}^{3})$}
        \EndFor
        \State\# Backwards Sampling
        \State $\tilde{\bm h}_T \gets \bm{m}^z_T;\; \tilde{\bm H}_T \gets \tau_T^{2} \bm M^z_T$
        \State Sample $\bm u_T \sim \mathcal{N}(\tilde{\bm h}_T, \tilde{\bm H}_T)$
        \Comment{$O(\tilde{S}^{3})$}
        \For{$t=T-1$ to $0$}
            \State\# Compute BS smoothing distribution moments
            \State $\tilde{\bm h}_t \gets \bm m^z_t + \tau^{2}_{t}\bm M^z_t \tilde{\bm A}^{-1}_{t+1}(\tilde{\bm h}_{t+1} - \bm m^z_{t});$
            \Comment{$O(\tilde{S}^{3})$}
            \State $\tilde{\bm H}_t \gets \tau^{2}_{t}\bm M^z_t - \tau_{t}^{4}\bm M^z_t \tilde{\bm A}_{t+1}^{-1}(\tilde{\bm A}_{t+1} - \tilde{\bm H}_{t+1})\tilde{\bm A}_{t+1}^{-1}\bm M^z_t$ 
            \Comment{$O(\tilde{S}^{3})$}
            \State Sample $\bm u_t \sim \mathcal{N}(\tilde{\bm h}_t,\tilde{\bm H}_t)$
            \Comment{$O(\tilde{S}^{3})$}
        \EndFor
        \State \Return $ \bm u_{0:T}$
        \EndFunction
        \Comment{$O(T\tilde{S}^{3})$}
    \end{algorithmic}
\end{algorithm}

\begin{algorithm}[ht]
    \caption{Log probability density of the full conditional $p(\bm \phi \mid \cdot)$.}\label{alg:target_density}
    \begin{algorithmic}[1]    
     \Function{\texttt{loglik}}{$\bm \phi := (\bm\rho, \bm\eta), \tau^{2}_{1:T}, \bm u_{1:T}, \bm z_{1:T},\tilde{\mathcal{S}}$}
    \State $\bm y_{1:T}, \bm B_{1:T}, \bm b_{1:T}$ $\gets$ \texttt{y\_post}$\left(\bm\eta, \tau^{2}_{1:T}, \bm z_{1:T}, \bm u_{1:T}, \tilde{\mathcal{S}}\right)$
    \Comment{Algorithm~\ref{alg:full_cond_y}}
    \State $\mathcal{J}(\bm\phi) \gets \left|\det\left(\frac{\partial g(\bm\phi)}{\partial\bm\phi }\right)\right|$
        \State $\log p(\bm \phi \mid \cdot) = \sum_{t=1}^{T}\left(\log \mathcal{N}\left(\bm y_t\mid \bm B_t\bm b_t, \bm B_t\right) + \sum_{i=1}^{\tilde{S}} \log \mathcal{N}\left(z_t(\bm s_i)\mid y_t(\bm s_i) + u_t(\bm s_i), \tau_t^{2}\right) \right.$ \\
        \qquad \qquad \qquad \qquad \qquad 
        $\left.+ \log \mathcal{N}\left(\bm u_t \mid \bm u_{t-1}, \tau_t^2\bm U(\bm\rho)\right) \right) + \log p(\bm\phi) - \log \mathcal{J}(\bm\phi)$
    \State \Return $\log p(\bm \phi \mid \cdot)$ 
    \EndFunction
    \Comment{$O(T\mathcal{\tilde{S}}^3)$}
\end{algorithmic}
\end{algorithm}

\begin{algorithm}[ht]
    \caption{Metropolis random walk algorithm (one step).}\label{alg:metrop}
    \begin{algorithmic}[1]
    \Function{\texttt{Metrop}}{$\bm \theta, p(), \bm\Upsilon$}
     \State Draw $\bm \theta^{\ast} \sim q(\cdot \mid \bm \theta) := N(\cdot \mid \bm \theta, \bm\Upsilon)$;
     \State 
      Set $\bm \theta = \bm \theta^{\ast}$ with probability $\displaystyle \alpha = \min\left\{1, \exp[\log p(\bm \theta^*\mid \cdot) - \log p(\bm \theta \mid \cdot)]\right\}$.
     \State \Return $\bm \theta$
     \EndFunction
    \end{algorithmic}
\end{algorithm}

\begin{algorithm}[ht]
    \caption{Sampler for estimating mechanistic system parameters using field data}\label{alg:calibration}
    \begin{algorithmic}[1]
    \State \textbf{Given:} Field data $\bm z_{1:T}$, emulation data $\bm Y_{1:T}$, $L$ posterior samples of $p(\bm\Theta_{0:T},\bm\Sigma|\bm{Y}_{1:T})$ from Algorithm \ref{alg:FFBS}, functions $\tilde{\bm F}_{1:T}$, $\bm J_{1:T}$, $\bm g$, hyperparameters $n_0$, $d_0$, $b$, $\tau^{2}_{0}$, $\bm u_{0}$, $\bm\rho^{(0)}$, $\bm\eta^{(0)}$, $\bm m^z_0$, $\bm M^z_0$, $\bm U$, $\bm \Upsilon$.
    \State \textbf{Initialize:} $\bm \phi^{(0)} = \bm g(\bm\rho^{(0)}, \bm\eta^{(0)})$, $\tau^{2,(0)}_{1:T}$, $\bm u^{(0)}_{1:T}$, $\bm y^{(0)}_{1:T}$.
    \For{$l=1$ to $L$}
        \State $\tau^{2,(l)}_{1:T}$ $\gets$ \texttt{TauSq\_post}$\left(\bm u_{0:T}^{(l-1)}, \bm y_{1:T}^{(l-1)}, \bm z_{1:T}, n_0, d_0, b, \bm U\right)$
        \Comment{Algorithm \ref{alg:full_cond_tau}}
        \State $\bm u^{(l)}_{1:T}$ $\gets$ \texttt{FFBS\_calibration}$\left(\tau_{0:T}^{2,(l)}, \bm y_{1:T}^{(l-1)}, \bm z_{1:T}, \bm m^z_0, \bm M^z_0, \bm U \right)$
        \Comment{Algorithm \ref{alg:FFBS_calib}}
        \State $\bm {\bm \phi}^{(l)} \gets$ \texttt{Metrop}$({\bm \phi}^{(l-1)}, \texttt{loglik}, \bm \Upsilon)$
        \Comment{Algorithm~\ref{alg:target_density},\ref{alg:metrop}}
         \State $\bm y_{1:T}^{(l)}$ $\gets$ \texttt{y\_post}$\left(\bm \eta^{(l)}, \tau^{2,(l)}_{1:T},\bm u_{1:T}^{(l)}, \bm z_{1:T}, \tilde{\mathcal{S}} \right)$
        \Comment{Algorithm \ref{alg:full_cond_y}}
    \EndFor
    \Comment{$O(LT\mathcal{\tilde{S}}^3)$}
\end{algorithmic}
\end{algorithm}

\section{Applications for calibration}\label{sec:app_calibration}

\subsection{Predator-prey analysis}\label{app:calibration_LV}

We estimate parameters in the Lotka-Volterra equations in \eqref{eq:lotka_volterra}. We fix $\bm\rho = 1.5$ and, analogous to Section \ref{sec:app_emulation_LV}, set the prior for $\bm\eta$ to be a multivariate lognormal. Then $g_{i}(\eta_i) = (\log(\eta_i) - \mu_i)/\sigma_i$ is the function that takes in $\bm\eta$ and outputs a 4-dimensional vector into $\mathbb{R}^{4}$ that is distributed $\mathcal{N}_{4}(0,I)$, and its Jacobian takes the simple form $\mathcal{J}(\bm\eta) = \prod_{i=1}^{4}(\eta_i\sigma_i)^{-1}$.

\begin{figure}[tbp]
    \centering
    \includegraphics[width=0.25\linewidth]{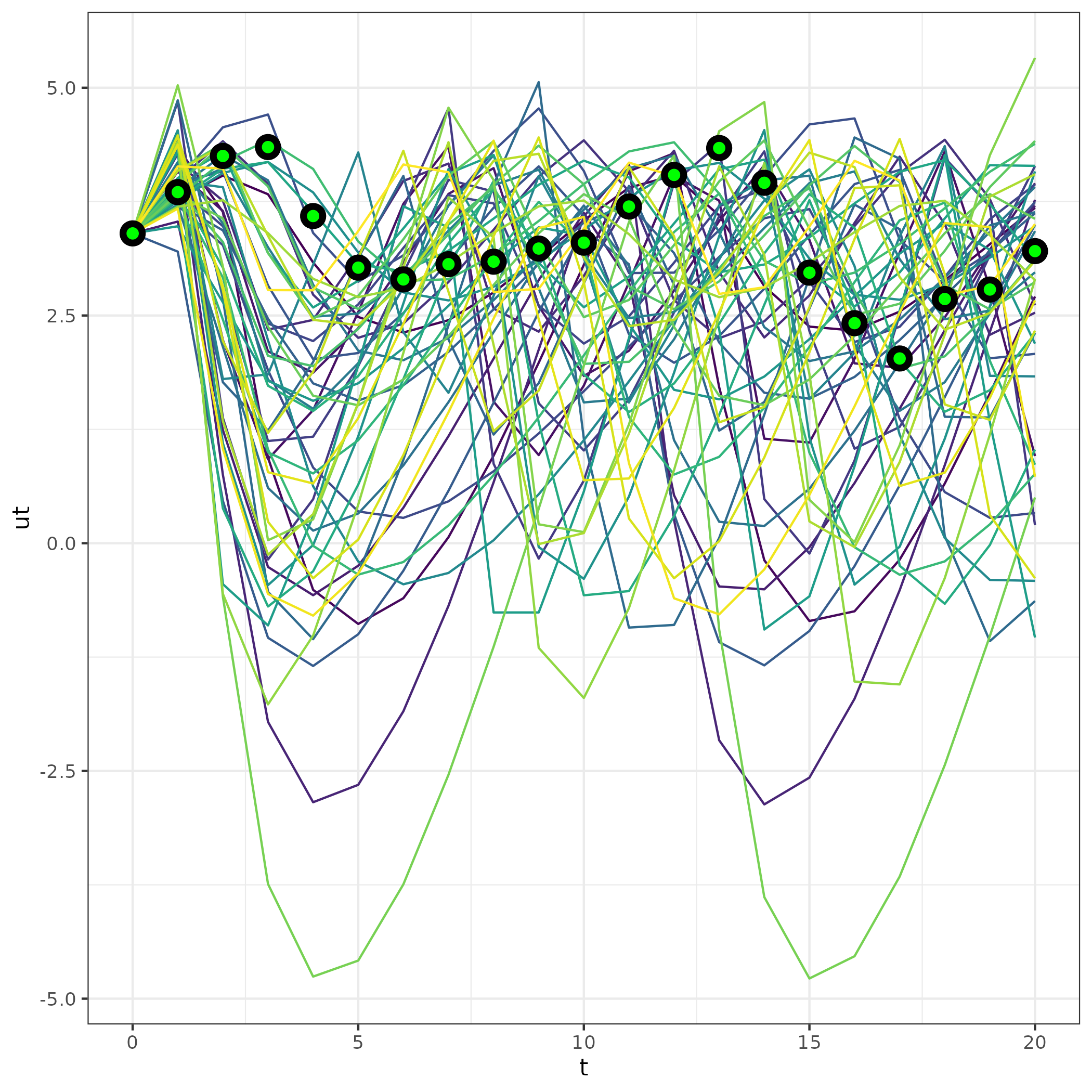}
    \includegraphics[width=0.25\linewidth]{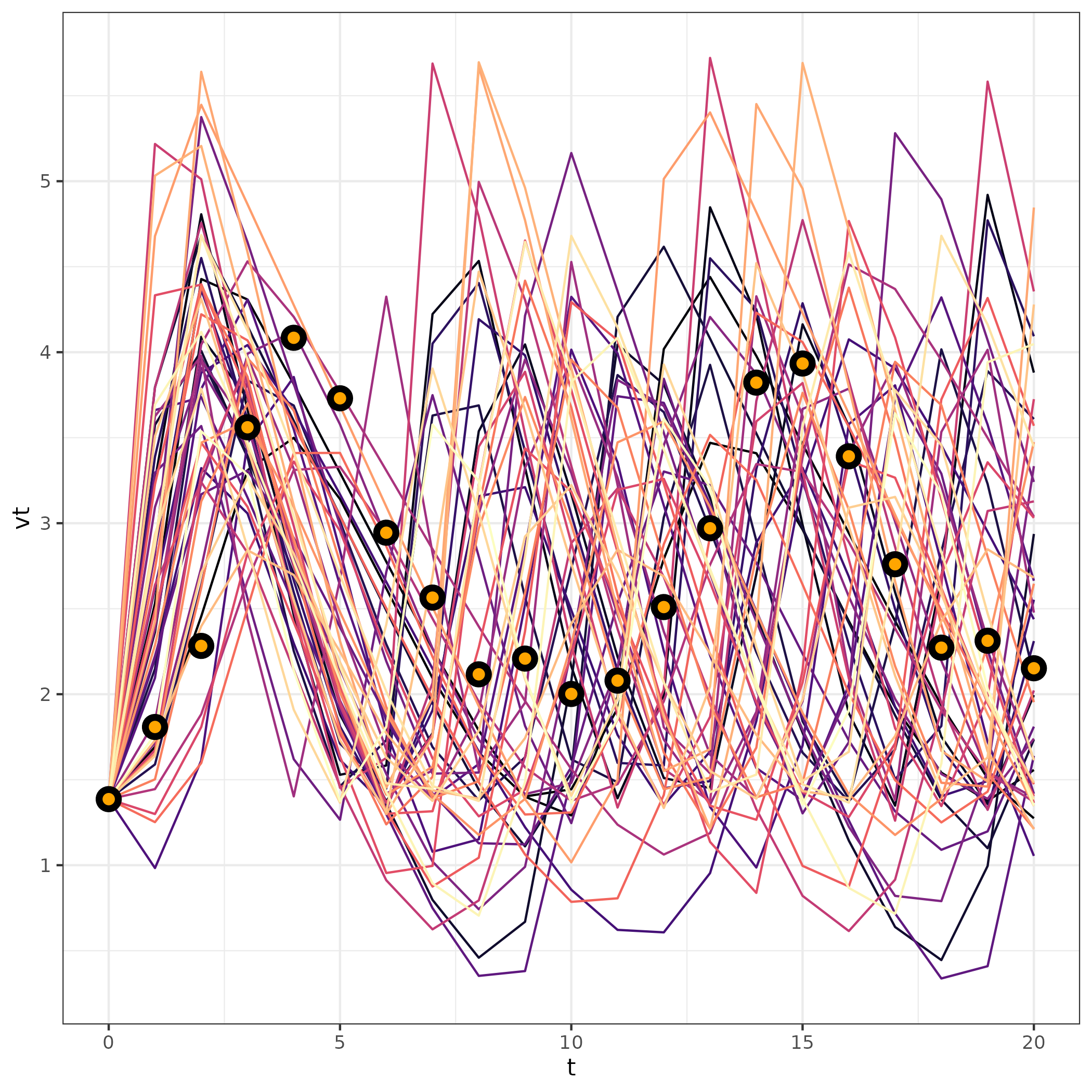}
    \caption{The simulated trajectories for $50$ different parameter values of the Lotka-Volterra equations (colored splines) along with the ground-truth log-populations of hares (left) and lynxes (right). The populations of the hares and lynxes are expressed in the points in each plot.}
    \label{fig:LV_sims_and_GT}
\end{figure}

For our noisy field observations, we utilize the data recorded on the Canadian lynx and snowshoe hare population sizes. \citep{Hewitt1921} We take our training data and emulation parameters from Section \ref{sec:app_emulation_LV} for calibration, with $N = 50$ lognormal $\bm\eta$ sampled with a latin square design and $T = 20$. Figure \ref{fig:LV_sims_and_GT} plots the simulated prey and predator log-populations with the ground-truth log-populations of the hares and lynxes respectively. The agreement between the collection of curves and the real data is remarkable especially given that only the initial population was used from the real data to generate the curves and the Lotka-Volterra parameters were independently generated.

\begin{figure}[!t]
    \centering
    \includegraphics[width=0.20\textwidth]{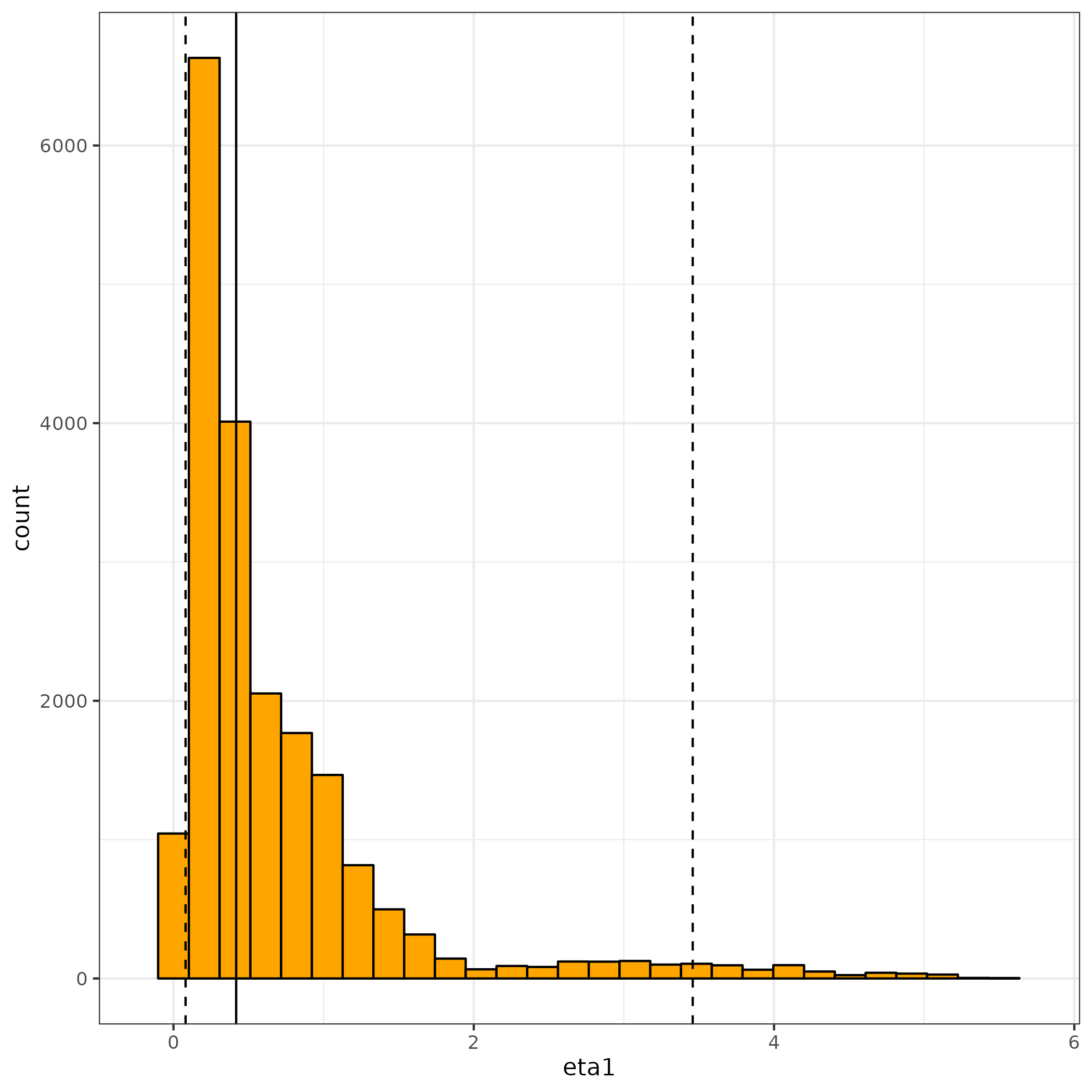}
    \includegraphics[width=0.20\textwidth]{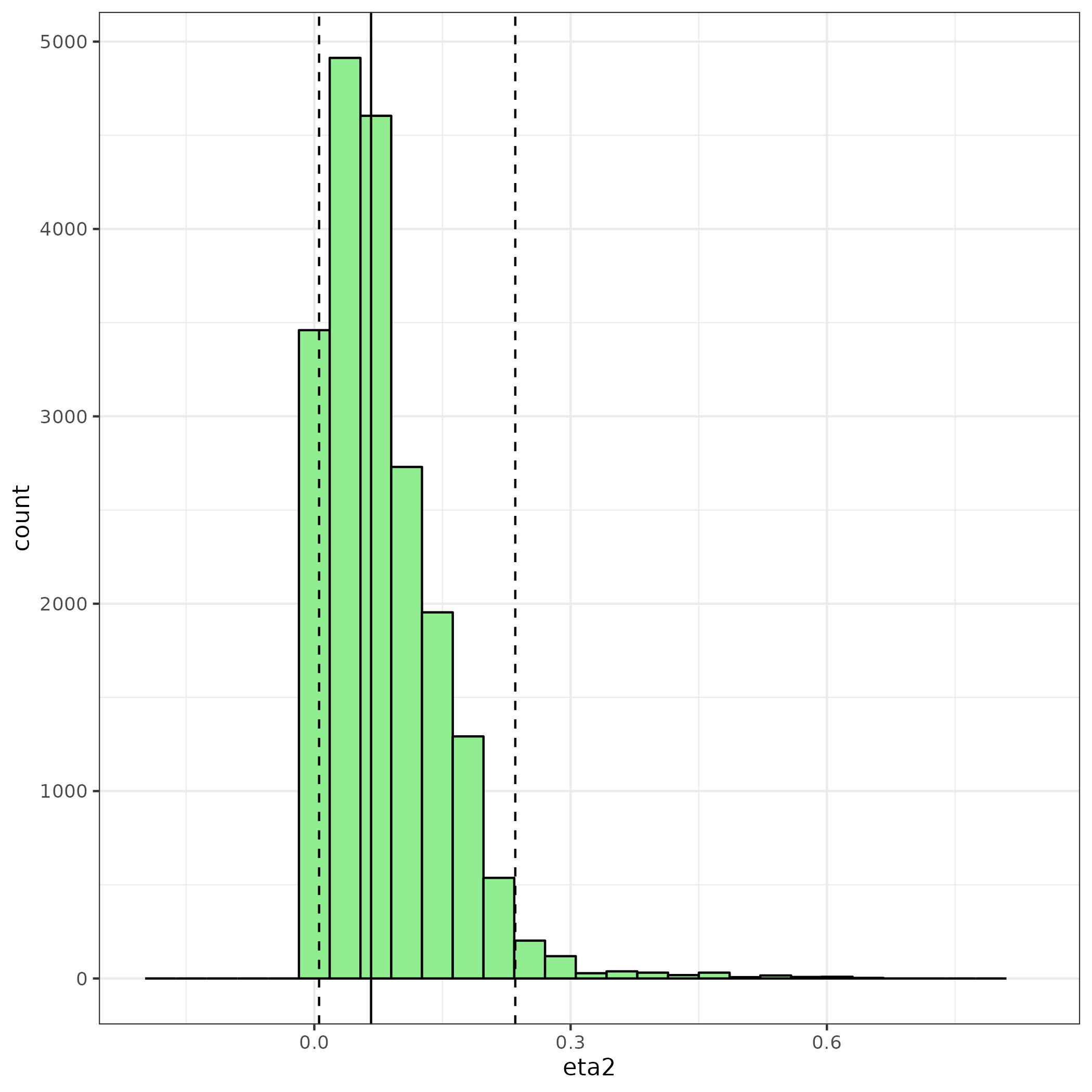}
    \includegraphics[width=0.20\textwidth]{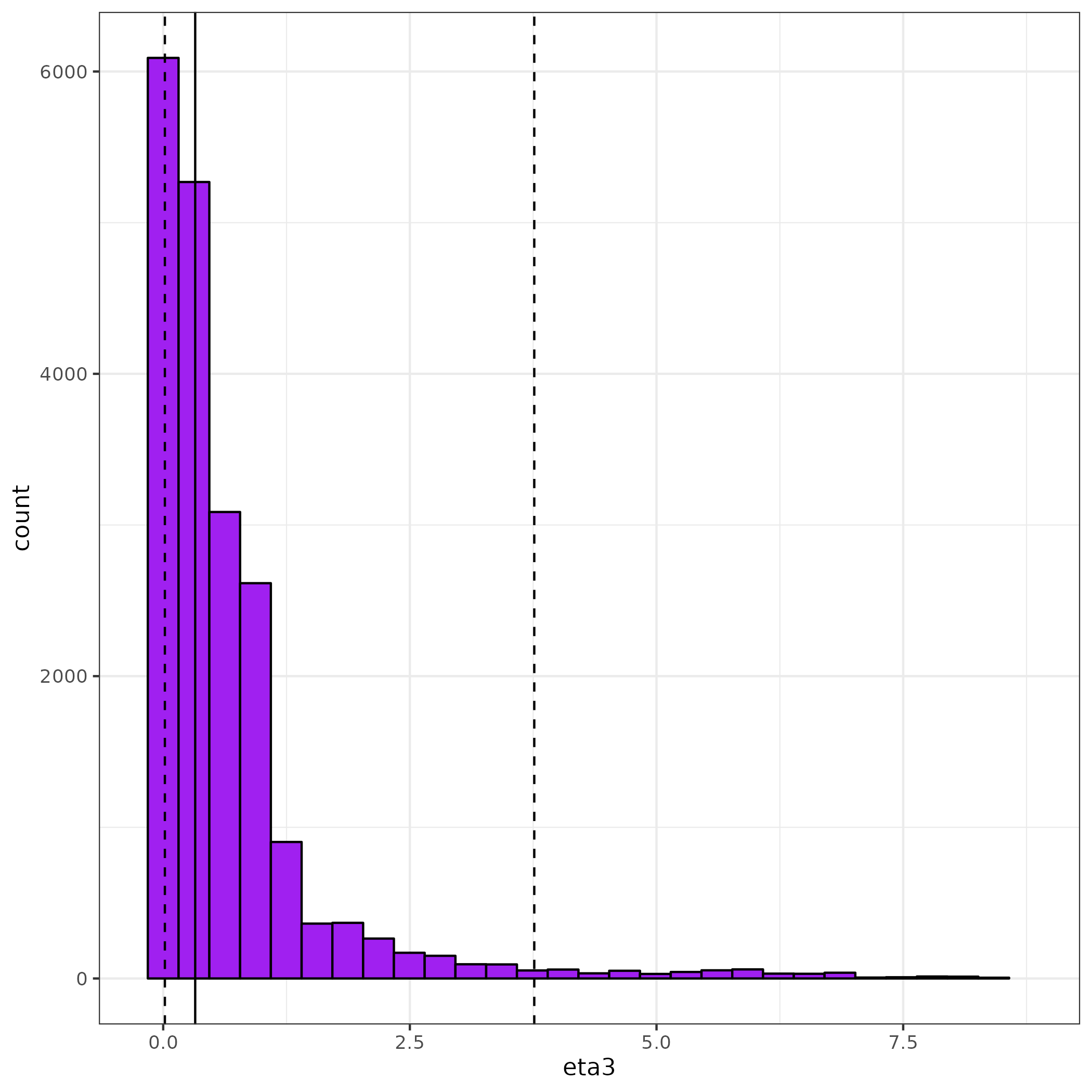}
    \includegraphics[width=0.20\textwidth]{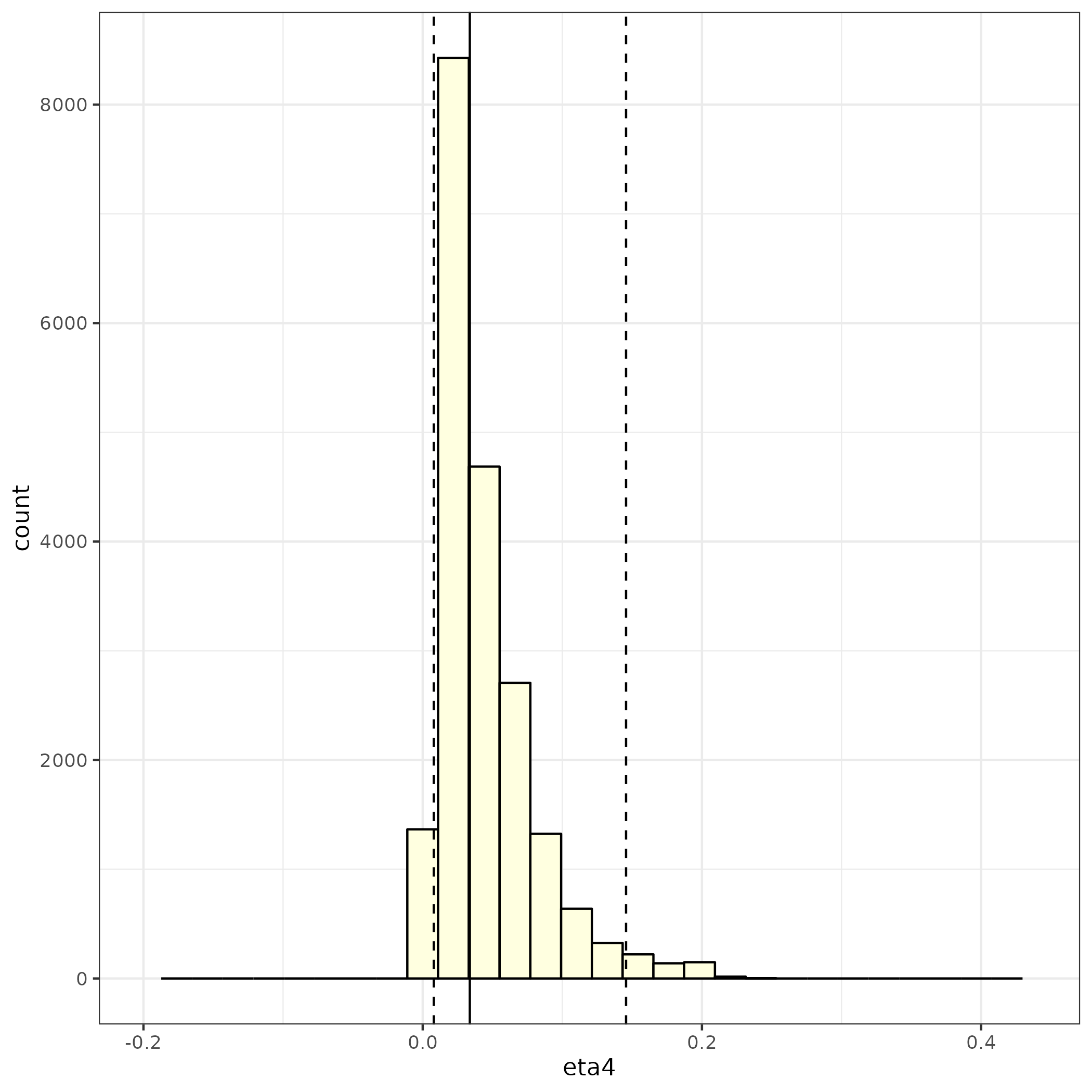}
    \caption{20,000 samples of $\bm\eta$ drawn with the calibrator on real Canadian lynx and snowshoe hare population data. The black vertical line denotes the median of the samples, while the dotted lines denote the $2.5$th and $97.5$th percentiles.}
    \label{fig:LV_eta_real}
\end{figure}

Figure \ref{fig:LV_eta_real} displays calibration samples on real lynx and hare population data collected over a 20-year period. The samples are generated using Algorithm \ref{alg:calibration} with $L = 20,000$. The resulting medians generated from the quantiles are: 0.308 $(0.036, 1.224)$, 0.054 $(0.008, 0.340)$, 0.305 $(0.09, 2.189)$, and 0.039 $(0.012, 0.097)$. Taking the medians as the point estimates for the calibrated solutions of the real-life lynx and hare population data, the number of hares grows at a rate of 0.308 times its population that year every year independent of the presence of any predators, but decreases at the rate of 0.054 times the populations of the hares and lynxes per year due to predation from the lynxes. At the same time, the number of lynxes decreases at the rate of 0.305 times its population that year, but grows at the rate of 0.039 times the populations of the hares and lynxes that year, due to eating the hares. 

\subsection{Nonlinear Partial Differential Equation Calibration}
Applying calibration to the SIR model equation \eqref{eq:sir_pde}, we fix $\bm\rho$ to be constant (approximately 0.85) and set the prior of $\bm\eta$ to be a multivariate uniform distribution $\prod_{i=1}^{5}U(a_i, b_i)$, where $a_i$ and $b_i$ are the limits of the uniform distribution for each parameter. Then $g_{i}(\eta_i) = \mathrm{logit}((\eta_i - a_i)/(b_i - a_i))$ and the Jacobian takes on the form: $\mathcal{J}(\bm\phi) = \prod_{i=1}^{l_{\bm\phi}}\left\{(\eta_i - a_i)^{-1} + (b_i - \eta_i)^{-1}\right\}$.

In this example, we consider a $6\times6$ grid with 36 locations and 26 time points, including time 0, and 50 inputs $\bm\eta$ from a deterministic system. Carrying over our analysis from emulation in Section \ref{sec:app_emulation}, we adopt the same AR2 structure for $\bm F_t$. We use 30,000 samples, with the transfer learning parameters specified in Section~\ref{sec:bayestransferlearning} set as $r=50$ and $c=2$. In calibration, we set $\alpha_1 = \alpha_3 = 0$ to simplify the analysis. 

\begin{figure}[bt]
    \centering
    \includegraphics[width=0.7\textwidth]{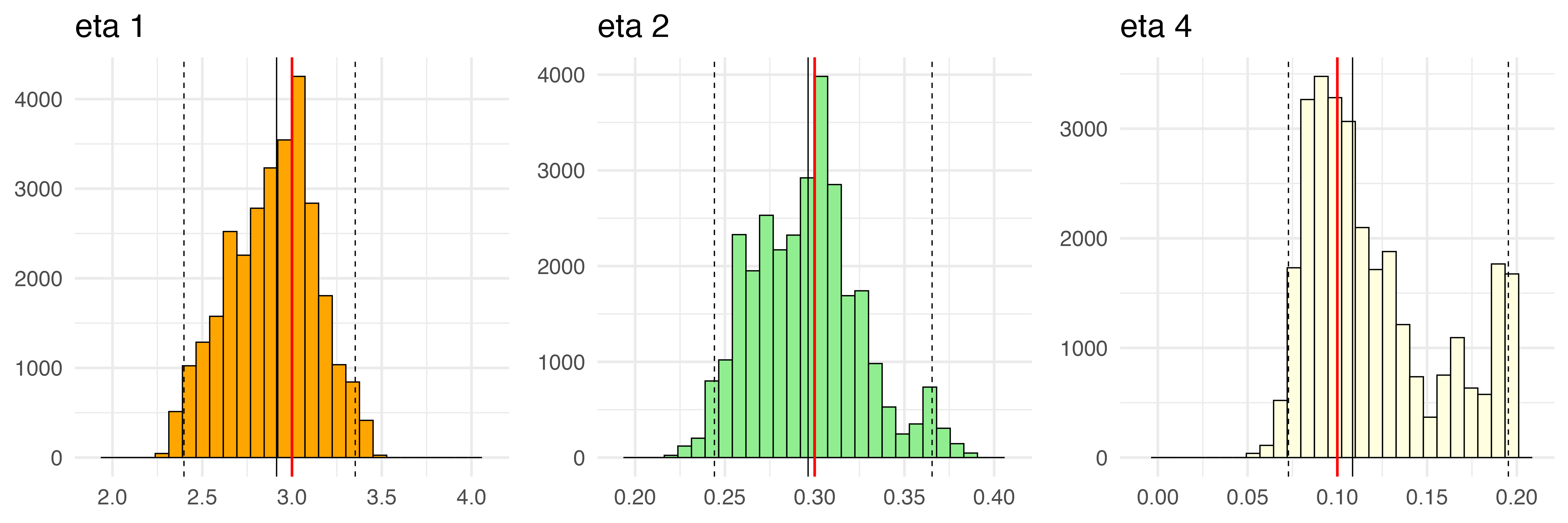}
    \caption{Histograms of posterior densities for elements of parameters $\bm \eta$. The red solid line represents the true value, the black solid line marks the median of the MCMC samples, and the black dotted lines indicate 95\% credible intervals.}
    \label{fig:cal_hist}
\end{figure}

\begin{figure}[bt]
    \centering
    \includegraphics[width=0.7\textwidth]{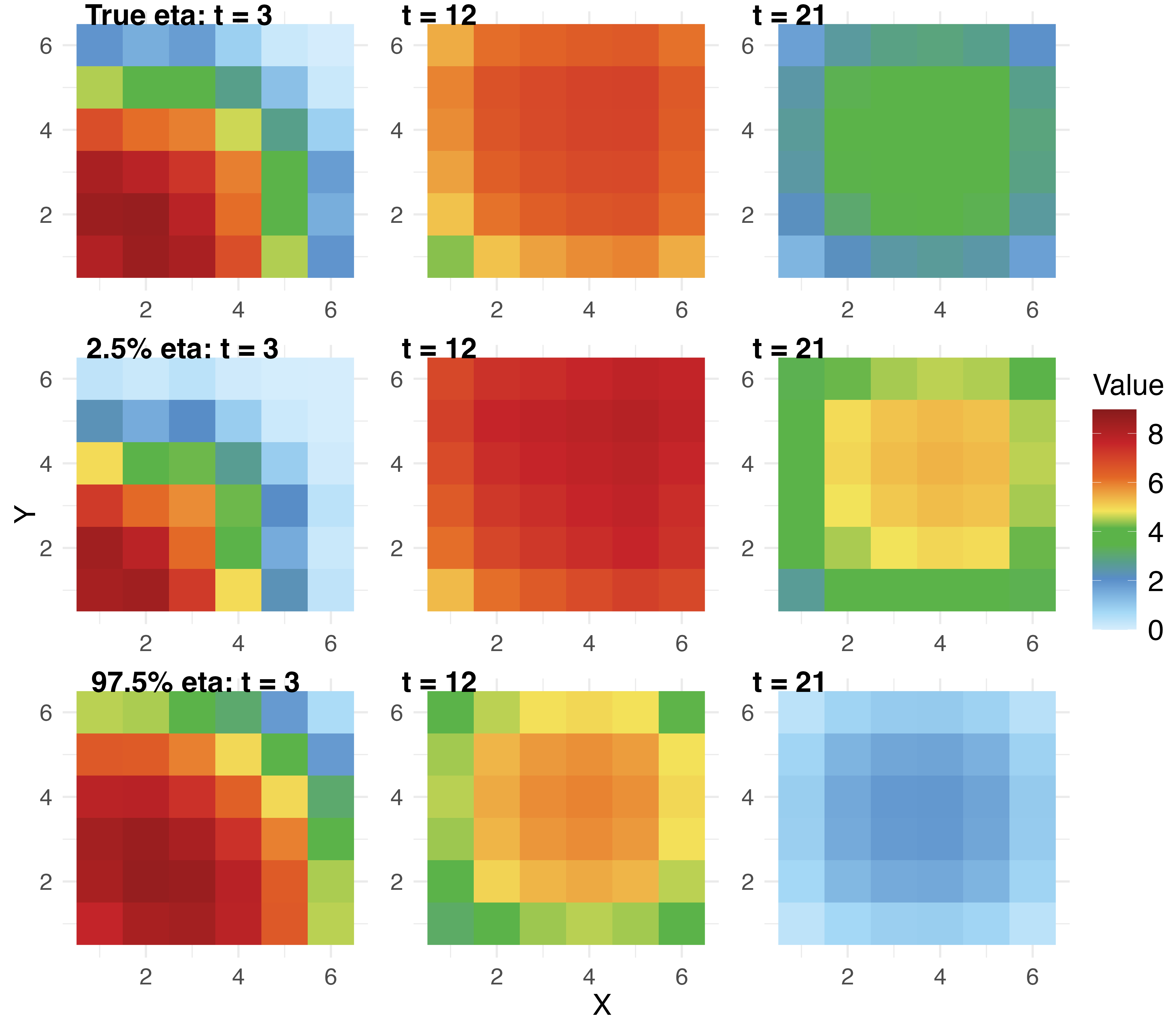}
    \caption{Heat maps of the spatial field generated by the PDE at $t=3,12,21$. The first row is generated from the true $\bm \eta$, while the second and third rows are generated from the 2.5\% and 97.5\% quantiles of posterior samples of $\bm \eta$.}
    \label{fig:cal_heat}
\end{figure}

\begin{figure}[bt]
    \centering
    \includegraphics[width=1.0\textwidth]{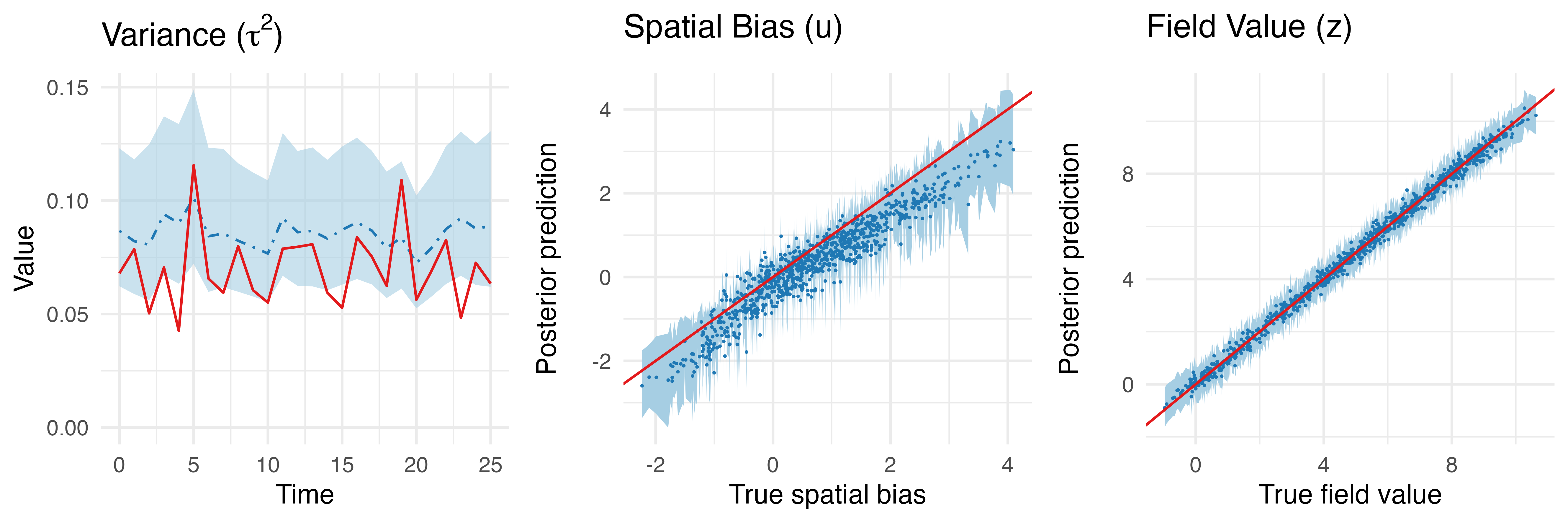}
    \caption{Joint visualization of posterior variability, spatial bias, and field predictions in calibration models, showing the median (blue dotted line and scatter points) with 95\% credible intervals (light blue bands). Red lines indicate a perfect match.}
    \label{fig:cal_panel}
\end{figure}

Figure~\ref{fig:cal_hist} shows the histograms of posterior densities for elements of parameters $\bm\eta$ indicating that 95\% credible intervals cover the true value and shrink our belief of distribution of $\bm\eta$ to a narrower interval. The posterior medians (95\% credible intervals) of $\bm \eta$'s are 2.91 (2.40, 3.35) for $\eta_1$, 0.30 (0.24, 0.37) for $\eta_2$, and 0.11 (0.07, 0.20) for $\alpha_2$, respectively. Figure~\ref{fig:cal_heat} compares the emulation field generated by true and 95\% credible intervals of $\bm\eta$. The 2.5\% percentile of posterior values of $\bm\eta$ shows a slower process in infection and recovery, while 97.5\% percentile shows a faster rate in infection and recovery over space. Figure~\ref{fig:cal_panel} shows the posterior variability, spatial bias, and field predictions in calibration models. The discrepancy in $\tau^2$ with the true value arises from using only a single realization. However, both the spatial bias and field values show good alignment with the true data.

\subsection{Calibrating a Computer Simulation with Network Output}
Network science studies how the macro-structure of interconnected systems such as telecommunication, economic, cognitive, or social networks affect global dynamics \citep{WattsStrogatz1998}. Within a network, nodes represent discrete entities, and edges represent relations. Of prominent interest in applied modeling is understanding how global structure affects spread of a quantity throughout the system, termed network activation or network diffusion. To date, there exist various statistical software packages implementing diffusion or contagion processes across networks, but less attention has been given to calibrating these models to real-world data \citep{netDiffuse, Siew2019}. In this example, we focus on calibrating a deterministic computer model popular within psychology and psycholinguistics communities \citep{Chan2009,Vitevitch2011}. The \texttt{spreadr} package \citep{Siew2019} implements a network diffusion simulation but ignores the calibration problem, justifying use of our methodology.

\begin{figure}[t]
    \centering
    \begin{tabular}{ccc}
    {\includegraphics[scale=.07]{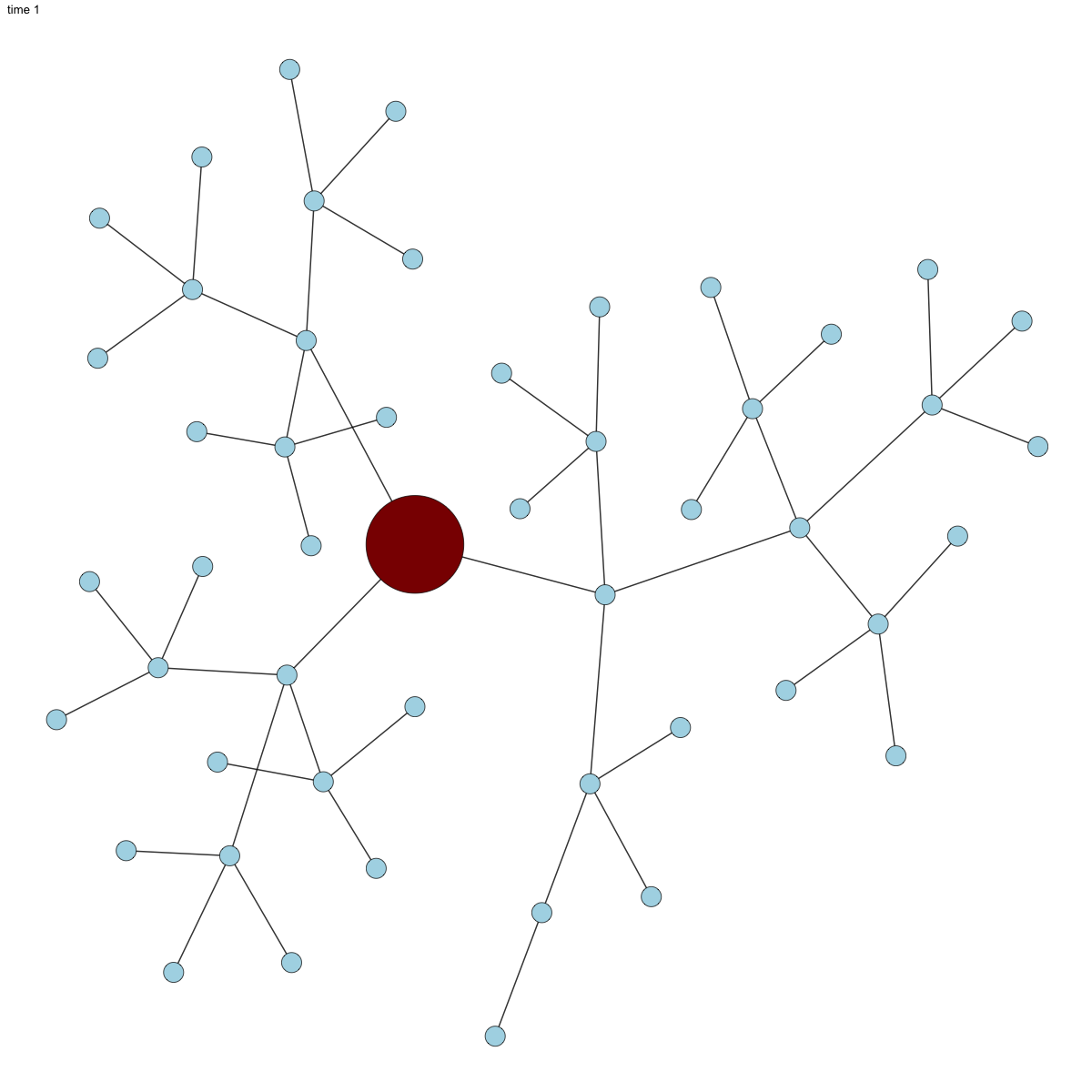}} &
    {\includegraphics[scale=.07]{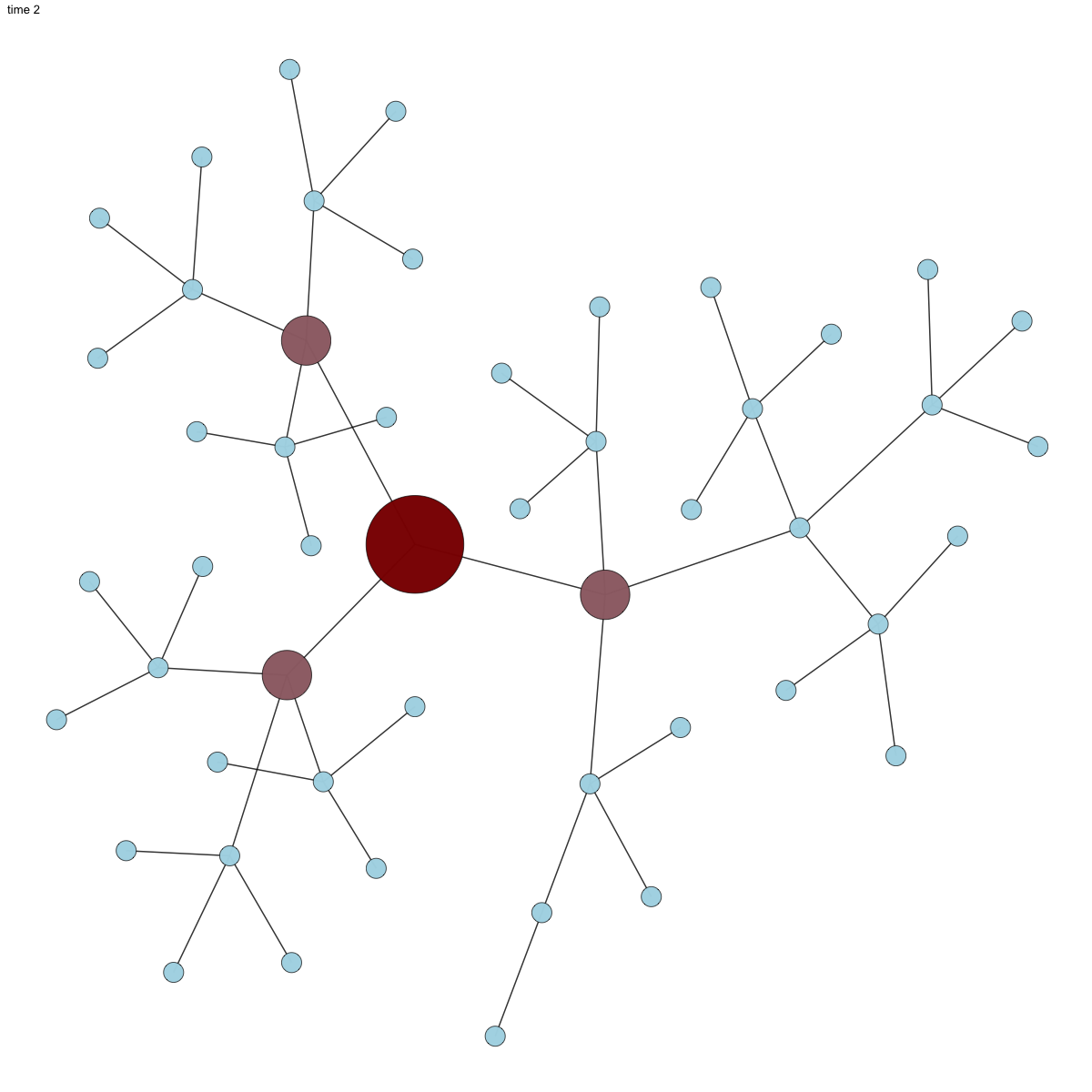}} &
    \includegraphics[scale=.07]{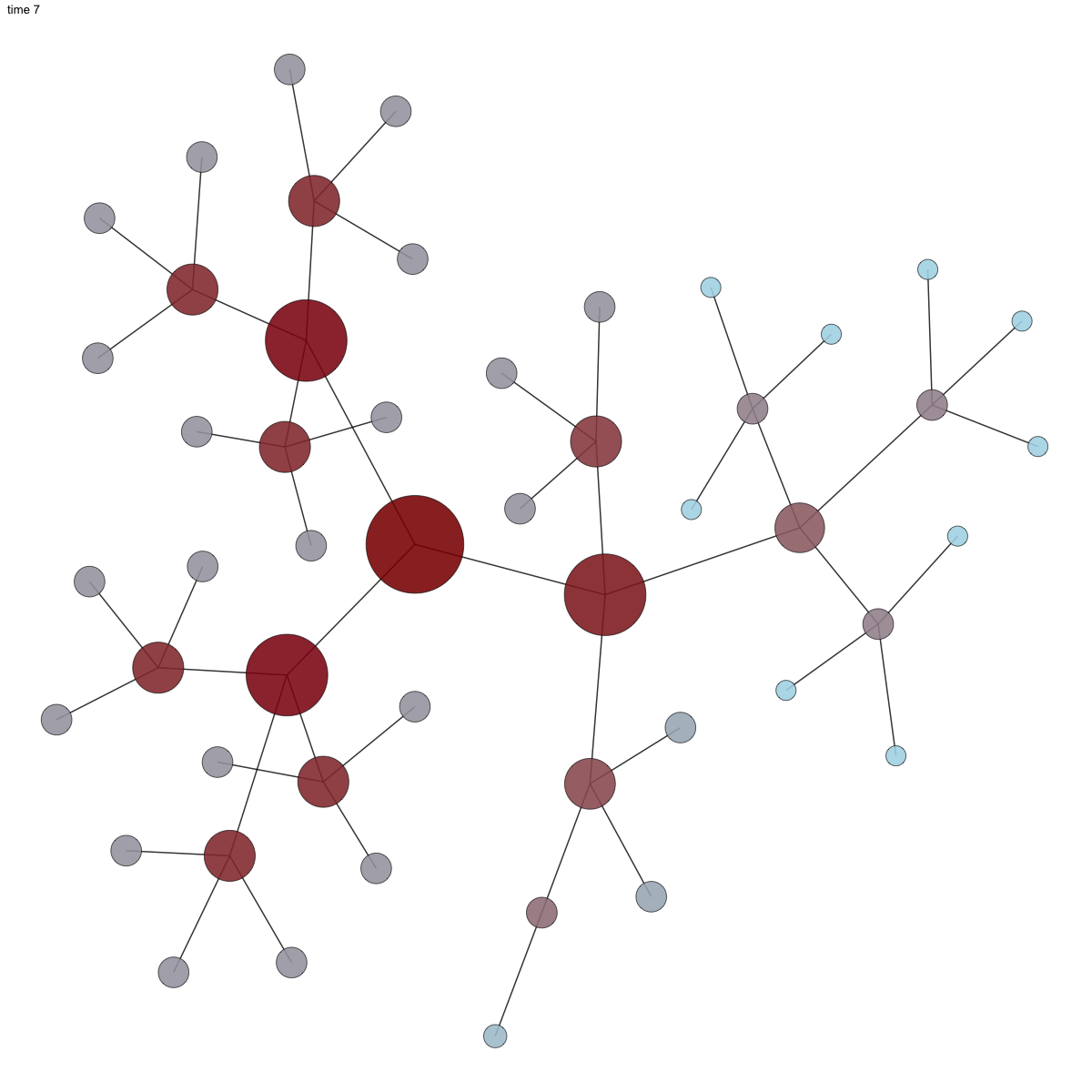}
    \end{tabular}
    \includegraphics[scale=.3]{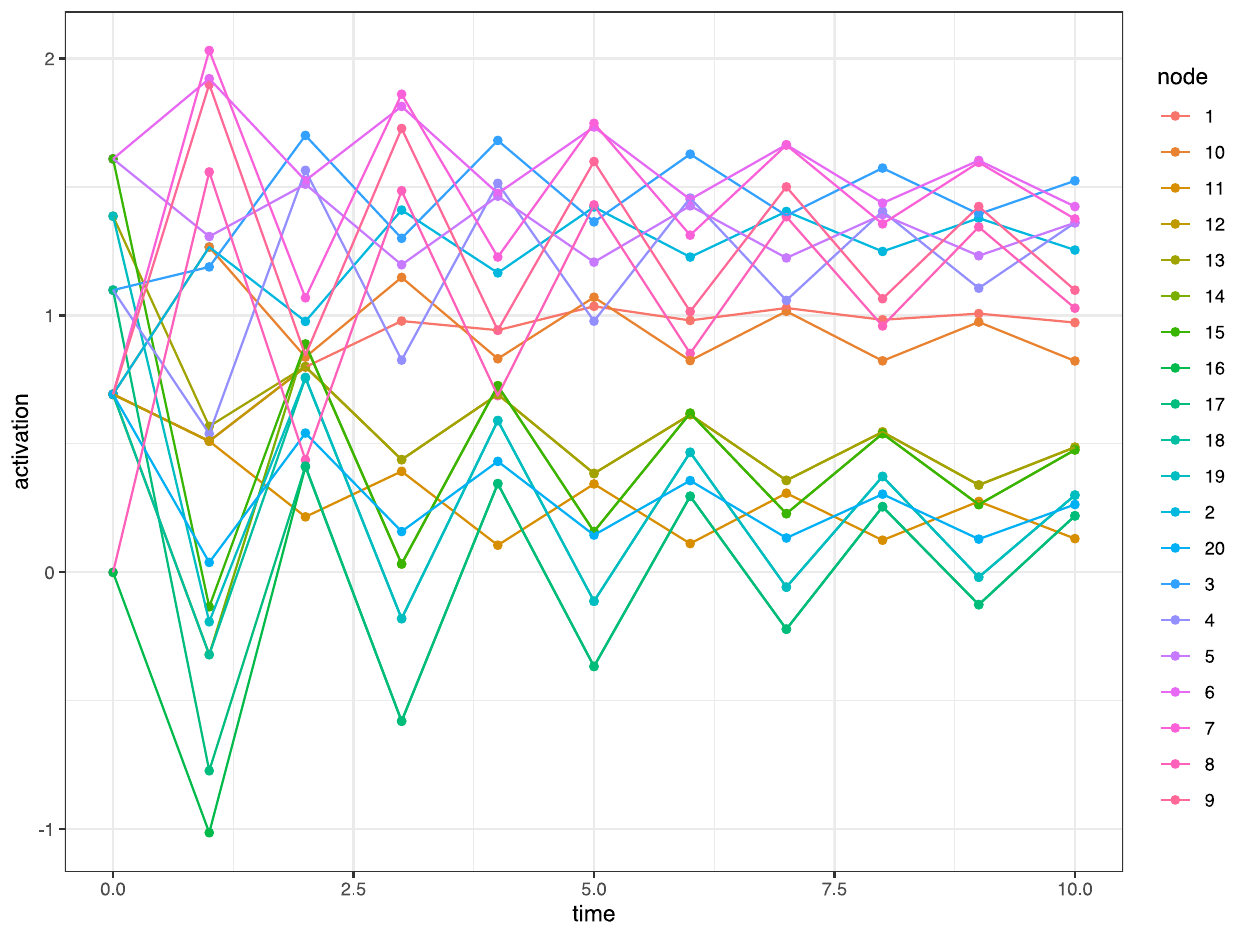}
    \caption{Individual node dynamics across the network. Each line corresponds to an activation level for a specific node over time.}
    \label{fig:networkSpread}
\end{figure}

We first define notation necessary for understanding the inputs to this computer simulation. For a network with $n$ nodes, two parameters control the activation spread, known as \textit{retention} and \textit{decay}, denoted as $r$ and $d$ respectively.
\begin{itemize}
    \item $r$ - a scalar or a vector of length $n$ that controls the proportion of activation  retained by a node at each time step of the simulation.
    \item $d$ - a scalar that controls the proportion of activation that is lost at each time step of the simulation.
\end{itemize}
There are three quantities of interest that define how activation spreads dynamically over time as functions of $r$ and $d$: \textit{reservoir}, \textit{outflow}, and \textit{inflow}. These are defined as follows:
\begin{itemize}
    \item $\text{reservoir}(t,n) = r\times \text{inflow}(t,n)$.
    \item $\text{outflow}(t,n) = \dfrac{(1-d)(1-r)\times \text{inflow}(t,n)}{\text{deg}(n)}$. 
    \item $\text{inflow}(t,n) = \sum_{i=1}^{\text{deg}(n)}\text{outflow}(t-1, n_i)+\text{reservoir}(t-1, n)$, where $\text{deg}(n)$ denotes the number of connections to node $n$.
\end{itemize}
The computer simulation then computes these quantities for each node in the network across time. A visualization of the spatiotemporal spreading dynamics is show in Figure \ref{fig:networkSpread}. 

\begin{figure}[t]
    \centering
    \includegraphics[scale=.50]{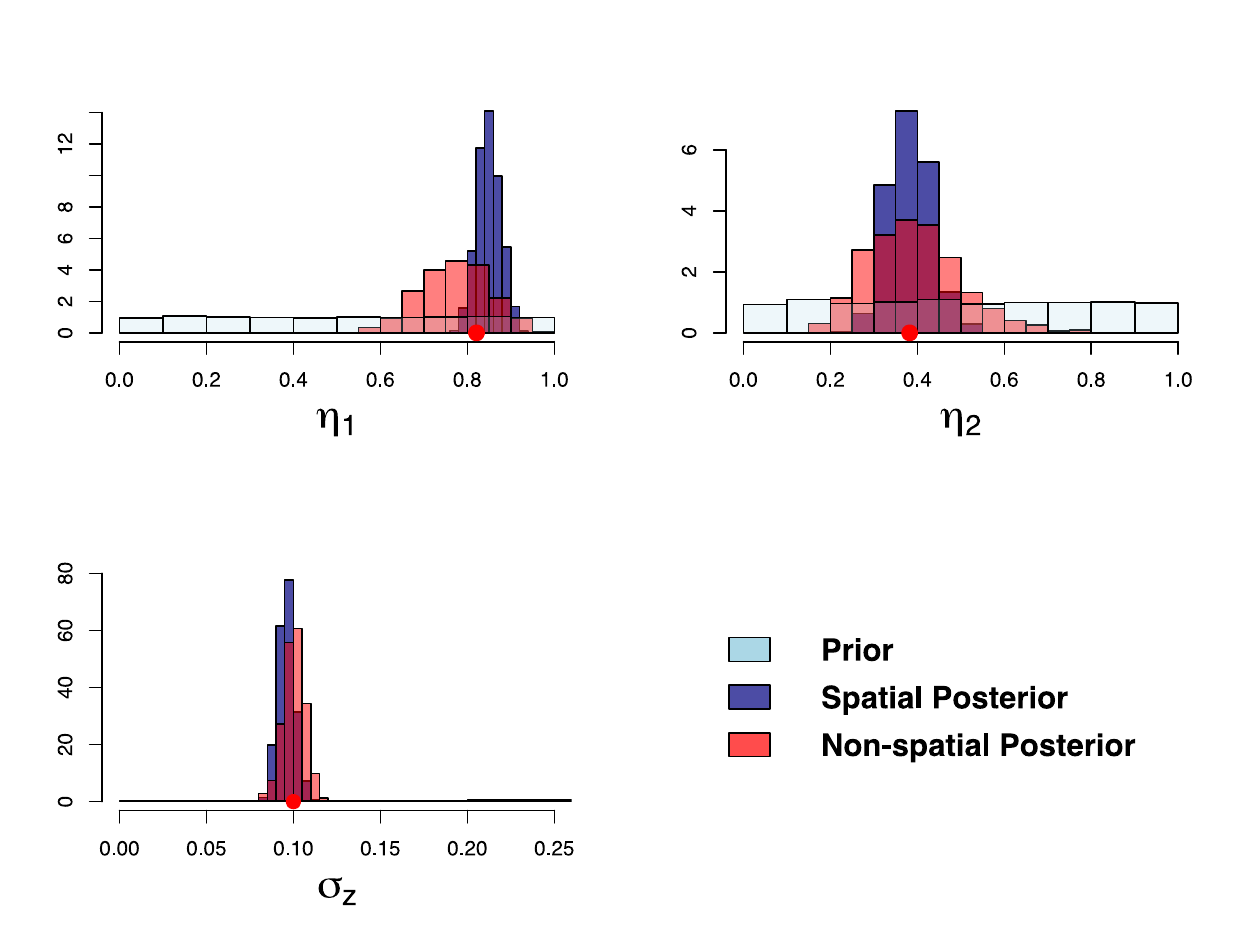}
    \caption{Network diffusion calibration posteriors. The lower uncertainty in the hierarchical model indicates pooling of information across node locations, lost in the heterogeneous model. Posterior mode point estimates are roughly similar.}
    \label{fig:networkPost}
\end{figure}

We employ a space-filling design to generate a small collection of plausible parameter values. Subsequently, we select a random validation point that is not included in the training set. 
After 10,000 posterior draws, the prior-to-posterior learning is evident in Figure~\ref{fig:networkPost} and demonstrates our methodology is capable of calibrating a computer model generating dynamics across a network. Although it is possible to define a Gaussian process over a graph \citep{Venkitaraman2020}, for this application we fix the correlation structure to arise from the adjacency matrix. Our results demonstrate that failing to account for spatial structure leads to information loss. Specifically, we observe a 13.7\% improvement in RMSE, decreasing from 0.110 to 0.095, when using the spatial model compared to the heterogeneous model, based on 25 computer model runs.

\section{Concluding remarks}\label{sec: discussion}
This manuscript demonstrates how statistical distributions can be used to devise a viable and effective probabilistic learning framework for emulating dynamically evolving spatiotemporal mechanistic systems without resorting to iterative inferential algorithms. Subsequently, we offer modularized calibration from field observations by regressing them on the emulated field. While this step requires MCMC for full probabilistic inference on the mechanistic model parameters, we do not need to recompute or update the emulated field. Building upon a rich literature on state-space extensions to nonlinear systems \citep[such as nonlinear PDEs explored in][but without melding of field observations]{Wang2021}, our approach can learn about mechanistic model parameters using more accurate and flexible emulators. Since we do not require accessibility to the mechanistic system (only its output is needed), this framework is equally applicable to systems built on mathematical models from physical principles and to agent-based models driven by computer programs (e.g., our network example). 

A key point of our framework is that we emulate the mechanistic system using analytically accessible matrix-variate statistical distributions without resorting to iterative algorithms that may consume resources until they converge. Furthermore, scaling to large number of locations and time points is also achieved in closed form using Bayesian transfer learning with the FFBS algorithm. Full uncertainty quantification for learning the mechanistic system's parameters will involve MCMC, but modularization again helps with computational feasibility. In this regard, our framework holds promise as an effective trainer for generative Artificially Intelligent engines such as ``BayesFlow'' \citep{radev2022ieee} or neural Bayes \citep{sainsbury2024likelihoodfree, zammit2024amortized}. Such developments pertaining to the use of our framework for amortized inference will constitute future research. 

In the interest of retaining exact inference, we fixed the covariance kernel hyper-parameters in the Gaussian processes used to emulate the system. This does not affect the emulated fields since, as we have remarked, the Gaussian process interpolates the field irrespective of the values of the hyper-parameters. Fixing these parameters can be achieved using exploratory methods for stochastic processes such as variograms \citep{zhang2019practical}. Alternatively, we can use predictive stacking to arrive at posterior distributions averaged over fixed values of the hyper-parameters \citep{zhang2024geostacking,presicce2024bayesian}. These approaches will still offer emulation without resorting to iterative MCMC algorithms. Alternatively, sequential screening procedure may be employed to identify the most influential kernel parameters \citep{Welch1992, Merrill2021}.     

Learning from massive amounts of high-resolution spatial-temporal field data has not been explicitly addressed here, but the rapidly evolving literature on scaling Gaussian processes to massive data is relevant. While traditional low-rank or inducing point methods such as \cite{snelson2005sparse}, \cite{wilson2015kernel} or the Gaussian predictive process \citep{Banerjee2008} remain feasible either over the set of spatial locations or over the space of inputs, it is well established that such methods can over-smooth when the number of outputs are very large \citep{Banerjee2017} and multi-resolution adaptations \citep{nychka2015, katzfussmultires} are needed to mitigate over-smoothing. Sparsity-inducing processes such as the Nearest-Neighbor Gaussian processes \citep{Datta2016, datta16b, finley2019efficient, Peruzzi2020} or the similarly themed Vecchia-based processes \citep{katzfuss2021general, gramacy2021deepGPactiveLearning, gramacy2022deepGP, cao2023vChol} offer greater scalability and can be used to model $\bm u_t(\cdot)$ in Section~\ref{sec:calibration}. Alternatively, we can extend the divide and conquer paradigm used for emulation (Section~\ref{sec:bayestransferlearning}) to calibration. However, the transfer learning framework may be hampered by irregular space-time coordinates. Future directions will explore spatial meta-kriging \citep{guhaniyogi2018meta, guhaniyogi2017divide} and predictive stacking \citep{presicce2024bayesian, pan2024bayesianinferencespatialtemporalnongaussian}.

Other extensions include treating outputs on different coordinates when adaptive grids are used to solve PDEs yielding misaligned outputs, i.e., not all runs are made over the same set of locations. Here, one may need to vectorize the incomplete output matrices and use multivariate normal distributions that would render learning using process-based misalignment models \citep[see, e.g., Chapter~7 in][]{Banerjee2014}. High-dimensional spatial outputs can be treated using dimension-reduction methods such as spatial factor models \citep{renBanerjee2013biocs,zhangbanerjee2021}. One could also generalize the model to have different variances at different spatial grids. In fact, richer inference is possible using MCMC algorithms that we have avoided (except to execute modularized calibration for the mechanistic parameters). The \texttt{RobustGaSP} package developed in \cite{gu2019robustgasp} is a viable candidate to achieve effective inference in these settings.  Extensions to generalized dynamic linear models 
\citep{West1985, Gamerman2013} will include analyzing non-Gaussian data and investigating richer specifications for $\bm G_t$ and $\bm W_t$. We can adapt our approach using vectorized multivariate Gaussian distributions to emulate mechanistic systems at unseen spatial locations and estimate (calibrate) in spatially misaligned settings, where the emulation locations and field measurements do not match \citep{Banerjee2014}. Lastly, the burgeoning field of sequential Monte Carlo analysis is applicable to build more general statistical emulators within this framework \citep{Hirt2019}. 

\section{Computing environments and timings}\label{sec: computing_environments}
We developed our methods in \texttt{C++} with functions available in \texttt{R} (version~4.4.1) employing the Rcpp package \citep{Dirk2011}. Computations for the predator-prey system were executed on a laptop running 64-bit Windows 11 with a 12th Gen Intel(R) Core(TM) i7-12700H, 2.30 GHz processor, 32.0 GB RAM at 4800 MHz, 6 GB Graphics Card equipped with multiple GPUs. Computations pertaining to the PDE and network diffusion examples were executed on a laptop running macOS 14.7.6, equipped with an Apple M3 Max processor (16-core CPU, 40-core GPU) and 64 GB of RAM. All computer programs required to reproduce the numerical results in this manuscript are available from the GitHub repository \href{https://github.com/xiangchen-stat/Bayesian_Modeling_Mechanistic_Systems}{https://github.com/xiangchen-stat/Bayesian\_Modeling\_Mechanistic\_Systems}. 

Empirical computation time required to solve and emulate the Lotka-Volterra ODE system, as described in Section~\ref{sec:app_emulation_LV}, was approximately 0.26 and 2.16 seconds, respectively. Computations described in Section~\ref{sec:app_emulation_pde} for the PDE system required approximately 9 minutes and 13 seconds for solving the system, and 21, 61, and 59 seconds to emulate the field for the three considered variance structures, respectively. Calibration experiments, as described in Section~\ref{sec:app_calibration}, delivered full posterior inference in 98 minutes for the predator-prey system, 59 minutes for the PDE system, and about 14~and~25 seconds for the non-spatial and spatial models in the network diffusion system, respectively.

\acks{Sudipto Banerjee acknowledges funding from the National Science Foundation through DMS-1916349 and DMS-2113778; and funding from the National Institute of Health through NIEHS-R01ES027027, NIEHS-R01ES030210 and NIGMS-R01GM148761. The work of all the authors were supported, in part, from these funds.}

\section{Appendix}\label{appendix}


\subsection{Distributions in the FFBS Algorithms~\ref{alg:KF},~\ref{alg:BS}~and~\ref{alg:FFBS}}\label{appendix: exact_theory}
The joint posterior distribution \eqref{matrix_normal_wishart_posterior} is obtained by exploiting familiar distribution theory from statistical linear models. The first two equations in \eqref{basic_dlm_matrix-variate} implies a well-defined joint distribution for $\bm Y_t$ and $\bm \Theta_t$ conditional on $\bm Y_{1:(t-1)}$ as
\begin{equation}
    \label{appendix: joint distribution_theta_y}
    \left. \begin{bmatrix}
        \mbox{vec}(\bm Y_t) \\
        \mbox{vec}(\bm \Theta_t)
    \end{bmatrix} \;\right |\; \bm Y_{1:(t-1)},\, \bm\Sigma \sim \mathcal{N}_{(N+p)S\times 1}\left(\begin{bmatrix}
        \mbox{vec}(\bm q_t) \\
        \mbox{vec}(\bm a_t)
    \end{bmatrix}, \begin{bmatrix}
        \bm\Sigma \otimes \bm Q_t & \bm\Sigma \otimes \left(\bm F_t\bm A_t\right) \\
        \bm\Sigma \otimes \left(\bm A_t\bm F_t^{\top}\right) & \bm\Sigma \otimes \bm A_t 
    \end{bmatrix}\right)\;. 
\end{equation}
Using familiar properties of the multivariate normal distribution, we obtain
\begin{equation}
    \label{appendix: posterior_theta}
    \mbox{vec}(\bm\Theta_t) \mid \bm Y_{1:t},\,\bm\Sigma \sim \mathcal{N}_{pS\times 1}\left(\mbox{vec}(\bm a_t) + \left(\bm I_S \otimes \left(\bm A_t\bm F_t^{\top}\bm Q_t^{-1}\right)\right)\mbox{vec}(\bm Y_t - \bm q_t), \bm\Sigma \otimes \bm M_t\right)\;,
\end{equation}
which implies $\displaystyle \bm\Theta_t \mid \bm Y_t, \bm\Sigma \sim \mathcal{MN}\left(\bm m_t, \bm M_t, \bm\Sigma\right)$, where $\bm m_t = \bm a_t + \bm A_t\bm F_t^{\top}\bm Q_t^{-1}(\bm Y_t - \bm q_t)$ and $\bm{M}_t = \bm{A}_t - \bm{A}_t\bm{F}_t^{\top}\bm{Q}_t^{-1}\bm{F}_t\bm{A}_t$. The posterior distribution of $\bm\Sigma \mid \bm Y_{1:t}$ is obtained from 
\begin{equation}
    \label{appendix: posterior_sigma}
    \begin{split}
        p(\bm\Sigma \mid \bm Y_{1:t}) &= \underbrace{\mathcal{IW}\left(\bm\Sigma \mid n_{t-1}, \bm D_{t-1}\right)}_{p(\bm\Sigma \mid \bm Y_{1:(t-1)})}\times \underbrace{\mathcal{MN}(\bm Y_t \mid \bm q_t, \bm Q_t, \bm \Sigma)}_{p(\bm Y_t \mid \bm Y_{1:(t-1)}, \bm\Sigma)} \\
        &\propto \frac{\exp\left(-\frac{1}{2}\mbox{tr}\left[\bm D_{t-1}\bm\Sigma^{-1}\right]\right)}{\left(\det(\bm\Sigma)\right)^{\frac{n_{t-1} + S + 1}{2}}} \times \frac{\exp\left(-\frac{1}{2}\mbox{tr}\left[(\bm Y_t - \bm q_t)^{\top}\bm Q_t^{-1}(\bm Y_t - \bm q_t)\bm\Sigma^{-1}\right]\right)}{\left(\det(\bm\Sigma)\right)^{\frac{N}{2}}} \\
        &\propto \frac{\exp\left(-\frac{1}{2}\mbox{tr}\left[\bm D_t \bm\Sigma^{-1}\right]\right)}{\det\left(\bm\Sigma\right)^{\frac{n_t + S+1}{2}}} \propto \mathcal{IW}\left(\bm\Sigma \mid n_t, \bm D_t\right)\;,
    \end{split}
\end{equation}
where $n_t = n_{t-1} + N$ and $\displaystyle \bm D_{t} = \bm D_{t-1} + (\bm Y_t - \bm q_t)^{\top}\bm Q_t^{-1}(\bm Y_t - \bm q_t)$. Forward sampling computes the above quantities recursively to arrive at \eqref{appendix: posterior_theta}~and~\eqref{appendix: posterior_sigma} for $t=T$ and draws $L$ samples of $\bm\Sigma$ from \eqref{appendix: posterior_sigma} followed by one draw of $\bm\Theta_T$ from \eqref{appendix: posterior_theta} for each of the $L$ values of $\bm\Sigma$. 

The forward sampling described above yields $L$ samples of $\bm\Theta_T, \bm\Sigma$ from \eqref{matrix_normal_wishart_posterior}. For each of the $L$ values, we will repeat an entire cycle of backward sampling. From \eqref{backwardsampling}, backward sampling will draw from $p(\bm\Theta_t \mid \bm\Sigma, \bm\Theta_{t+1}, \bm Y_{1:t})$, which can be achieved by drawing one value of $\bm\Theta_t$ for each of the $L$ sampled values of $(\bm\Theta_{t+1},\bm\Sigma)$ sequentially from $t=T-1,\ldots,0$. Alternatively, we can draw the samples from $p(\bm\Theta_t \mid \bm \Sigma, \bm Y_{1:t})$ directly as described in \eqref{BS:conj}. To derive this distribution, we make use of the Markovian structure in the model to note that 
\[
p(\bm\Theta_{t} \mid \bm\Theta_{t+1:T}, \bm Y_{1:T}, \bm\Sigma) \propto p(\bm\Theta_{t} \mid \bm Y_{1:t}, \bm\Sigma) \times p(\bm\Theta_{t+1} \mid \bm \Theta_t, \bm\Sigma) \propto p(\bm\Theta_{t} \mid \bm\Theta_{t+1}, \bm Y_{1:t}, \bm\Sigma)\;,
\]
where we have used the facts that $\bm\Theta_t$ is conditionally independent of $\bm Y_{t+1:T}$ given $\bm Y_{1:t}$ and that $\bm\Theta_{t+1}$ is conditionally independent of $\bm Y_{1:T}$ given $\bm\Theta_t$. The joint distribution of $\bm\Theta_t$ and $\bm\Theta_{t+1}$ conditional on $\bm Y_{1:t}$ is given by
\begin{equation*}
    \label{appendix: joint_theta_given y}
    \left. \begin{bmatrix}
        \mbox{vec}(\bm\Theta_{t+1}) \\
        \mbox{vec}(\bm\Theta_{t})
    \end{bmatrix} \;\right|\; \bm Y_{1:t}, \bm\Sigma \sim \mathcal{N}_{2pS\times 1}\left(\begin{bmatrix} \mbox{vec}(\bm a_{t+1}) \\ \mbox{vec}(\bm m_t) \end{bmatrix}, \begin{bmatrix} \bm\Sigma\otimes\bm A_{t+1} & \bm\Sigma\otimes \left(\bm G_{t+1}\bm M_t\right) \\ \bm\Sigma\otimes \left(\bm M_t\bm G_{t+1}^{\top}\right) & \bm\Sigma\otimes \bm M_t\end{bmatrix}\right)\;. 
\end{equation*}
Therefore, the conditional distribution of $\bm\Theta_{t} $ given $\bm\Theta_{t+1}, \bm Y_{1:t}, \bm\Sigma$ is
\begin{multline}
    \label{appendix: conditional_BS_conj_1}
    \mbox{vec}(\bm\Theta_{t}) \mid \bm\Theta_{t+1}, \bm Y_{1:t}, \bm\Sigma \sim \mathcal{N}_{pS\times 1}\left(\mbox{vec}(\bm m_t) + \left(\bm I_S\otimes \bm M_t\bm G_{t+1}^{\top}\bm A_{t+1}^{-1}\right)\mbox{vec}(\bm\Theta_{t+1} - \bm a_{t+1}), \right. \\ 
    \left. \bm\Sigma\otimes\left(\bm M_t - \bm M_t\bm G_{t+1}^{\top}\bm A_{t+1}^{-1}\bm G_{t+1}\bm M_t\right)\right)\;,
\end{multline}
which implies that $p(\bm \Theta_t \mid \bm\Theta_{t+1} \bm Y_{1:t}, \bm\Sigma)$ is equal to
\[
 \mathcal{MN}\left(\bm\Theta_t \mid \bm m_t + \bm M_t\bm G_{t+1}^{\top}\bm A_{t+1}^{-1}\left(\bm \Theta_{t+1} - \bm a_{t+1}\right), \bm M_t - \bm M_t\bm G_{t+1}^{\top}\bm A_{t+1}^{-1}\bm G_{t+1}\bm M_t, \bm\Sigma\right)\;.
\]
This also equals $p(\bm\Theta_t \mid \bm\Theta_{t+1}, \bm Y_{1:T}, \bm\Sigma)$ because $\bm\Theta_{t}$ and $\bm Y_{t+1:T}$ are conditionally independent given $\bm\Theta_{t+1}$. Denoting $\bm\Theta_{t+1} \mid \bm Y_{1:T}, \bm\Sigma \sim \mathcal{MN}\left(\bm h_{t+1}, \bm H_{t+1}, \bm\Sigma\right)$, where $\bm h_{T} = \bm m_{T}$ and $\bm H_{T} = \bm M_T$, we derive the joint distribution $p(\bm \Theta_t, \bm\Theta_{t+1} \mid \bm Y_{1:T}, \bm\Sigma) = p(\bm\Theta_{t+1} \mid \bm Y_{1:T}, \bm\Sigma) \times p(\bm \Theta_t \mid \bm\Theta_{t+1} \bm Y_{1:T}, \bm\Sigma)$, which is a composition of two matrix-variate linear models with independent error matrices constructed over the same probability space (conditional on $\bm Y_{1:T}$ and $\bm\Sigma$),
\begin{equation}
    \label{appendix: BS_conj_linear_model}
    \begin{split}
        \bm \Theta_t = \bm m_t + \bm M_t\bm G_{t+1}^{\top}\bm A_{t+1}^{-1}\left(\bm \Theta_{t+1} - \bm a_{t+1}\right) + \bm{E}_t\quad \mbox{and}\quad \bm \Theta_{t+1} = \bm h_{t+1} + \tilde{\bm E}_{t+1}, 
    \end{split}   
\end{equation}
where $\bm{E}_t \sim \mathcal{MN}\left(\bm O, \bm M_t - \bm M_t\bm G_{t+1}^{\top}\bm A_{t+1}^{-1}\bm G_{t+1}\bm M_t, \bm\Sigma\right)$ is distributed independently of $\tilde{\bm E}_{t+1} \sim \mathcal{MN}\left(\bm O, \bm H_{t+1}, \bm\Sigma\right)$. Substituting the model for $\bm\Theta_{t+1}$ into the first equation in \eqref{appendix: BS_conj_linear_model} followed by familiar multivariate normal calculations yields $p(\bm \Theta_t \mid \bm Y_{1:T}, \bm \Sigma) = \mathcal{MN}\left(\bm h_t, \bm H_t, \bm\Sigma\right)$, where $\bm h_t$ and $\bm H_{t}$ are defined recursively as in \eqref{BS:conj}. This completes the required closed-form derivations for all the distributions in the exact FFBS algorithm. 

\subsection{Calculus-free derivation of $\mathcal{HT}$ density}

Recall that in equations \eqref{eq:hyper_t} and \eqref{eq:hyper-T-scalar} we were able to compute their distributions by means of marginalizing, that is, integrating, out the undesired variance terms. We demonstrate an approach to this problem without using integration. Begin with the identity
\begin{equation}\label{eq:bayesrule}
    p(A\mid B)p(B) = p(A,B)
\end{equation}
where $A$ and $B$ denote random variables. Let $p(A\mid B) = p(\bm{\Sigma}\mid \bm{\Theta}_t, \bm{Y}_{1:T})$, $p(B) = p(\bm{\Theta}_{t}\mid \bm{Y}_{1:T})$, and $p(A,B) = p(\bm{\Theta}_t, \bm{\Sigma} \mid \bm{Y}_{1:T})$. Dividing both sides of \eqref{eq:bayesrule} by $p(\bm{\Sigma}\mid \bm{\Theta}_t, \bm{Y}_{1:T})$, we get:
\begin{equation}\label{eq:marginalizebybayes}
    p(\bm{\Theta}_{t}\mid\bm{Y}_{1:T}) = \frac{p(\bm{\Theta}_t, \bm{\Sigma} \mid \bm{Y}_{1:T})}{p(\bm{\Sigma} \mid \bm{\Theta}_t, \bm{Y}_{1:T})}
\end{equation}
Recall that $p(\bm{\Theta}_t, \bm{\Sigma} \mid \bm{Y}_{1:T})$ is matrix-normal-inverse-Wishart with density (follows from \eqref{matrix_normal_wishart_prior}:)
\begin{equation}\label{eq:dmniw}
    \begin{split}
        p(\bm\Theta_t, \bm\Sigma_T\mid \bm{Y}_{1:T})
        &= \frac{\left(\det(\bm D_T)\right)^{n_T/2}\exp\left(-\frac{1}{2}\mbox{tr}\left\{\left[\bm{D}_T + (\bm\Theta_t - \bm h_t)^{\top}\bm H_t^{-1}(\bm \Theta_t - \bm h_t)\right]\bm\Sigma^{-1}\right\}\right)}{2^{n_T S/2}\left(2\pi\right)^{pS/2}\Gamma_{S}(\frac{n_T}{2})\left(\det(\bm H_t)\right)^{S/2}\left(\det(\bm\Sigma)\right)^{\frac{n_T + S + p + 1}{2}}}\;,
    \end{split}
\end{equation}

We are also given that $p(\bm{\Sigma} \mid \bm{\Theta}_t, \bm{Y}_{1:T})$ is an inverse-Wishart. More specifically, $p(\bm{\Sigma} \mid \bm{\Theta}_t, \bm{Y}_{1:T}) \propto p(\bm{\Theta}_t, \bm{\Sigma}\mid \bm{Y}_{1:T})$, which means they share product terms in their respective expressions that contain $\bm{\Sigma}$ and $\bm{\Theta}_t$ (and $\bm{Y}_{1:T}$). Hence:
\begin{equation}\label{eq:iwsigmagivthetaprop}
    p(\bm{\Sigma} \mid \bm{\Theta}_t, \bm{Y}_{1:T}) \propto \frac{\exp\left(-\frac{1}{2}\mbox{tr}\left\{\left[\bm{D}_T + (\bm\Theta_t - \bm h_t)^{\top}\bm H_t^{-1}(\bm \Theta_t - \bm h_t)\right]\bm\Sigma^{-1}\right\}\right)}{\left(\det(\bm\Sigma)\right)^{\frac{n_T + S + p + 1}{2}}}
\end{equation}

Introducing the constant (non-$\bm\Sigma$) multiplier for \eqref{eq:iwsigmagivthetaprop}, given it is Inverse-Wishart, yields:
\begin{equation}\label{eq:iwsigmagivthetaexact}
    \begin{split}
        p(\bm{\Sigma} \mid \bm{\Theta}_t, \bm{Y}_{1:T}) =& \frac{\exp\left(-\frac{1}{2}\mbox{tr}\left\{\left[\bm{D}_T + (\bm\Theta_t - \bm h_t)^{\top}\bm H_t^{-1}(\bm \Theta_t - \bm h_t)\right]\bm\Sigma^{-1}\right\}\right)}{2^{(n_T + p)S/2}\Gamma_{S}\left(\frac{n_T + p}{2}\right)\left(\det(\bm\Sigma)\right)^{\frac{n_T + S + p + 1}{2}}}\\
        &\times \left(\det\left(\bm{D}_T + (\bm\Theta_t - \bm h_t)^{\top}\bm H_t^{-1}(\bm \Theta_t - \bm h_t)\right)\right)^{(n_T + p)/2}
    \end{split}
\end{equation}
Finally, we take the quotient to obtain the exact form of $p(\bm\Theta_t \mid \bm Y_{1:T})$:
\begin{equation}\label{eq:hypert_nocalc}
\begin{split}
    p(\bm\Theta_t \mid \bm Y_{1:T}) =& \frac{\left(\det(\bm D_T)\right)^{n_T/2}\exp\left(-\frac{1}{2}\mbox{tr}\left\{\left[\bm{D}_T + (\bm\Theta_t - \bm h_t)^{\top}\bm H_t^{-1}(\bm \Theta_t - \bm h_t)\right]\bm\Sigma^{-1}\right\}\right)}{2^{n_T S/2}\left(2\pi\right)^{pS/2}\Gamma_{S}(\frac{n_T}{2})\left(\det(\bm H_t)\right)^{S/2}\left(\det(\bm\Sigma)\right)^{\frac{n_T + S + p + 1}{2}}}\\
    &\times\frac{2^{(n_T + p)S/2}\Gamma_{S}\left(\frac{n_T + p}{2}\right)\left(\det(\bm\Sigma)\right)^{\frac{n_T + S + p + 1}{2}}}{\exp\left(-\frac{1}{2}\mbox{tr}\left\{\left[\bm{D}_T + (\bm\Theta_t - \bm h_t)^{\top}\bm H_t^{-1}(\bm \Theta_t - \bm h_t)\right]\bm\Sigma^{-1}\right\}\right)}\\
        &\times \left(\det\left(\bm{D}_T + (\bm\Theta_t - \bm h_t)^{\top}\bm H_t^{-1}(\bm \Theta_t - \bm h_t)\right)\right)^{-(n_T + p)/2}\\
    =& \pi^{-pS/2}\frac{\Gamma_{S}\left(\frac{n_T + p}{2}\right)}{\Gamma_{S}\left(\frac{n_T}{2}\right)}\left(\det(\bm D_T)\right)^{-p/2}\left(\det(\bm H_t)\right)^{S/2}\\
    &\times \left(\det\left(\bm{I}_S + \bm{D}_{T}^{-1}(\bm\Theta_t - \bm h_t)^{\top}\bm H_t^{-1}(\bm \Theta_t - \bm h_t)\right)\right)^{-(n_T + p)/2}
\end{split}
\end{equation}
This is the Hyper-T density in \eqref{eq:hyper_t}, specifically $\mathcal{HT}(\bm{\Theta}_{t}\mid \bm{h}_t, \bm{H}_t, n_T, \bm{D}_T)$. We have thus demonstrated a derivation of the Hyper-T density without using integration.

\subsection{Distributions in the FFBS algorithm for the model in \eqref{dlm_gp_double}}
The joint posterior distribution corresponding to the model in \eqref{dlm_gp_double} is obtained by modifying \eqref{appendix: joint distribution_theta_y}. The first two equations in \eqref{dlm_gp_double} implies the joint distribution for $\bm Y_t$ and $\bm \Theta_t$ conditional on $\bm Y_{1:(t-1)}$ as
\begin{equation}
    \label{appendix: joint distribution_theta_y_R}
    \left. \begin{bmatrix}
        \mbox{vec}(\bm Y_t) \\
        \mbox{vec}(\bm \Theta_t)
    \end{bmatrix} \;\right |\; \bm Y_{1:(t-1)},\, \sigma^2 \sim \mathcal{N}_{(N+p)S\times 1}\left(\begin{bmatrix}
        \mbox{vec}(\bm q_t) \\
        \mbox{vec}(\bm a_t)
    \end{bmatrix}, \sigma^2\begin{bmatrix}
        \bm R \otimes \bm Q_t & \bm R \otimes \left(\bm F_t\bm A_t\right) \\
        \bm R \otimes \left(\bm A_t\bm F_t^{\top}\right) & \bm R \otimes \bm A_t 
    \end{bmatrix}\right)\;. 
\end{equation}
We obtain the analog of \eqref{appendix: posterior_theta}
\begin{equation}
    \label{appendix: posterior_theta_R}
    \mbox{vec}(\bm\Theta_t) \mid \bm Y_{1:t},\,\sigma^2 \sim \mathcal{N}_{pS\times 1}\left(\mbox{vec}(\bm a_t) + \left(\bm I_S \otimes \left(\bm A_t\bm F_t^{\top}\bm Q_t^{-1}\right)\right)\mbox{vec}(\bm Y_t - \bm q_t), \sigma^2\bm R \otimes \bm M_t\right)\;,
\end{equation}
which implies $\displaystyle \bm\Theta_t \mid \bm Y_t, \sigma^2 \sim \mathcal{MN}\left(\bm m_t, \bm M_t, \sigma^2\bm R\right)$, where $\bm m_t = \bm a_t + \bm A_t\bm F_t^{\top}\bm Q_t^{-1}(\bm Y_t - \bm q_t)$ and $\bm{M}_t = \bm{A}_t - \bm{A}_t\bm{F}_t^{\top}\bm{Q}_t^{-1}\bm{F}_t\bm{A}_t$. The posterior distribution of $\sigma^2 \mid \bm Y_{1:t}$ is obtained as 
\begin{equation}
    \label{appendix: posterior_sigma_R}
    \begin{split}
        p(\sigma^2 \mid \bm Y_{1:t}) &= \underbrace{\mathcal{IG}\left(\sigma^2 \mid n_{t-1}, d_{t-1}\right)}_{p(\sigma^2 \mid \bm Y_{1:(t-1)})}\times \underbrace{\mathcal{MN}(\bm Y_t \mid \bm q_t, \bm Q_t, \sigma^2\bm R)}_{p(\bm Y_t \mid \bm Y_{1:(t-1)}, \sigma^2)} \\
        &\propto \frac{1}{{\sigma}^{2(n_{t-1}+1)}}
        \times \exp \left(-\frac{d_{t-1}}{\sigma^2}\right) 
        \times \frac{\exp\left(-\frac{1}{2\sigma^2}\mbox{tr}\left[(\bm Y_t - \bm q_t)^{\top}\bm Q_t^{-1}(\bm Y_t - \bm q_t)\bm R^{-1}\right]\right)}{\sigma^{2(\frac{NS}{2})}} \\
        &\propto \frac{1}{{\sigma}^{2(n_{t-1}+\frac{NS}{2}+1)}}
        \times \exp\left(-\frac{1}{\sigma^2}\left(d_{t-1}+\frac{1}{2}\mbox{tr}\left[(\bm Y_t - \bm q_t)^{\top}\bm Q_t^{-1}(\bm Y_t - \bm q_t)\bm R^{-1}\right]\right)\right) \\
        &\propto \mathcal{IG}\left(\sigma^2 \mid n_{t}, d_{t}\right)\;,
    \end{split}
\end{equation}
where $n_t = n_{t-1} + \frac{NS}{2}$ and $d_{t} = d_{t-1}+\frac{1}{2}\mbox{tr}\left[(\bm Y_t - \bm q_t)^{\top}\bm Q_t^{-1}(\bm Y_t - \bm q_t)\bm R^{-1}\right]$. Forward sampling computes the above quantities recursively to arrive at \eqref{appendix: posterior_theta_R}~and~\eqref{appendix: posterior_sigma_R} for $t=T$ and draws $L$ samples of $\sigma^2$ from \eqref{appendix: posterior_sigma_R} followed by one draw of $\bm\Theta_T$ from \eqref{appendix: posterior_theta_R} for each of the $L$ values of $\sigma^2$. 

\bibliography{bib}

\begin{thebibliography}{100}
\providecommand{\natexlab}[1]{#1}
\providecommand{\url}[1]{\texttt{#1}}
\expandafter\ifx\csname urlstyle\endcsname\relax
  \providecommand{\doi}[1]{doi: #1}\else
  \providecommand{\doi}{doi: \begingroup \urlstyle{rm}\Url}\fi

\bibitem[Abdalla et~al.(2020)Abdalla, Banerjee, Ramachandran, and Arnold]{AbdallaEtAl2020}
N.~Abdalla, S.~Banerjee, G.~Ramachandran, and S.~Arnold.
\newblock Bayesian state space modeling of physical processes in industrial hygiene.
\newblock \emph{Technometrics}, 62\penalty0 (2):\penalty0 147--160, 2020.
\newblock \doi{10.1080/00401706.2019.1630009}.
\newblock URL \url{https://doi.org/10.1080/00401706.2019.1630009}.

\bibitem[Banerjee(2017)]{Banerjee2017}
S.~Banerjee.
\newblock {High-dimensional {B}ayesian geostatistics}.
\newblock \emph{Bayesian analysis}, 12\penalty0 (2):\penalty0 583, 2017.

\bibitem[Banerjee and Roy(2014)]{banerjee2014linear}
S.~Banerjee and A.~Roy.
\newblock \emph{Linear algebra and matrix analysis for statistics}.
\newblock CRC Press, Boca Raton, FL, 2014.

\bibitem[Banerjee et~al.(2008)Banerjee, Gelfand, Finley, and Sang]{Banerjee2008}
S.~Banerjee, A.~E. Gelfand, A.~O. Finley, and H.~Sang.
\newblock Gaussian predictive process models for large spatial data sets.
\newblock \emph{Journal of the Royal Statistical Society Series B: Statistical Methodology}, 70\penalty0 (4):\penalty0 825--848, 2008.

\bibitem[Banerjee et~al.(2014)Banerjee, Carlin, and Gelfand]{Banerjee2014}
S.~Banerjee, B.~Carlin, and A.~Gelfand.
\newblock \emph{{Hierarchical Modeling and Analysis for Spatial Data}}.
\newblock CRC Press, 2014.

\bibitem[Bayarri et~al.(2007{\natexlab{a}})Bayarri, Berger, Cafeo, Garcia-Donato, Liu, Palomo, Parthasarathy, Paulo, Sacks, and Walsh]{Bayarri2007}
M.~J. Bayarri, J.~O. Berger, J.~Cafeo, G.~Garcia-Donato, F.~Liu, J.~Palomo, R.~J. Parthasarathy, R.~Paulo, J.~Sacks, and D.~Walsh.
\newblock Computer model validation with functional output.
\newblock \emph{The Annals of Statistics}, 35\penalty0 (5):\penalty0 1874--1906, 2007{\natexlab{a}}.

\bibitem[Bayarri et~al.(2007{\natexlab{b}})Bayarri, Berger, Paulo, Sacks, Cafeo, Cavendish, Lin, and Tu]{Bayarri2007tech}
M.~J. Bayarri, J.~O. Berger, R.~Paulo, J.~Sacks, J.~A. Cafeo, J.~Cavendish, C.-H. Lin, and J.~Tu.
\newblock A framework for validation of computer models.
\newblock \emph{Technometrics}, 49\penalty0 (2):\penalty0 138--154, 2007{\natexlab{b}}.

\bibitem[Bayarri et~al.(2009)Bayarri, Berger, and Liu]{Bayarri2009}
M.~J. Bayarri, J.~O. Berger, and F.~Liu.
\newblock {Modularization in {B}ayesian analysis, with emphasis on analysis of computer models}.
\newblock \emph{Bayesian Analysis}, 4\penalty0 (1):\penalty0 119 -- 150, 2009.

\bibitem[Blei et~al.(2017)Blei, Kucukelbir, and McAuliffe]{bleiEtAl2017jasaReview}
D.~M. Blei, A.~Kucukelbir, and J.~D. McAuliffe.
\newblock Variational inference: A review for statisticians.
\newblock \emph{Journal of the American Statistical Association}, 112\penalty0 (518):\penalty0 859--877, 2017.
\newblock \doi{10.1080/01621459.2017.1285773}.
\newblock URL \url{https://doi.org/10.1080/01621459.2017.1285773}.

\bibitem[Cao et~al.(2023)Cao, Kang, Jimenez, Sang, Schaefer, and Katzfuss]{cao2023vChol}
J.~Cao, M.~Kang, F.~Jimenez, H.~Sang, F.~T. Schaefer, and M.~Katzfuss.
\newblock Variational sparse inverse cholesky approximation for latent {G}aussian processes via double kullback-leibler minimization.
\newblock In A.~Krause, E.~Brunskill, K.~Cho, B.~Engelhardt, S.~Sabato, and J.~Scarlett, editors, \emph{Proceedings of the 40th International Conference on Machine Learning}, volume 202 of \emph{Proceedings of Machine Learning Research}, pages 3559--3576. PMLR, 23--29 Jul 2023.
\newblock URL \url{https://proceedings.mlr.press/v202/cao23b.html}.

\bibitem[Carter and Kohn(1994)]{Carter1994}
C.~Carter and R.~Kohn.
\newblock {On Gibbs Sampling for State Space Models}.
\newblock \emph{Biometrika}, 1994.

\bibitem[Chan and Vitevitch(2009)]{Chan2009}
K.~Y. Chan and M.~S. Vitevitch.
\newblock The influence of the phonological neighborhood clustering coefficient on spoken word recognition.
\newblock \emph{Journal of Experimental Psychology: Human Perception and Performance}, 35\penalty0 (6):\penalty0 1934, 2009.

\bibitem[Chan-Golston et~al.(2020)Chan-Golston, Banerjee, and Handcock]{ChanGolston_Banerjee_Handcock_2020}
A.~M. Chan-Golston, S.~Banerjee, and M.~S. Handcock.
\newblock {Bayesian inference for finite populations under spatial process settings}.
\newblock \emph{Environmetrics}, 31, 2020.

\bibitem[Conti and O’Hagan(2010)]{conti2010bayesian}
S.~Conti and A.~O’Hagan.
\newblock Bayesian emulation of complex multi-output and dynamic computer models.
\newblock \emph{Journal of Statistical Planning and Inference}, 140\penalty0 (3):\penalty0 640--651, 2010.

\bibitem[Cressie and Johannesson(2008)]{cressie2008frk}
N.~Cressie and G.~Johannesson.
\newblock Fixed rank kriging for very large spatial data sets.
\newblock \emph{Journal of the Royal Statistical Society: Series B (Statistical Methodology)}, 70\penalty0 (1):\penalty0 209--226, 2008.
\newblock \doi{https://doi.org/10.1111/j.1467-9868.2007.00633.x}.
\newblock URL \url{https://rss.onlinelibrary.wiley.com/doi/abs/10.1111/j.1467-9868.2007.00633.x}.

\bibitem[Datta et~al.(2016{\natexlab{a}})Datta, Banerjee, Finley, and Gelfand]{Datta2016}
A.~Datta, S.~Banerjee, A.~O. Finley, and A.~E. Gelfand.
\newblock Hierarchical nearest-neighbor {G}aussian process models for large geostatistical datasets.
\newblock \emph{Journal of the American Statistical Association}, 111\penalty0 (514):\penalty0 800--812, 2016{\natexlab{a}}.
\newblock \doi{10.1080/01621459.2015.1044091}.
\newblock URL \url{https://doi.org/10.1080/01621459.2015.1044091}.
\newblock PMID: 29720777.

\bibitem[Datta et~al.(2016{\natexlab{b}})Datta, Banerjee, Finley, Hamm, and Schaap]{datta16b}
A.~Datta, S.~Banerjee, A.~O. Finley, N.~A.~S. Hamm, and M.~Schaap.
\newblock {Nonseparable dynamic nearest neighbor {G}aussian process models for large spatio-temporal data with an application to particulate matter analysis}.
\newblock \emph{The Annals of Applied Statistics}, 10\penalty0 (3):\penalty0 1286 -- 1316, 2016{\natexlab{b}}.

\bibitem[Dey et~al.(2022)Dey, Datta, and Banerjee]{dey2022graphical}
D.~Dey, A.~Datta, and S.~Banerjee.
\newblock Graphical {G}aussian process models for highly multivariate spatial data.
\newblock \emph{Biometrika}, 109\penalty0 (4):\penalty0 993--1014, 2022.
\newblock ISSN 1464-3510.
\newblock \doi{10.1093/biomet/asab061}.
\newblock URL \url{https://doi.org/10.1093/biomet/asab061}.

\bibitem[Eddelbuettel and Fran\c{c}ois(2011)]{Dirk2011}
D.~Eddelbuettel and R.~Fran\c{c}ois.
\newblock {Rcpp}: Seamless {R} and {C++} integration.
\newblock \emph{Journal of Statistical Software}, 2011.

\bibitem[Eleftheriadis et~al.(2017)Eleftheriadis, Nicholson, Deisenroth, and Hensman]{Eleftheriadis2017}
S.~Eleftheriadis, T.~Nicholson, M.~Deisenroth, and J.~Hensman.
\newblock Identification of {G}aussian process state space models.
\newblock \emph{Advances in neural information processing systems}, 30, 2017.

\bibitem[Fang et~al.(2005)Fang, Li, and Sudjianto]{Fang2005}
K.-T. Fang, R.~Li, and A.~Sudjianto.
\newblock \emph{Design and Modeling for Computer Experiments}.
\newblock Chapman and Hall/CRC, oct 2005.

\bibitem[Farah et~al.(2014)Farah, Birrell, Conti, and Angelis]{Farah2014}
M.~Farah, P.~Birrell, S.~Conti, and D.~D. Angelis.
\newblock Bayesian emulation and calibration of a dynamic epidemic model for a/h1n1 influenza.
\newblock \emph{Journal of the American Statistical Association}, 109\penalty0 (508):\penalty0 1398--1411, 2014.

\bibitem[Finley et~al.(2019)Finley, Datta, Cook, Morton, Andersen, and Banerjee]{finley2019efficient}
A.~O. Finley, A.~Datta, B.~C. Cook, D.~C. Morton, H.~E. Andersen, and S.~Banerjee.
\newblock {Efficient algorithms for {B}ayesian Nearest Neighbor {G}aussian Processes}.
\newblock \emph{Journal of Computational and Graphical Statistics}, 28\penalty0 (2):\penalty0 401--414, 2019.

\bibitem[Fr{\"u}hwirth-Schnatter(2006)]{fruhwirth2006}
S.~Fr{\"u}hwirth-Schnatter.
\newblock \emph{{Finite Mixture and Markov Switching Models}}.
\newblock Springer New York, 2006.

\bibitem[Gamerman and Lopes(2006)]{Gamerman2006book}
D.~Gamerman and H.~F. Lopes.
\newblock \emph{Markov chain Monte Carlo}.
\newblock Chapman \& Hall/CRC Texts in Statistical Science. Chapman \& Hall/CRC, Philadelphia, PA, 2 edition, May 2006.

\bibitem[Gamerman et~al.(2013)Gamerman, dos Santos, and Franco]{Gamerman2013}
D.~Gamerman, T.~R. dos Santos, and G.~C. Franco.
\newblock A non-{G}aussian family of state-space models with exact marginal likelihood.
\newblock \emph{Journal of Time Series Analysis}, 34\penalty0 (6):\penalty0 625--645, 2013.

\bibitem[Gel et~al.(2004)Gel, Raftery, and Gneiting]{GelEtAl2004}
Y.~Gel, A.~E. Raftery, and T.~Gneiting.
\newblock Calibrated probabilistic mesoscale weather field forecasting.
\newblock \emph{Journal of the American Statistical Association}, 99\penalty0 (467):\penalty0 575--583, 2004.
\newblock \doi{10.1198/016214504000000872}.
\newblock URL \url{https://doi.org/10.1198/016214504000000872}.

\bibitem[Gelfand and Ghosh(1998)]{Gelfand_Ghosh_1998}
A.~E. Gelfand and S.~K. Ghosh.
\newblock {Model Choice: A Minimum Posterior Predictive Loss Approach}.
\newblock \emph{Biometrika}, 85\penalty0 (1):\penalty0 1--11, June 1998.

\bibitem[Gramacy and Apley(2015)]{gram14}
R.~B. Gramacy and D.~W. Apley.
\newblock Local {G}aussian process approximation for large computer experiments.
\newblock \emph{Journal of Computational and Graphical Statistics}, 24\penalty0 (2):\penalty0 561--578, 2015.
\newblock \doi{10.1080/10618600.2014.914442}.
\newblock URL \url{http://dx.doi.org/10.1080/10618600.2014.914442}.

\bibitem[Gramacy and Lee(2008)]{Gramacy2007}
R.~B. Gramacy and H.~K.~H. Lee.
\newblock Bayesian treed {G}aussian process models with an application to computer modeling.
\newblock \emph{Journal of the American Statistical Association}, 103\penalty0 (483):\penalty0 1119--1130, 2008.

\bibitem[Gu(2022)]{gu2022robustcalibration}
M.~Gu.
\newblock Robustcalibration: Robust calibration of computer models in r.
\newblock \emph{arXiv preprint arXiv:2201.01476}, 2022.

\bibitem[Gu and Berger(2016)]{gu2016parallel}
M.~Gu and J.~O. Berger.
\newblock Parallel partial {G}aussian process emulation for computer models with massive output.
\newblock \emph{The Annals of Applied Statistics}, pages 1317--1347, 2016.

\bibitem[Gu and Li(2022)]{gu2022gaussian}
M.~Gu and H.~Li.
\newblock Gaussian orthogonal latent factor processes for large incomplete matrices of correlated data.
\newblock \emph{Bayesian Analysis}, 17\penalty0 (4):\penalty0 1219--1244, 2022.

\bibitem[Gu and Shen(2020)]{gu2020generalized}
M.~Gu and W.~Shen.
\newblock Generalized probabilistic principal component analysis of correlated data.
\newblock \emph{Journal of Machine Learning Research}, 21\penalty0 (13):\penalty0 1--41, 2020.

\bibitem[Gu et~al.(2019)Gu, Palomo, and Berger]{gu2019robustgasp}
M.~Gu, J.~Palomo, and J.~O. Berger.
\newblock Robustgasp: Robust {G}aussian stochastic process emulation in r.
\newblock \emph{R Journal}, 11\penalty0 (1), 2019.

\bibitem[Guhaniyogi and Banerjee(2018)]{guhaniyogi2018meta}
R.~Guhaniyogi and S.~Banerjee.
\newblock Meta-kriging: Scalable {B}ayesian modeling and inference for massive spatial datasets.
\newblock \emph{Technometrics}, 60\penalty0 (4):\penalty0 430--444, 2018.

\bibitem[Guhaniyogi et~al.(2017)Guhaniyogi, Li, Savitsky, and Srivastava]{guhaniyogi2017divide}
R.~Guhaniyogi, C.~Li, T.~D. Savitsky, and S.~Srivastava.
\newblock A divide-and-conquer {B}ayesian approach to large-scale kriging.
\newblock \emph{arXiv preprint arXiv:1712.09767}, 2017.

\bibitem[Gupta and Nagar(1999)]{Gupta_Nagar_1999}
A.~K. Gupta and D.~K. Nagar.
\newblock \emph{Matrix variate distributions}.
\newblock CRC Press, 1999.

\bibitem[Haylock and O'Hagan(1996)]{Haylock1996}
R.~Haylock and A.~O'Hagan.
\newblock \emph{Bayesian Statistics}, volume~5.
\newblock Oxford University Press, New York, 1996.

\bibitem[Heaton et~al.(2019)Heaton, Datta, Finley, Furrer, Guinness, Guhaniyogi, Gerber, Gramacy, Hammerling, Katzfuss, Lindgren, Nychka, Sun, and Zammit‐Mangion]{Heaton2019}
M.~J. Heaton, A.~Datta, A.~O. Finley, R.~Furrer, J.~Guinness, R.~Guhaniyogi, F.~Gerber, R.~B. Gramacy, D.~M. Hammerling, M.~Katzfuss, F.~Lindgren, D.~W. Nychka, F.~Sun, and A.~Zammit‐Mangion.
\newblock A case study competition among methods for analyzing large spatial data.
\newblock \emph{Journal of Agricultural, Biological, and Environmental Statistics}, 24\penalty0 (3):\penalty0 398--425, 2019.

\bibitem[Hewitt(1917)]{Hewitt1921}
C.~G. Hewitt.
\newblock Conservation of wild life in canada.
\newblock \emph{Nature}, 99:\penalty0 246--247, 1917.

\bibitem[Higdon et~al.(2004)Higdon, Kennedy, Cavendish, Cafeo, and Ryne]{Higdon2004}
D.~Higdon, M.~Kennedy, J.~C. Cavendish, J.~A. Cafeo, and R.~D. Ryne.
\newblock Combining field data and computer simulations for calibration and prediction.
\newblock \emph{SIAM Journal on Scientific Computing}, 26\penalty0 (2):\penalty0 448--466, 2004.

\bibitem[Higdon et~al.(2008)Higdon, Gattiker, Williams, and Rightley]{Higdon2008}
D.~Higdon, J.~Gattiker, B.~Williams, and M.~Rightley.
\newblock Computer model calibration using high-dimensional output.
\newblock \emph{Journal of the American Statistical Association}, 103\penalty0 (482):\penalty0 570--583, 2008.

\bibitem[Hirt and Dellaportas(2019)]{Hirt2019}
M.~Hirt and P.~Dellaportas.
\newblock Scalable {B}ayesian learning for state space models using variational inference with smc samplers.
\newblock In \emph{The 22nd International Conference on Artificial Intelligence and Statistics}, pages 76--86. PMLR, 2019.

\bibitem[Kang and Deng(2020)]{kangDeng2020wires}
X.~Kang and X.~Deng.
\newblock Design and analysis of computer experiments with quantitative and qualitative inputs: A selective review.
\newblock \emph{WIREs Data Mining and Knowledge Discovery}, 10\penalty0 (3):\penalty0 e1358, 2020.
\newblock \doi{https://doi.org/10.1002/widm.1358}.
\newblock URL \url{https://wires.onlinelibrary.wiley.com/doi/abs/10.1002/widm.1358}.

\bibitem[Katzfuss(2017)]{katzfussmultires}
M.~Katzfuss.
\newblock A multi-resolution approximation for massive spatial datasets.
\newblock \emph{Journal of the American Statistical Association}, 112:\penalty0 201--214, 2017.
\newblock \doi{10.1080/01621459.2015.1123632}.
\newblock URL \url{http://dx.doi.org/10.1080/01621459.2015.1123632}.

\bibitem[Katzfuss and Guinness(2021)]{katzfuss2021general}
M.~Katzfuss and J.~Guinness.
\newblock A general framework for {V}ecchia approximations of {G}aussian processes.
\newblock \emph{Statistical Science}, 36\penalty0 (1):\penalty0 124--141, 2021.
\newblock \doi{10.1214/19-STS755}.
\newblock URL \url{https://doi.org/10.1214/19-STS755}.

\bibitem[Keeling and Rohani(2008)]{keeling2008}
M.~J. Keeling and P.~Rohani.
\newblock \emph{{Modeling Infectious Diseases in Humans and Animals}}.
\newblock Princeton University Press, 2008.

\bibitem[Kennedy and O'Hagan(2001)]{Kennedy2001}
M.~C. Kennedy and A.~O'Hagan.
\newblock Bayesian calibration of computer models.
\newblock \emph{Journal of the Royal Statistical Society: Series B (Statistical Methodology)}, 63\penalty0 (3):\penalty0 425--464, 2001.

\bibitem[Krock et~al.(2023)Krock, Kleiber, Hammerling, and Becker]{Krock2021modeling}
M.~L. Krock, W.~Kleiber, D.~Hammerling, and S.~Becker.
\newblock Modeling massive highly multivariate nonstationary spatial data with the basis graphical lasso.
\newblock \emph{Journal of Computational and Graphical Statistics}, 0\penalty0 (0):\penalty0 1--16, 2023.
\newblock \doi{10.1080/10618600.2023.2174126}.
\newblock URL \url{https://doi.org/10.1080/10618600.2023.2174126}.

\bibitem[Liu and West(2009)]{LiuWest2009}
F.~Liu and M.~West.
\newblock {A Dynamic Modelling Strategy for {B}ayesian Computer Model Emulation}.
\newblock \emph{Bayesian Analysis}, 4\penalty0 (2):\penalty0 393--412, 2009.

\bibitem[Merrill et~al.(2021)Merrill, Fern, Fern, and Dolatnia]{Merrill2021}
E.~Merrill, A.~Fern, X.~Fern, and N.~Dolatnia.
\newblock An empirical study of {B}ayesian optimization: acquisition versus partition.
\newblock \emph{The Journal of Machine Learning Research}, 22\penalty0 (1):\penalty0 200--224, 2021.

\bibitem[Noble(1974)]{Nobel1974}
J.~V. Noble.
\newblock Geographic and temporal development of plagues.
\newblock \emph{Nature}, 250\penalty0 (5469):\penalty0 726--729, 1974.

\bibitem[Nychka et~al.(2015)Nychka, Bandyopadhyay, Hammerling, Lindgren, and Sain]{nychka2015}
D.~Nychka, S.~Bandyopadhyay, D.~Hammerling, F.~Lindgren, and S.~Sain.
\newblock A multiresolution {G}aussian process model for the analysis of large spatial datasets.
\newblock \emph{Journal of Computational and Graphical Statistics}, 24\penalty0 (2):\penalty0 579--599, 2015.

\bibitem[Oakley and O'Hagan(2004)]{Oakley2004}
J.~E. Oakley and A.~O'Hagan.
\newblock Probabilistic sensitivity analysis of complex models: a {B}ayesian approach.
\newblock \emph{Journal of the Royal Statistical Society Series B: Statistical Methodology}, 66\penalty0 (3):\penalty0 751--769, 2004.

\bibitem[O'Hagan(1992)]{OHagan1992}
A.~O'Hagan.
\newblock \emph{Bayesian Statistics}, volume~4.
\newblock Oxford University Press, New York, 1992.

\bibitem[Pan et~al.(2024)Pan, Zhang, Bradley, and Banerjee]{pan2024bayesianinferencespatialtemporalnongaussian}
S.~Pan, L.~Zhang, J.~R. Bradley, and S.~Banerjee.
\newblock Bayesian inference for spatial-temporal non-{G}aussian data using predictive stacking, 2024.
\newblock URL \url{https://arxiv.org/abs/2406.04655}.

\bibitem[Peruzzi et~al.(2022)Peruzzi, Banerjee, and Finley]{Peruzzi2020}
M.~Peruzzi, S.~Banerjee, and A.~Finley.
\newblock {Highly Scalable {B}ayesian Geostatistical Modeling via Meshed {G}aussian Processes on Partitioned Domains}.
\newblock \emph{Journal of the American Statistical Association}, 117\penalty0 (538):\penalty0 969--982, 2022.

\bibitem[Petris et~al.(2009)Petris, Petrone, and Campagnoli]{Petris2009}
G.~Petris, S.~Petrone, and P.~Campagnoli.
\newblock \emph{{Dynamic Linear Models with R}}.
\newblock Springer New York, 2009.

\bibitem[Poole and Raftery(2000)]{Poole2000}
D.~Poole and A.~E. Raftery.
\newblock Inference for deterministic simulation models: the {B}ayesian melding approach.
\newblock \emph{Journal of the American Statistical Association}, 95\penalty0 (452):\penalty0 1244--1255, 2000.

\bibitem[Presicce and Banerjee(2024)]{presicce2024bayesian}
L.~Presicce and S.~Banerjee.
\newblock Bayesian transfer learning for artificially intelligent geospatial systems: A predictive stacking approach.
\newblock \emph{arXiv preprint arXiv:2410.09504}, 2024.

\bibitem[Quinonero-Candela and Rasmussen(2005)]{Quinoner2005}
J.~Quinonero-Candela and C.~E. Rasmussen.
\newblock A unifying view of sparse approximate {G}aussian process regression.
\newblock \emph{The Journal of Machine Learning Research}, 6:\penalty0 1939--1959, 2005.

\bibitem[Radev et~al.(2022)Radev, Mertens, Voss, Ardizzone, and K\"{o}the]{radev2022ieee}
S.~T. Radev, U.~K. Mertens, A.~Voss, L.~Ardizzone, and U.~K\"{o}the.
\newblock Bayesflow: Learning complex stochastic models with invertible neural networks.
\newblock \emph{IEEE Tranacttions on Neural Networks ad Learning Systems}, 33\penalty0 (4):\penalty0 1452--1466, 2022.

\bibitem[Rasmussen and Williams(2005)]{GP_ML}
C.~E. Rasmussen and C.~K.~I. Williams.
\newblock \emph{{Gaussian Processes for Machine Learning (Adaptive Computation and Machine Learning)}}.
\newblock The MIT Press, 2005.

\bibitem[Rauch et~al.(1965)Rauch, Tung, and Striebel]{rauch1965maximum}
H.~E. Rauch, F.~Tung, and C.~T. Striebel.
\newblock Maximum likelihood estimates of linear dynamic systems.
\newblock \emph{AIAA journal}, 3\penalty0 (8):\penalty0 1445--1450, 1965.

\bibitem[Ren and Banerjee(2013)]{renBanerjee2013biocs}
Q.~Ren and S.~Banerjee.
\newblock Hierarchical factor models for large spatially misaligned data: A low-rank predictive process approach.
\newblock \emph{Biometrics}, 69\penalty0 (1):\penalty0 19--30, 2013.
\newblock \doi{10.1111/j.1541-0420.2012.01832.x}.
\newblock URL \url{https://onlinelibrary.wiley.com/doi/abs/10.1111/j.1541-0420.2012.01832.x}.

\bibitem[Ren et~al.(2011)Ren, Banerjee, Finley, and Hodges]{renBanerjeeEtAl2011csda}
Q.~Ren, S.~Banerjee, A.~O. Finley, and J.~S. Hodges.
\newblock Variational {B}ayesian methods for spatial data analysis.
\newblock \emph{Computational Statistics \& Data Analysis}, 55\penalty0 (12):\penalty0 3197--3217, 2011.
\newblock ISSN 0167-9473.
\newblock \doi{https://doi.org/10.1016/j.csda.2011.05.021}.
\newblock URL \url{https://www.sciencedirect.com/science/article/pii/S0167947311002003}.

\bibitem[Robert and Casella(2005)]{Robert2005}
C.~Robert and G.~Casella.
\newblock \emph{Monte Carlo statistical methods}.
\newblock Springer Texts in Statistics. Springer, New York, NY, 2 edition, Jul 2005.

\bibitem[Rue et~al.(2009)Rue, Martino, and Chopin]{Rue2009}
H.~Rue, S.~Martino, and N.~Chopin.
\newblock Approximate {B}ayesian inference for latent {G}aussian models by using integrated nested {L}aplace approximations.
\newblock \emph{Journal of the Royal Statistical Society: Series B (Statistical Methodology)}, 71\penalty0 (2):\penalty0 319--392, 2009.

\bibitem[Sacks et~al.(1989)Sacks, Welch, Mitchell, and Wynn]{Sacks1989}
J.~Sacks, W.~J. Welch, T.~J. Mitchell, and H.~P. Wynn.
\newblock {Design and Analysis of Computer Experiments}.
\newblock \emph{Statistical Science}, 4\penalty0 (4):\penalty0 409 -- 423, 1989.
\newblock \doi{10.1214/ss/1177012413}.
\newblock URL \url{https://doi.org/10.1214/ss/1177012413}.

\bibitem[Sainsbury-Dale et~al.(2024)Sainsbury-Dale, Zammit-Mangion, and Huser]{sainsbury2024likelihoodfree}
M.~Sainsbury-Dale, A.~Zammit-Mangion, and R.~Huser.
\newblock Likelihood-free parameter estimation with neural bayes estimators.
\newblock \emph{The American Statistician}, 78\penalty0 (1):\penalty0 1--14, 2024.

\bibitem[Santner et~al.(2019)Santner, Williams, and Notz]{Santner2019}
T.~J. Santner, B.~J. Williams, and W.~I. Notz.
\newblock \emph{The design and analysis of computer experiments the design and analysis of computer experiments}.
\newblock Springer series in statistics. Springer, New York, NY, 2 edition, Jan. 2019.

\bibitem[S{\"a}rkk{\"a} and Svensson(2013)]{Sarkka2013}
S.~S{\"a}rkk{\"a} and L.~Svensson.
\newblock \emph{Bayesian filtering and smoothing}, volume~17.
\newblock Cambridge university press, 2013.

\bibitem[Sauer et~al.(2023{\natexlab{a}})Sauer, Cooper, and Gramacy]{gramacy2022deepGP}
A.~Sauer, A.~Cooper, and R.~B. Gramacy.
\newblock Vecchia-approximated deep {G}aussian processes for computer experiments.
\newblock \emph{Journal of Computational and Graphical Statistics}, 32\penalty0 (3):\penalty0 824--837, 2023{\natexlab{a}}.
\newblock \doi{10.1080/10618600.2022.2129662}.
\newblock URL \url{https://doi.org/10.1080/10618600.2022.2129662}.

\bibitem[Sauer et~al.(2023{\natexlab{b}})Sauer, Gramacy, and Higdon]{gramacy2021deepGPactiveLearning}
A.~Sauer, R.~B. Gramacy, and D.~Higdon.
\newblock Active learning for deep {G}aussian process surrogates.
\newblock \emph{Technometrics}, 65\penalty0 (1):\penalty0 4--18, 2023{\natexlab{b}}.
\newblock \doi{10.1080/00401706.2021.2008505}.
\newblock URL \url{https://doi.org/10.1080/00401706.2021.2008505}.

\bibitem[Siew(2019)]{Siew2019}
C.~Siew.
\newblock {spreadr: An R package to simulate spreading activation in a network}.
\newblock \emph{Behavior Research Methods}, 2019.

\bibitem[Snelson and Ghahramani(2005)]{snelson2005sparse}
E.~Snelson and Z.~Ghahramani.
\newblock Sparse {G}aussian processes using pseudo-inputs.
\newblock \emph{Advances in neural information processing systems}, 18, 2005.

\bibitem[Soetaert et~al.(2010)Soetaert, Petzoldt, and Setzer]{Karline2010}
K.~Soetaert, T.~Petzoldt, and R.~W. Setzer.
\newblock Solving di erential equations in r: Package desolve.
\newblock \emph{Journal of Statistical Software}, 33\penalty0 (9), 2010.

\bibitem[Stroud et~al.(2010)Stroud, Stein, Barry M.~Lesht, and Beletsky]{Stroud2010}
J.~R. Stroud, M.~L. Stein, D.~J.~S. Barry M.~Lesht, and D.~Beletsky.
\newblock An ensemble kalman filter and smoother for satellite data assimilation.
\newblock \emph{Journal of the American Statistical Association}, 105\penalty0 (491):\penalty0 978--990, 2010.
\newblock \doi{10.1198/jasa.2010.ap07636}.
\newblock URL \url{https://doi.org/10.1198/jasa.2010.ap07636}.

\bibitem[Turner et~al.(2010)Turner, Deisenroth, and Rasmussen]{Turner2010}
R.~Turner, M.~Deisenroth, and C.~Rasmussen.
\newblock {State-Space Inference and Learning with {G}aussian Processes}.
\newblock In Y.~W. Teh and M.~Titterington, editors, \emph{Proceedings of the Thirteenth International Conference on Artificial Intelligence and Statistics}, Proceedings of Machine Learning Research, 2010.

\bibitem[Vega~Yon and Valente(2021)]{netDiffuse}
G.~Vega~Yon and T.~Valente.
\newblock netdiffuser: Analysis of diffusion and contagion processes on networks.
\newblock \emph{R package version}, 1\penalty0 (3), 2021.

\bibitem[Vehtari et~al.(2017)Vehtari, Gelman, and Gabry]{Vehtari_Gelman_Gabry_2017}
A.~Vehtari, A.~Gelman, and J.~Gabry.
\newblock {Practical {B}ayesian model evaluation using leave-one-out cross-validation and WAIC}.
\newblock \emph{Statistical Computing}, 27:\penalty0 1413--1432, 2017.

\bibitem[Venkitaraman et~al.(2020)Venkitaraman, Chatterjee, and Handel]{Venkitaraman2020}
A.~Venkitaraman, S.~Chatterjee, and P.~Handel.
\newblock Gaussian processes over graphs.
\newblock In \emph{ICASSP 2020-2020 IEEE International Conference on Acoustics, Speech and Signal Processing (ICASSP)}, pages 5640--5644. IEEE, 2020.

\bibitem[Vitevitch et~al.(2011)Vitevitch, Ercal, and Adagarla]{Vitevitch2011}
M.~Vitevitch, G.~Ercal, and B.~Adagarla.
\newblock {Simulating Retrieval from a Highly Clustered Network: Implications for Spoken Word Recognition}.
\newblock \emph{Frontiers in psychology}, 2011.

\bibitem[Wang et~al.(2021)Wang, Cockayne, Chkrebtii, Sullivan, and Oates]{Wang2021}
J.~Wang, J.~Cockayne, O.~Chkrebtii, T.~Sullivan, and C.~Oates.
\newblock Bayesian numerical methods for nonlinear partial differential equations.
\newblock \emph{Statistics and Computing}, 2021.

\bibitem[Watts and Strogatz(1998)]{WattsStrogatz1998}
D.~J. Watts and S.~H. Strogatz.
\newblock Collective dynamics of ‘small-world’networks.
\newblock \emph{Nature}, 393\penalty0 (6684):\penalty0 440--442, 1998.

\bibitem[Welch et~al.(1992)Welch, Buck, Sacks, Wynn, Mitchell, and Morris]{Welch1992}
W.~J. Welch, R.~J. Buck, J.~Sacks, H.~P. Wynn, T.~J. Mitchell, and M.~D. Morris.
\newblock Screening, predicting, and computer experiments.
\newblock \emph{Technometrics}, 34\penalty0 (1):\penalty0 15--25, 1992.

\bibitem[West and Harrison(1997)]{HarrisonWest1997}
M.~West and J.~Harrison.
\newblock \emph{{Bayesian Forecasting and Dynamic Models (2nd Ed.)}}.
\newblock Springer-Verlag, 1997.

\bibitem[West et~al.(1985)West, Harrison, and Migon]{West1985}
M.~West, P.~J. Harrison, and H.~S. Migon.
\newblock Dynamic generalized linear models and {B}ayesian forecasting.
\newblock \emph{Journal of the American Statistical Association}, 80\penalty0 (389):\penalty0 73--83, 1985.

\bibitem[Wikle(2003)]{wikle2003hierarchical}
C.~K. Wikle.
\newblock Hierarchical {B}ayesian models for predicting the spread of ecological processes.
\newblock \emph{Ecology}, 84\penalty0 (6):\penalty0 1382--1394, 2003.

\bibitem[Wikle(2010)]{Wikle2011}
C.~K. Wikle.
\newblock Low-rank representations for spatial processes.
\newblock \emph{Handbook of Spatial Statistics}, pages 107--118, 2010.
\newblock Gelfand, A. E., Diggle, P., Fuentes, M. and Guttorp, P., editors, Chapman and Hall/CRC, pp. 107-118.

\bibitem[Wikle and Berliner(2007)]{WikleBerliner2007}
C.~K. Wikle and L.~M. Berliner.
\newblock A {B}ayesian tutorial for data assimilation.
\newblock \emph{Physica D: Nonlinear Phenomena}, 230\penalty0 (1):\penalty0 1--16, 2007.
\newblock ISSN 0167-2789.
\newblock \doi{https://doi.org/10.1016/j.physd.2006.09.017}.
\newblock URL \url{https://www.sciencedirect.com/science/article/pii/S016727890600354X}.
\newblock Data Assimilation.

\bibitem[Wikle and Hooten(2010)]{WikleHooten2010}
C.~K. Wikle and M.~B. Hooten.
\newblock A general science-based framework for dynamical spatio-temporal models.
\newblock \emph{TEST}, 19\penalty0 (3):\penalty0 417--451, Nov 2010.
\newblock ISSN 1863-8260.
\newblock \doi{10.1007/s11749-010-0209-z}.
\newblock URL \url{https://doi.org/10.1007/s11749-010-0209-z}.

\bibitem[Wilson and Nickisch(2015)]{wilson2015kernel}
A.~Wilson and H.~Nickisch.
\newblock Kernel interpolation for scalable structured {G}aussian processes (kiss-gp).
\newblock In \emph{International conference on machine learning}, pages 1775--1784. PMLR, 2015.

\bibitem[Zammit-Mangion et~al.(2024)Zammit-Mangion, Sainsbury-Dale, and Huser]{zammit2024amortized}
A.~Zammit-Mangion, M.~Sainsbury-Dale, and R.~Huser.
\newblock Neural methods for amortized inference.
\newblock \emph{Annual Review of Statistics and Its Application}, 12, 2024.

\bibitem[Zhang and Banerjee(2022)]{zhangbanerjee2021}
L.~Zhang and S.~Banerjee.
\newblock {Spatial factor modeling: A {B}ayesian matrix‐normal approach for misaligned data}.
\newblock \emph{Biometrics}, 78\penalty0 (2):\penalty0 560--573, June 2022.

\bibitem[Zhang et~al.(2019)Zhang, Datta, and Banerjee]{zhang2019practical}
L.~Zhang, A.~Datta, and S.~Banerjee.
\newblock Practical {B}ayesian modeling and inference for massive spatial data sets on modest computing environments.
\newblock \emph{Statistical Analysis and Data Mining: The ASA Data Science Journal}, 12\penalty0 (3):\penalty0 197--209, 2019.

\bibitem[Zhang et~al.(2024)Zhang, Tang, and Banerjee]{zhang2024geostacking}
L.~Zhang, W.~Tang, and S.~Banerjee.
\newblock Bayesian geostatistics using predictive stacking.
\newblock \emph{arXiv:2304.12414v2}, 2024.
\newblock URL \url{https://arxiv.org/abs/2304.12414}.

\bibitem[Zhang and Wang(2014)]{Zhang2014}
T.~Zhang and W.~Wang.
\newblock Existence of traveling wave solutions for influenza model with treatment.
\newblock \emph{Journal of Mathematical Analysis and Applications}, 419\penalty0 (1):\penalty0 469--495, 2014.

\bibitem[Zhu et~al.(2007)Zhu, Yu, and Gong]{NIPS2007_061412e4}
S.~Zhu, K.~Yu, and Y.~Gong.
\newblock Predictive matrix-variate t models.
\newblock In J.~Platt, D.~Koller, Y.~Singer, and S.~Roweis, editors, \emph{Advances in Neural Information Processing Systems}, volume~20. Curran Associates, Inc., 2007.
\newblock URL \url{https://proceedings.neurips.cc/paper_files/paper/2007/file/061412e4a03c02f9902576ec55ebbe77-Paper.pdf}.

\end{thebibliography}
\end{document}